\documentclass[12pt,preprint,longnamesfirst]{aastex}

\newcommand{\kms}{\ifmmode \mathrm{km~s^{-1}}\else km~s$^{-1}$\fi}
\newcommand{\smpy}{\ifmmode M_\sun~\mathrm{yr}^{-1}\else M$_\sun$~yr$^{-1}$\fi}
\newcommand{\lir}{\ifmmode L_\mathrm{IR}\else $L_\mathrm{IR}$\fi}
\newcommand{\lsun}{\ifmmode L_\sun\else $L_\sun$\fi}
\newcommand{\msun}{\ifmmode M_\sun\else $M_\sun$\fi}
\newcommand{\nags}{\ion{Na}{1}}
\newcommand{\nad}{\ion{Na}{1}~D}

\newcommand{\ot}{[\ion{O}{3}]}
\newcommand{\otl}{[\ion{O}{3}] $\lambda5007$}
\newcommand{\nt}{[\ion{N}{2}]}
\newcommand{\ntl}{[\ion{N}{2}] $\lambda6583$}
\newcommand{\ntll}{[\ion{N}{2}] $\lambda\lambda$6548,~6583}
\newcommand{\cf}{\ifmmode C_f\else $C_f$\fi}
\newcommand{\co}{\ifmmode C_\Omega\else $C_\Omega$\fi}
\newcommand{\dvmax}{\ifmmode \Delta v_{max}\else $\Delta v_{max}$\fi}
\newcommand{\dvtau}{\ifmmode \Delta v_{maxN}\else $\Delta v_{maxN}$\fi}

\shorttitle{Outflows in AGN ULIRGs}
\shortauthors{Rupke, Veilleux, \& Sanders}

\begin{document}

\title{Outflows in AGN/Starburst-Composite Ultraluminous Infrared Galaxies\footnotemark[1] \footnotemark[2] \footnotemark[3]}

\author{David S. Rupke, Sylvain Veilleux}
\affil{Department of Astronomy, University of Maryland, College Park, MD 20742}
\email{drupke@astro.umd.edu, veilleux@astro.umd.edu}
\and
\author{D.~B. Sanders}
\affil{Institute for Astronomy, University of Hawaii, 2680 Woodlawn Drive, Honolulu, HI 96822}
\email{sanders@ifa.hawaii.edu}
\footnotetext[1]{Some of the observations reported here were obtained at the W. M. Keck Observatory, which is operated as a scientific partnership among the California Institute of Technology, the University of California, and the National Aeronautics and Space Administration. The Observatory was made possible by the generous financial support of the W. M. Keck Foundation.}
\footnotetext[2]{Some of the observations reported here were obtained at the MMT Observatory, which is a joint facility of the Smithsonian Institution and the University of Arizona.}
\footnotetext[3]{Some of the observations reported here were obtained at the Kitt Peak National Observatory, National Optical Astronomy Observatory, which is operated by the Association of Universities for Research in Astronomy, Inc. (AURA) under cooperative agreement with the National Science Foundation.}

\begin{abstract}

  Galactic superwinds  occur in almost  all infrared-luminous galaxies
  with star formation rates (SFRs) above 10~\smpy, as shown by studies
  of the \nad\ interstellar absorption line.  We demonstrate that this
  result  also  applies to  ultraluminous  infrared galaxies  (ULIRGs)
  which host  an active  galactic nucleus (AGN)  embedded in  a strong
  starburst (SFR $\ga$  100 \smpy) by studying a  sample of 26 Seyfert
  ULIRGs  in  \nad.  The  infrared  luminosity  of  these galaxies  is
  powered jointly  by the AGN and  starburst.  We find  that there are
  hints of the influence of the AGN on outflows in Seyfert~2/starburst
  composites, but  the evidence  is not yet  statistically conclusive.
  The  evidence we  find is  lower  wind detection  rates (i.e.,  wind
  opening angles)  in Seyfert~2 ULIRGs than in  galaxies of comparable
  \lir,  higher velocities  than in  galaxies of  comparable  SFR, and
  correlations  between the  neutral gas  and the  ionized gas  in the
  extended narrow-line region.  Though the AGN probably contributes to
  the outflows in Seyfert~2  ULIRGs, its momentum and energy injection
  is  equal to or  less than  that of  the starburst.   Similarly, the
  outflow  mechanical  luminosity   (energy  outflow  rate)  per  unit
  radiative luminosity is the same for starburst and Seyfert~2 ULIRGs.
  In the  nuclei of Seyfert~1s,  we observe small-scale  outflows that
  are powered solely by the AGN.  However, in Mrk~231, we observe both
  a  high-velocity, small-scale  and  low-velocity, extended  outflow.
  The  latter may  be  powered by  a  starburst or  radio jet.   These
  large-scale,  lower-velocity  outflows   certainly  exist  in  other
  Seyfert~1  ULIRGs, but  they  are washed  out  by the  light of  the
  nucleus.

\end{abstract}

\keywords{galaxies: Seyfert --- galaxies: active --- galaxies: absorption lines --- infrared: galaxies --- ISM: jets and outflows --- ISM: kinematics and dynamics}

\section{INTRODUCTION} \label{intro}

Mechanical and radiative feedback from active galactic nuclei (AGN) are beginning to gain attention as important astrophysical processes.  By definition, active galactic nuclei have strong radiation fields.  They typically possess outflows in the form of spatially-resolved jets on parsec to kiloparsec scales, in both radio-loud \citep[e.g.,][]{z97,wb04} and radio-quiet \citep[e.g.,][]{m_ea99,nfw05} objects.  UV and X-ray absorption-line probes also point to wide-angle outflows originating from AGN on scales of tens of pc or less \citep[][and references therein]{ckg03,ck05,myr05}.  This radiation and outflowing gas deposit mass and energy into the surroundings of the AGN, including the ISM of the galaxy in which the AGN resides and the larger-scale intergalactic and intracluster media.

The possible effects of AGN feedback are numerous; we here list a few important ones.  Radiation from quasars almost certainly helped to reionize the universe at high redshift, though the strength of their contribution is yet uncertain \citep{f_ea04}.  Outflows from AGN have recently been invoked as global heat (or entropy) sources for the intracluster medium (ICM) at both small and large radii \citep[e.g.,][and references therein]{rhb02,r_ea04}.  On more local scales, jets from AGN at both low and high redshift have been shown to induce star formation by shocking dense clouds of gas \citep[e.g.,][]{b_ea00,m_ea00,r_ea02,fm_ea04}.  Very recently, \citet{sdh05} have suggested that AGN feedback can quench star formation in massive, gas-rich mergers, creating a population of very red ellipticals and possibly explaining the observed bimodal color distribution observed in deep galaxy surveys \citep[e.g.,][]{k_ea03,w_ea05}.

This work will focus on wide-angle outflows (though this will also lead us to some discussion of jets).  By `wide-angle outflows', we mean outflows that subtend large solid angles as seen from the wind's origin.  They are thus structurally distinct from jets, in that jets are highly collimated.  The narrow-line regions (NLRs) and host galaxies of AGN often host large-scale, wide-angle outflows.  (For a recent summary, see \citealt{vcb05}.)  Radio continuum, optical emission-line, and X-ray emission probes reveal extended structures and kinematic evidence for these types of outflows in many local Seyfert galaxies.  A number of surveys have shown that outflows are common, though they disagree on some of the details and on whether or not the AGN drives the outflow if a starburst is also present (\citealt{b_ea93,c_ea96a,c_ea96b,cb_ea98,lwh01a,lwh01b}).  More intensive studies of individual objects are quite numerous (e.g., \citealt{cbt90} [NGC~1068]; \citealt{l99,v_ea03} [NGC~1365]; \citealt{vsm01,gam01} [NGC~2992]; \citealt{v_ea99a,i_ea03} [NGC~4388]).  The scales of these structures often extend outside of the NLR (i.e., $r \ga 1$ kpc) and are comparable to those found in starburst galaxies.

However, large-scale, wide-angle winds in Seyferts are not identical to those found in starbursts.  For instance, the large-scale radio structures in Seyferts do not typically line up with the galaxy's minor axis \citep{c_ea96b}.  However, neither do they align with the nuclear `linear' radio structure \citep{b_ea93}.  Since the large-scale structures are wide-angle winds, and the nuclear structures are jet-like, this implies that the jets from the supermassive black hole accretion disk do not directly drive the large-scale winds (though they may indirectly by injecting energy into the ISM; e.g., \citealt{s85}).  Furthermore, the X-ray emission in these galaxies is thermal, rather than non-thermal synchrotron as expected in a relativistic plasma that could accompany an AGN wind or jet \citep{cb_ea98,lwh01a,lwh01b}.

Different authors have drawn different conclusions about whether or not the AGN powers the large-scale winds in Seyfert~2s.  \citet{lwh01a,lwh01b} argue for a starburst origin from X-ray imaging and spectroscopy, as do \citet{b_ea93} from their radio survey.  However, the former sample contains only Seyfert/starburst composites.  In a less biased, multi-wavelength survey, \citet{c_ea96a,c_ea96b,cb_ea98} argue for AGN jet-heating of the galaxy ISM as the driving force of the outflow.  AGN influence is also invoked in various individual cases (see above references for individual galaxies).

In galaxies containing both a starburst and a Seyfert nucleus, the confusion over the source of the outflow arises in the fact that large-scale, wide-angle outflows are also common features of starbursting galaxies.  Outflows of neutral gas and dust stretching over kiloparsec scales are found in almost all massive, starbursting galaxies in the local universe (\citealt{rvs05a,rvs05b}; hereafter Papers I and II).  These outflows reach projected velocities up to 600~\kms\ (and $>$1000~\kms\ in one galaxy), though most of the gas is at lower velocities ($100-200$~\kms).  A large fraction (43/78) of the galaxies we studied in Papers I and II were starburst-dominated ultraluminous infrared galaxies (ULIRGs), which have total infrared (and bolometric) luminosities greater than $10^{12}$~\lsun.  Many ULIRGs also host a strong AGN along with a strong starburst.  Mid-infrared spectroscopy with the {\it Infrared Space Observatory} ({\it ISO}) suggests that these AGN account for $60-80\%$ of the galaxies' bolometric luminosities \citep{lvg99}.

These starburst/Seyfert composites are the focus of the current paper.  Based on our study of pure starbursts in Papers I and II, we hypothesize that most Seyfert ULIRGs, with star formation rates (SFRs) greater than 100 \smpy, will also host starburst-driven winds.  However, the dominant AGN in these galaxies could have a measurable impact on the outflow.  By comparing the properties of outflows in Seyfert~1s and 2s with those in starburst-dominated ULIRGs, we can search for evidence that these outflows are partially powered by the central AGN.

The 26-galaxy sample we discuss in this paper consists of 17 Seyfert~2 ULIRGs, 3 Seyfert~2 LIRGs (with $10^{11} < \lir/\lsun < 10^{12}$), and 6 Seyfert~1 ULIRGs observed in the \nad\ $\lambda\lambda5890,~5896$ doublet absorption feature at moderately high spectral resolution ($65-85$~\kms).  This feature probes neutral gas, due to its low ionization potential (5.1~eV).  For $z < 0.5$, it is found in the optical.  Its high interstellar abundance makes it a good probe of the ISM.  Blueshifted velocity components in \nad\ unambiguously indicate the presence of outflowing gas.

The organization of this paper is as follows.  In \S\ref{sampobs}, we discuss our sample selection and observations and summarize the data analysis procedures.  We discuss the outflow properties of Seyfert~2 ULIRGs in \S\ref{sy2} and compare to the properties of outflows in infrared-luminous starbursts.  We also study correlations between emission and absorption lines.  In \S\ref{sy1} we discuss outflows in Seyfert~1 ULIRGs.  Mrk~231 is presented as a special case in \S\ref{mrk231}, since it possesses both a small-scale and large-scale wind.  We discuss further issues in \S\ref{discuss}, including the location of the absorbing gas, outflow/galaxy correlations, the global covering factor of the outflows, the gas escape fraction, and the scientific context.  \S\ref{summary} summarizes and concludes.

For all calculations, we assume present-day values for the cosmological parameters of $H_0 = 75$~\kms~Mpc$^{-1}$ and the standard $\Omega_m = 0.3$, $\Omega_{\Lambda} = 0.7$ cosmology.  All wavelengths quoted are vacuum wavelengths (except those used as labels for spectral lines) and are generally taken from the NIST Atomic Spectra Database\footnote{\texttt{http://physics.nist.gov/cgi-bin/AtData/main\_asd}}.  (The vacuum wavelengths of \nad\ are 5891.58 and 5897.55~\AA.)

\section{SAMPLE, OBSERVATIONS, AND ANALYSIS} \label{sampobs}

\subsection{Sample}

The primary selection criteria for our sample are that the galaxies be optically classified as Seyfert galaxies and have high infrared luminosities ($\lir\ga10^{12}\lsun$).  Our sample of infrared-selected Seyferts consists of 23 ULIRGs and 3 LIRGs.  The only redshift criterion is $z < 0.50$, so that \nad\ is not redshifted into the near-infrared and the galaxy is not too faint.  All but four galaxies have redshifts less than 0.25; the others have redshifts in the range $0.25 < z < 0.50$.

Most galaxies are selected from 1~Jy survey \citep{ks98}, which is a complete flux-limited sample of local ULIRGs observed by the {\it Infrared Astronomical Satellite} ({\it IRAS}).  Spectral types are available from low-resolution spectra \citep{vks99b}.  Objects were chosen from this catalog purely on the basis of observability.  For Papers I and II we also observed galaxies from the FIRST/FSC sample \citep{ssvd00}; by selection, these galaxies have been detected in both the infrared and the radio.  They also have faint infrared fluxes.  We selected galaxies from this sample that were neither too compact (to decrease the chances of observing a Seyfert~1) nor too faint (based on their $K$-band fluxes).  We classified several of these galaxies as Seyfert~2s (see below) and thus include them in the present sample.  One object (F04210$+$0401, a galaxy with a known radio jet) is neither a 1~Jy or FIRST/FSC source; it is, however, infrared-luminous.

Table~\ref{objprop} lists the basic properties of the galaxies in our sample.  These properties are measured or taken from other sources as described in Paper~I (see also the table caption).

\subsection{Observations} \label{obs}

Our observations were obtained during several observing runs at three different telescopes.  The observing runs, exposure times, and slit position angles are listed for each object in Table~\ref{objprop}.

Several faint galaxies were observed with the Echellette Spectrograph and Imager (ESI; \citealt{s_ea02}) on Keck II.  Using echellette mode, we covered the wavelength range $4000-11000$ \AA\ in one exposure at a spectral resolution of 65~\kms\ FWHM.  Data were obtained for two targets with the Red Channel Spectrograph on the MMT, also with an echellette grating.  We used a different setup for each run; during the 2002 December run, we covered the wavelengths $4500-10200$ \AA, and during the 2003 June run, we covered the wavelengths $4300-8800$ \AA.  In each case we used a 1$\arcsec$ slit to achieve a resolution of 87~\kms.

Much of our data were obtained at the Kitt Peak 4m using the R-C Spectrograph and a moderate-resolution grating (the KPC-18C, in first order).  With this spectrograph we observed 1700 \AA\ in one exposure, allowing us to obtain both the \nad\ line and the H$\alpha$/\nt\ complex at once.  We used the GG-475 or GG-495 blocking filters on the R-C Spectrograph and obtained an average resolution of 85~\kms\ with a 1$\farcs$25 slit.

\subsection{Star Formation Rates} \label{sfr}

The galaxies in our sample possess both an AGN and a strong starburst.  For comparison with infrared-luminous galaxies that only contain a strong starburst, we compute the star formation rates of the Seyfert ULIRGs.  To do so, we modify the relation between infrared luminosity and star formation rate \citep{k98} to include a factor $\alpha$ that is equal to the fraction of the infrared luminosity powered by star formation:
\begin{equation}
\mathrm{SFR}= \alpha~\frac{\lir}{5.8\times10^9~\lsun}.
\end{equation}
We recognize that $\alpha$ is not a well-constrained quantity, either globally or for individual sources.  However, infrared spectroscopy suggests that the AGN is dominant in Seyfert ULIRGs \citep{g_ea98,lvg99}.  Based on these observations, we assign different values of $\alpha$ for different subsets of galaxies.  For the Seyfert~1 galaxies in our sample, and a few Seyfert~2s that show either broad lines in the near-infrared or faint broad lines in the optical (see the next section), we assume $\alpha = 0.3$.  For the rest of the Seyfert~2s, we assume $\alpha = 0.4$.  For reference, we assumed $\alpha = 0.8$ for starburst ULIRGs and $\alpha = 1$ for starbursts with log$[\lir/\lsun] = 10-12$ (which we label IRGs, or infrared galaxies; Paper~II).  Small increases in $\alpha$ ($\la$50\%) for the Seyfert~2s will not change the conclusions of this study.

\subsection{Spectral Types} \label{spectype}

Each of the galaxies in our sample has a Seyfert~1 or 2 optical spectral classification.  These classifications are based on low-dispersion spectroscopy \citep{kvs98,vks99b} or on our spectra.  In confirmation of this classification, we note that four of the galaxies in this subsample were observed in the mid-infrared with {\it ISO} and are classified as AGN \citep{lvg99}.  Five to ten Seyfert~2 galaxies in our sample also show evidence for an AGN in the near-infrared, either by the presence of a broad line in Pa$\alpha$ or strong [\ion{Si}{6}] 1.962~\micron\ emission \citep{vsk99c}.

Brief comments are in order regarding the spectral types of a few galaxies.  F08526$+$3720 possesses a faint broad-line region (BLR) in H$\alpha$ but not H$\beta$, suggesting it is a Seyfert~1.9 galaxy.  We also observe a faint BLR in H$\alpha$ in F04210$+$0401 (a galaxy with a large radio jet), though the galaxy is classified as a Seyfert~2 in the literature.  We put these galaxies in the Seyfert~2 category despite the visible broad lines.

A final special case is F05189$-$2524.  This optical Seyfert~2 has an optically-obscured BLR and has been shown to host a dominant AGN \citep{vsk99c}.  Furthermore, the strength of its 7.7 \micron\ PAH feature \citep{l_ea00} is identical to that of Mrk~231 \citep{g_ea98}, an optically-classified Seyfert~1.  However, for the purposes of this paper we keep the optical spectral type of Seyfert~2.

In all, our sample contains six bona-fide Seyfert~1s, two Seyfert~1.9s with faint BLRs in H$\alpha$ (which we label Seyfert~2s in this paper), and eighteen other Seyfert~2s.

Three of these Seyferts have double nuclei, and one is a triple nucleus system (F13443$+$0802).  In each case, only one of the nuclei is a Seyfert~1 or 2, and we have derived spectral classes for the other nuclei when the \otl\ line is available.  For simplicity, we subsume the absorption properties of these non-Seyfert nuclei into our analysis.

\subsection{Analysis} \label{analysis}

Paper~I  describes the details  of the  spectral extraction  and \nad\
fitting procedure.  From the spectra, we extract as much of the galaxy
continuum light  as possible,  stopping when further  extraction harms
the signal-to-noise  ratio of the extracted spectra.   We fit multiple
velocity components  to the \nad\  feature in each object.   We assume
Gaussians   in  optical  depth,   which  translates   into  observable
non-Gaussian intensity profiles for optical depths greater than unity.
We also fit a constant  covering fraction for each velocity component.
The results of the fitting are listed in Table~\ref{compprop}.

A  wind `detection'  is a  velocity  component with  $\Delta v  \equiv
v_{comp} - v_{sys}  < -50$.  We also require  a 2$\sigma$ threshold in
the measurement  uncertainty, such that $|\Delta  v| > 2~\delta(\Delta
v)$; this  excludes only  one source (F17179$+$5444).   These criteria
are chosen because of uncertainty in both the outflow velocity and the
galaxy's systemic velocity.  We compute the `maximum' velocity in each
galaxy, \dvmax, equal to the  central velocity of the most blueshifted
component  plus  one-half its  full-width  at  half-maximum.  We  also
compute  the central  velocity of  the outflowing  component  with the
highest column density, \dvtau.

Our physical  model is  described in detail  in Papers~I and  II.  For
each galaxy, we compute hydrogen column densities from those of \nags\
assuming  standard Galactic  depletion and  an ionization  fraction of
0.90.  Rather than assume  solar metallicity, we use the near-infrared
luminosity-metallicity  relation \citep{s_ea05};  this  results in  an
average metallicity  in our sample of  two times solar.   (Most of the
$K^\prime$-band  magnitudes  used   to  compute  the  metallicity  are
corrected  for nuclear  point source  contribution, and  thus  are not
contaminated  by  the  AGN   \citep{vks02}.)   A  simple  model  of  a
mass-conserving free wind is used  to estimate the gas mass, momentum,
and energy in the wind, as well as their outflow rates.

There are two differences  between the starburst and Seyfert analyses.
One  is the  assumed global  covering factor,  $\Omega/4\pi$.   In our
model, we assume  $\Omega/4\pi = D~C_f$ \citep[Paper~II and][]{ckg03},
where D is the detection rate for the sample and $C_f$ is the covering
fraction for  each velocity component.   For the Seyferts, $D  = 0.5$,
which differs  from the values of  0.4 and 0.8 used  for the starburst
IRGs and  ULIRGs, respectively.  The second difference  is the assumed
radius of the absorbers.  For the Seyfert~2s, we use the same value of
$r = 5$ kpc as for the starbursts.  However, we assume $r = 10$ pc for
the Seyfert~1s.  See \S\S\ref{radius} and \ref{gcf} for more details.

\section{OUTFLOWS IN SEYFERT 2 ULIRGS} \label{sy2}

\subsection{Outflow Properties} \label{sy2_ofprop}

In Figure~\ref{spec_s2}, we display the \nad\ lines in the 20 Seyfert~2 galaxies in our sample with our fits to the lines superimposed.  Tables~\ref{compprop} and \ref{ofprop} list the measured properties of each \nad\ velocity component and of the outflow in each galaxy, respectively.  Table~\ref{avgprop} lists the average outflow properties in this subsample alongside those of starburst IRGs and $z < 0.25$ ULIRGs from Paper~II.  (We ignore the 13 ULIRGs from Paper~II with $0.25 < z < 0.50$, since they are typically higher in luminosity than the other ULIRGs and may have slightly different properties.)

\subsubsection{Detection Rate} \label{sy2_detrate}

We detect absorbers blueshifted by more than 50~\kms\ in half of Seyfert~2s ($45\pm11$\%).  As we discuss in \S\ref{gcf}, this detection rate reflects the geometry of the outflows, and thus most or all Seyfert~2 ULIRGs contain large-scale, wide-angle outflows.

Other studies produce widely varying detection rates.  Applying our threshold to the sample of \citet{hlsa00}, there are winds in 3 of 6 Seyfert~2s, a result identical to ours.  Optical investigations show a detection rate of $\ga$25\%\ \citep{c_ea96a}, while radio imaging studies show much higher incidences ($\sim$60$-$90\%; \citealt{b_ea93,c_ea96b}).  Low-resolution X-ray images of starburst/Seyfert~2 composites also show evidence of extended emission in most galaxies \citep{lwh01a,lwh01b}.  If outflows really do occur in all Seyfert~2 ULIRGs, these differences are attributable to the different phases of the ISM that these studies probe, as well as their varying sensitivities.

The median star formation rate of our Seyfert~2 ULIRGs (120~\smpy) is in between those of the IRG subsample (40~\smpy) and low-$z$ starburst ULIRG subsample (225~\smpy) from Papers I and II.  The detection rate in Seyfert~2s is much closer to that of the IRGs, however (43\%, 45\%, and 80\% for the IRGs, Seyfert~2 ULIRGs, and low-$z$ starburst ULIRGs, respectively, each with a 10\% error).

\subsubsection{Velocities}

In Seyfert~2s, the median velocities of the highest column density gas, \dvtau, and maximum velocity, $\dvmax \equiv \Delta v - \mathrm{FWHM}/2$ (computed for the most blueshifted component), are 220 and 456~\kms, respectively.  There is one Seyfert~2 with \dvtau\ and \dvmax\ $>$ 1000~\kms: F05024$-$1941.

Tables~\ref{avgprop} and \ref{avgprop_hls} compare average properties of outflows in Seyfert~2s with those in starburst galaxies from Paper~II.  In Table~\ref{avgprop}, we classify the galaxies according to SFR and AGN activity.  In Table~\ref{avgprop_hls}, we subdivide by spectral type.  (The Seyfert~2 velocities in Table~\ref{avgprop} are different from those in Table~\ref{avgprop_hls} because in the former we list galaxy properties, which may contain non-Seyfert nuclei, while in the latter we limit ourselves to Seyfert nuclei.)

Our subsamples reveal a hierarchy in the median values of \dvmax\ and \dvtau, such that as one moves from IRGs to low-$z$ ULIRGs to Seyfert~2s, or from \ion{H}{2} galaxies to LINERs to Seyfert~2s, the velocities and variances increase.  In Figures~\ref{histdv} and \ref{histdvhls}, we compare the full distributions of \dvmax\ and \dvtau\ among our subsamples.  Statistical comparisons of these distributions do not fully confirm this hierarchy of velocities.  In Table~\ref{pnull}, we list the results from comparing each subsample with the others using both Kolmogorov-Smirnov and Kuiper tests.  The K-S test is weighted in favor of differences in the mean, while the Kuiper test is less biased.  We print in bold face those comparisons for which the likelihood of two distributions sharing the same parent distribution is less than 10\%\ as determined from {\it both} tests.

UV data may also indicate higher outflow velocities in some Seyfert~2s than in starbursts, but the evidence is not conclusive due to the number of galaxies observed.  (Though many Seyfert~1s have been studied in the UV with the {\it Hubble Space Telescope} [{\it HST}] and the {\it Far-Ultraviolet Spectroscopic Explorer} [{\it FUSE}], few Seyfert~2s have been.)   Outflowing gas has also been observed in the ultraviolet in three Seyfert~2s that are infrared-luminous (log$[\lir/\lsun] = 10.6-11.4$) using blueshifted interstellar absorption lines and redshifted Ly$\alpha$ emission lines \citep{gd_ea98b}.  The observed central velocities relative to systemic are $200-700$~\kms.  These velocities are similar to those observed in the UV in three infrared-luminous starbursts (log$[\lir/\lsun] = 10.3-11.5$), whose blueshifted absorption lines and redshifted Ly$\alpha$ lines have central velocities of $200-500$~\kms\ \citep{gd_ea98a}.

Only sample is as yet too small to fully confirm or deny the existence of velocity trends. We conclude that the evidence for higher neutral gas velocities in infrared-luminous Seyfert~2s than in starbursts is highly suggestive but not yet compelling.  However, comparing the Seyfert~2 ULIRGs with the starburst IRGs may demonstrate the influence of AGN more conclusively -- see \S\ref{sb_v_agn}.

\subsubsection{Other Outflow Properties}

Our  outflow  model  and  the  relevant equations  for  computing  the
outflow's column density, mass,  mass outflow rate, momentum, momentum
outflow rate, energy, and energy  outflow rate are discussed in detail
in  Papers~I  and  II.   We   assume  an  absorber  radius  of  5  kpc
(\S\S\ref{analysis} and \ref{radius}), an ionization fraction of 0.90,
and  standard Galactic depletion,  and use  the luminosity-metallicity
relation to  estimate metallicity.  These quantities  and their median
values,  as well as  \nad\ equivalent  width, optical  depth, covering
fraction,  Doppler  parameter,  and  column  density,  are  listed  in
Tables~$\ref{objprop}-\ref{avgprop}$.

We find no statistically significant differences in these outflow properties between infrared-luminous Seyfert~2s and starbursts.  The median mass, mass outflow rate, and energy are large, as in starburst galaxies.  We measure median values of $M = 10^{8.8}$~\msun, $dM/dt = 18$~\smpy, and $E = 10^{57.0}$~ergs.  However, these values can go as high as $10^{9.3}$~\msun, 300~\smpy, and $\sim$10$^{58}$~ergs.  We have chosen a simple model and assumed several quantities; these values should be treated with some caution.

The  median energy  outflow  rate, or  mechanical  luminosity, of  the
neutral   gas  in   Seyfert~2  ULIRGs   ($10^{42.2}$~erg~s$^{-1}$)  is
comparable to that  in starburst ULIRGs ($10^{42.2}$~erg~s$^{-1}$) and
starburst  IRGs  ($10^{41.6}$~erg~s$^{-1}$).   The average  mechanical
luminosity ($dE/dt$) per unit  radiative luminosity (\lir) is thus the
same for each subsample, at $\sim$5$\times10^{-4}$.

\subsection{Emission-Line Properties \label{emlprop}}

\subsubsection{Blue Emission-Line Asymmetries}

In AGN, emission-line profiles with blue-asymmetric wings are a common and well-documented phenomenon \citep[e.g.,][]{do84,do86,v91a,v91b,v91c,w92a,w92b,w92c}.  The atoms producing these lines are located in the narrow-line region (the NLR, within a few hundred pc of the nucleus; \citealt{v91c}).  The galaxy's, and especially the bulge's, gravitational potential clearly influences the dynamics of the NLR \citep{v91b,w92a,w92b,nw96}.  However, outflowing gas is also a common feature of NLR models \citep[e.g.,][and references therein]{v91c}, which may help to explain the observed blue-asymmetric emission-line wings.  There are several lines of evidence for NLR outflow.  High-ionization lines tend to show more blueward asymmetry (and higher velocities) than low-ionization lines \citep{do84,v91c}, and the base of these blue-asymmetric profiles are less correlated with galaxy properties than the core of the profiles \citep{nw96}.  Furthermore, in galaxies with strong radio emission and linear radio structures (i.e., jets), there is an additional high-velocity kinematic component seen in \otl\ that indicates the influence of the collimated outflow of radio plasma \citep{v91c,w92b,w92c,nw96,vcb05}.

In emission lines, however, outflow is not always distinguishable from radial inflow.  Blueshifted absorption lines are thus useful in that they are an unambiguous indicator of outflow.  Finding a distinct correlation between neutral absorbing and ionized emitting lines, the latter arising in the NLR in Seyferts, would illuminate not only NLR models but also show that the AGN is playing a role in the neutral outflow. 

In Figure~\ref{emlspec}, we plot the \nad\ line along with \ntll\ and/or \otl\ (when available) for the galaxies in which we observe emission lines whose blue and red wings have asymmetric profiles (i.e., one has a higher maximum velocity and/or more flux than the other).  As expected, we observe emission lines in which the blue wing has a higher maximum velocity and/or more flux than the red wing in \otl\ and/or \ntll\ in 15 of 20 Seyfert~2 nuclei, or 75\%.  We hereafter refer to this blueward asymmetry as BELA, for blue emission-line asymmetry.  (Note that for four nuclei, we used low-dispersion spectra from \citealt{vks99b} to look for asymmetries in \otl.)  This compares favorably with the result of \citet{v91c}, who observe this phenomenon in 10 of 16 Seyfert galaxies, or 63\%.  It is much larger than the number of infrared-luminous starburst nuclei with BELA (16 of 87, or 18\%; Paper~II).  We also observe redward asymmetries in the emission-line wings of two galaxies (F01436$+$0120 and Z03150$-$0219).  In F01436$+$0120, there is a blue wing which has much higher velocities than the red wing, although it carries less flux.  We thus include it in the BELA category, as well.

The percentage of nuclei with neutral outflows that also show BELA is high in Seyfert~2s, at 88\% (7 of 8 nuclei), much higher than the percentage of starburst nuclei with winds and BELA (24\%, or 11 of 45).  Note that these percentages are close to the detection rate of BELA in each case.  Conversely, the percentage of galaxies with BELA which also have winds is comparable to the detection rate of winds in both cases (69\% for starbursts, 47\% for Seyfert~2s).  This zeroth-order test indicates no strong association of BELA in emission lines and neutral outflows.

\subsubsection{Neutral and Ionized Gas} \label{o3corr}

To further explore the relation between neutral outflowing and ionized gas, in Figures~\ref{fwhmcorr} and \ref{fwtmcorr} we plot the full-width at half-maximum (FWHM) and full-width at 20\% of maximum (FW20) of the \otl\ line as a function of \dvmax\ and \dvtau\ in Seyfert~2s and starbursts.  We plot these quantities against each other for all nuclei with winds and only for those in which we observe winds and BELA.  Finding correlations between these would suggest that BELA and large-scale, neutral outflows are, in fact, associated.  We judge those correlations to be significant for which the slope ($a\pm\delta a$) satisfies $a > 3\delta a$ and for which Pearson's (parametric) and Spearman's (non-parametric) correlation coefficients each show a correlation at $\geq$90\% confidence.  Noting that there is one galaxy with high \dvtau/\dvmax\ and FWHM/FW20 (F05024$-$1941) which could drive a correlation by itself, we require that the correlation exist with and without this galaxy present and that the slopes be consistent.

We find no significant correlations in the starburst galaxies alone (38 nuclei with \ot\ data).  When the Seyfert~2s are added (8 nuclei), correlations emerge between FWHM(\ot) and \dvtau, both in all the galaxies and in only those nuclei with BELA.  Though the number statistics are low ($7-8$ nuclei), we also observe correlations between FWHM(\ot) and \dvtau\ in {\it only} the Seyferts with BELA.  The fits to these data and correlation coefficients are listed in Table~\ref{emlcorr}.

The observation of significant correlations between \dvtau\ and FWHM(\ot) may suggest a link between the neutral outflow and the NLR in Seyfert galaxies.  The starburst galaxies alone do not exhibit these correlations; it appears that the presence of a Seyfert nucleus is necessary.  Furthermore, there are no strong correlations between the high-velocity neutral and ionized gas (the latter is traced by FW20).  Since FWHM(\ot) probes the outer part of the NLR ($r \la 1$ kpc) in Seyferts, while the higher-velocity gas probably probes much nearer the nucleus \citep{v91c}, this may imply that the neutral gas in Seyferts is generally located in the `extended' NLR of the galaxy.

A caveat is that the dynamics of the core of the \ot\ line are strongly affected by the gravity of the bulge \citep{v91b,w92a,w92b,nw96}, and that the gas that contributes to the line emission is not purely outflowing.   However, as Figure~\ref{fwhmcorr} demonstrates, when galaxies without BELA are removed, the scatter in the correlation decreases quite dramatically.  This lends support to the idea that the blueshifted neutral and ionized gas are somehow linked, especially in Seyfert ULIRGs.

\subsubsection{Individual Galaxies}

The emission-line profiles in a few of these galaxies are worthy of individual mention.  Note that we observe extended gas that is photoionized by the AGN in several galaxies.

{\bf Z03150$-$0219}.  This galaxy has a red emission-line component with velocities up to 800~\kms\ above systemic and a higher \otl/H$\beta$ flux ratio than the gas at systemic by a large factor ($\geq$10).  No detectable H$\beta$ emission is observed at this velocity.

{\bf F04210$+$0401}.  This galaxy possesses a large radio jet.  We observed this galaxy with the spectroscopic slit aligned with the position angle of the jet, and observe spectacular emission-line structures that have been studied in previous works \citep[e.g.,][]{shp96a,shp96b}.  We observe a narrow and low signal-to-noise ratio \nad\ component at systemic.

{\bf F05024$-$1941}.  This galaxy has the highest absorption-line velocity $\Delta v$ of all starbursts and Seyfert~2s in our survey.  A low-velocity blue wing is present, as well as an optically thick component with low covering fraction at $\Delta v = 1550$~\kms.  This galaxy also has a redshifted absorbing component whose velocity is identical to that of an emission-line feature extending $\sim$15~kpc to the south.  The velocities of the \ot\ line do not extend as far as $\Delta v = 1550$~\kms, but those of the \ntll\ line do.  The peak of \ot\ is blueshifted from systemic by $\sim$470~\kms.

{\bf F05189$-$2524}.  This galaxy has numerous emission lines of various stages of ionization in its spectrum (superimposed on a continuum with deep stellar absorption lines; see also \citealt{f_ea05}).  The high-ionization lines (e.g., \ot, [\ion{Ar}{3}] $\lambda\lambda7137,~7752$, and [\ion{S}{3}] $\lambda\lambda9069,~9531$) are blueshifted by 510~\kms\ with respect to systemic (as determined by the stellar absorption lines and CO emission), while the low-ionization lines are at systemic.  Some of the high-ionization lines also have blue wings extending even farther from systemic.  The centroid of the blue component of \nad\ is within 100~\kms\ of the peak of the high-ionization lines.

{\bf F08526$+$3720}.  There is extended emission $\sim$15~kpc to the north and south of this galaxy that has the same velocity as the outflowing \nad; the high \ot/H$\beta$ flux ratio indicates this material may be ionized by the AGN.

{\bf F08559$+$1053}.   The peak of \ot\ is located 200~\kms\ blueward of systemic, and \nt\ has a blue component at this velocity.  There is extended emission 5 kpc to the south of this galaxy at a velocity near systemic; this gas has \ot/H$\beta\sim5-6$, which implies ionization by the AGN.  There are substantial variations between \ot/H$\beta$ in both position and velocity space, perhaps probing variations in density (i.e., substructure).

{\bf F13428$+$5608 (Mrk~273)}.  This well-known Seyfert~2 shows extended line-emitting structures and complex velocity profiles on scales of up to $30-40$~kpc on both sides of the nucleus.  Variations in H$\alpha$/\nt\ are visible, including likely shock-excited regions several kpc to the NE and SW that also exhibit line-splitting of $300-400$~\kms.  Larger line-splitting ($\sim$500~\kms) occurs at larger radii \citep[see also][]{cab99}.

\section{OUTFLOWS IN SEYFERT 1 ULIRGS} \label{sy1}

\subsection{Outflow Properties} \label{sy1_ofprop}

In Figure~\ref{spec_s1}, we display the \nad\ lines in the 6 Seyfert~1 ULIRGs in our sample with our fits to the lines superimposed.  Tables~\ref{compprop} and \ref{ofprop} list the measured properties of each \nad\ velocity component and of the outflow in each galaxy, respectively.  Table~\ref{avgprop} lists the average outflow properties in this subsample.

We measure an outflow detection rate in \nad\ of $\sim$50\%\ (3 of 6 galaxies).  On the surface, this is discrepant from the result of \citet{bm92}, who found blueshifted \nad\ absorption lines in $3-4$ of 19 infrared-selected quasars and Seyfert~1s with warm infrared colors.  However, if the galaxies in \citet{bm92} are segregated by luminosity, the detection rate is 0\%\ in non-ULIRGs and $40-60$\%\ ($3-4$ of 7 galaxies) in ULIRGs.  Our detection rate is similar to that of intrinsic UV and X-ray absorbers in local Seyfert~1s ($50-70$\%; \citealt{ckg03}).  It has interesting implications for the geometry of the \nad\ absorbers in Seyfert~1s (\S\ref{gcf}) and for the relation of ULIRGs to low-ionization broad absorption-line quasars (\S\ref{balqsos}).

The properties of the nuclear absorbers in these Seyfert~1 galaxies show clear differences from those found in Seyfert~2 and starburst galaxies, as we discuss below.  They are thus powered by the AGN, rather than a starburst or starburst$+$AGN (\S\ref{sb_v_agn}).  However, Seyfert~1s possess strong starbursts as well as AGN, and thus must contain an extended, starburst-driven outflow alongside a nuclear AGN outflow.  The detection rate of extended outflows is 15\%, since we observe them in only Mrk~231 (\S\ref{mrk231}).  Extended outflows are difficult to detect in Seyfert~1s due to the high luminosity of the nucleus.

Two galaxies (F07599$+$6508 and F12540$+$5708 [Mrk~231]) are broad-absorption-line quasars, and thus have very broad (FWHM~$\ga 2000$~\kms) and deep absorption complexes with high maximum velocities (up to $\sim$10$^4$~\kms).  A third Seyfert~1 (F11119$+$3257) also possesses high-velocity, outflowing gas; the three components of the \nad\ feature have central velocities of $\Delta v \sim 700 - 1400$~\kms.  Velocities of this magnitude are found in one Seyfert~2 (F05024$-$1941) and one starburst (F10378$+$1108).  However, the outflow properties in these two galaxies are different from those in F11119$+$3257 in that the covering fraction of the high-velocity gas is smaller ($C_f = 0.2-0.4$ vs. $0.7-1.0$ in F11119$+$3257), there is only a single high-velocity component (vs. three in F11119$+$3257), and there is low-velocity gas (there is none detected in F11119$+$3257).  The properties of the absorbers in F11119$+$3257 are consistent with those of UV and X-ray intrinsic absorbers found in local AGN (\S\ref{intabs}; \citealt{ckg03}).

As we discuss below (\S\ref{radius}), the nuclear absorption in Seyfert~1s is likely to arise on scales of tens of pc or less.  In calculating the masses, momenta, and energies listed in Tables~\ref{objprop} and \ref{avgprop}, we assume an absorbing radius of 10 pc.  This yields masses, momenta, and energies of $10^{4.3}$ \msun, $10^{46}$ dyne s, and $10^{55}$ erg on average, values that are much smaller than those observed in Seyfert~2 and starburst galaxies.  However, the outflow rates of these quantities are comparable to or larger than those in Seyfert~2s and starbursts due to the high observed velocities.  On average, we compute $dM/dt = 12$ \smpy, $dp/dt = 10^{36}$ dyne, and $dE/dt = 10^{44}$ erg s$^{-1}$.  The energy outflow rate of the neutral gas is surprisingly large, at $\sim$1\%\ of the radiative luminosity of the galaxy on average.

\subsection{Broad Absorption-Line Modeling}

In \citet{rvs02}, we discuss in detail our method for modeling the broad absorption line (BAL) in Mrk~231.  We have only slightly modified the results of this modeling for the current work.  F07599$+$6508 is a more difficult case since, unlike Mrk~231, it has strong \ion{Fe}{2} emission lines overlapping \nad\ \citep[e.g.,][]{bm92} that are complex to model and remove.  Rather than attempt to remove these lines, we have simply fit the continuum around the broad complex.  This partially removes the emission feature that affects the blue half of the \nad\ line \citep{bm92}.  However, we are insensitive to the very blueshifted absorption that \citet{bm92} claim to see in this galaxy at $\sim$16000~\kms\ because of our decision not to model the \ion{Fe}{2} lines.

The number of components that we should fit to each of the broad features is uncertain.  In Mrk~231, we fit many narrow components to the broad profile.  We use as a starting point the components suggested by \citet{frm95}, who vary the number of components in the \nad\ line and choose the number which gives the best fit.  For F07599$+$6508, however, we fit only three very broad components to the deep \nad\ feature; without doing a detailed analysis similar to that of \citet{frm95}, it is unclear whether we should use few or many components.  The conclusions of this paper are not sensitive to the details of the \nad\ fits in the BAL Seyfert~1s, so we have chosen not to do more sophisticated modeling.

The \nad\ absorption lines in F07599$+$6508 have been observed previously by other authors \citep[e.g.,][]{bm92}.  However, ours is the first attempt to model them using detailed profile fits.  We note here that F07599$+$6508 also has BALs (both low- and high-ionization) in the UV with outflow velocities that range from $5000-22000$~\kms\ \citep{lts93,l94,hw95}.  These velocities are larger than those seen in \nad\ ($4000-11000$~\kms\ in our spectra, or up to 16000~\kms\ in the spectra of \citealt{bm92}).  Furthermore, the broad \nad\ absorption feature is not seen in polarized light \citep{hw95}, suggesting that the global covering factor of the absorbing gas is less than unity (\S\ref{gcf}).

\section{A CASE STUDY: SMALL- AND LARGE-SCALE OUTFLOWS IN MRK~231} \label{mrk231}

\subsection{Background}

We observe an AGN-driven wind arising in the broad-line region of Mrk~231, a Seyfert~1.  This conclusion is based on the observation of broad, high-velocity absorption lines in the nucleus, as well as on the time variability of the bluest component (\S\ref{radius}).  However, it was proposed by \citet{hk87} that there is also a wide-angle, large-scale wind in this galaxy.  They observe emission lines with blue asymmetries in a region that extends several kpc from the galaxy's nucleus in multiple directions.  Line-splitting is also present in \otl\ a few kpc to the south of the nucleus (at $v_\sun=12100$ and 12600~\kms; \citealt{hk87}) and in [\ion{O}{2}] $\lambda3727$ in the nuclear spectrum (at 11900 and 12700~\kms; \citealt{b_ea77}).  This implies blueshifts of $500-800$~\kms\ with respect to systemic (12642~\kms, from CO and \ion{H}{1} data; \citealt{sss91,cwu98}).

These blueshifts and blue wings surrounding the nucleus are strongly suggestive of a large-scale outflow.  However, given the ambiguous nature of emission-line kinematics in an interacting system where inflow could be present, it would be good to confirm this claim by looking for blueshifted absorption lines.

\subsection{Present Observations}

Mrk~231 is the closest galaxy (by a factor of 2.5) in our Seyfert~1 sample.  We observed this galaxy in a short (5-minute!) exposure with ESI on Keck II.  We later observed it again at Kitt Peak in a 4800~s exposure.  The former instrument has a much better point-spread function (PSF) and spatial sampling, so we used this observation to decompose the near-nuclear profile.  However, we note that the KPNO spectrum shows similar behavior.  The long slit in the latter observation also reveals a faint emission-line region $10-17$~kpc south of the nucleus, which has a narrow linewidth of 135~\kms, is blueshifted from systemic by $150-200$~\kms, and corresponds to a stellar tidal feature or spiral arm in continuum and emission-line images \citep{hk87}.  (The velocity of this feature is misprinted in \citealt{hk87} as being 700~\kms\ {\it above} systemic.)

Along the north-south slit, we extracted thirteen, $0\farcs8$-wide regions and fit the \nad\ feature in each one.  We detect low-velocity blueshifted \nad\ absorption lines ($\Delta v \ga -2100$~\kms) in most of these spectra, over a projected extent of 6~kpc.  In Figure~\ref{mrk231_spec}, we plot the spectrum of each region in which we fit the low-velocity absorption.  This absorption arises in the nuclear spectrum, but the lines are difficult to see here because they have low equivalent width and arise at the juncture of what may be overlapping broad emission lines (\ion{He}{1} $\lambda5876$ and \nad).  In the two northernmost spectra we did not detect low-velocity \nad\ to the limit of our signal-to-noise ratio.  From these spectra, we see that the velocities and profiles of the absorbers vary significantly across the disk, and are rather clumpy in the northern regions.  We also see that the gas has much higher velocities north of the nucleus than south of the nucleus.  

As further illustration, in Figure~\ref{mrk231_vel} we plot, as a function of slit position, the velocities of (a) the low-velocity \nad\ components; (b) the narrow components of H$\alpha$ and [\ion{S}{2}] $\lambda\lambda$6716,~6731; and (c) the \ion{Ca}{2} triplet (where available).  Along the north-south slit, the stellar absorption-line and gaseous narrow-line velocities are consistent with the systemic velocity of the nuclear disk (to within $\pm$50~\kms).  In the 8 extra-nuclear bins where we fit the low-velocity \nad\ line, the blueshifted interstellar gas reaches a velocity of at least 300~\kms\ with respect to systemic (and up to 2100~\kms\ in one bin).

The nuclear spectrum spills into the spectra of the extended regions due to the wings of the PSF.  We confirmed that the newly-discovered, low-velocity absorption is not an artifact of PSF smearing by computing the ratio of the equivalent width of the broad, high-velocity component to that of the low-velocity component.  This ratio decreases dramatically away from the three nuclear bins (by a factor $>$100), meaning that the low-velocity absorption occurs over an extended region.

We detect two extended continuum components besides the nuclear component; these serve as the background illumination for the extended absorbers.  There is an extended baseline component, as well as a peak in continuum flux $3-4\arcsec$ south of the nucleus.  A bright emission-line region with a LINER spectrum is also present near the southern peak, offset slightly towards the nucleus.  In general, the extranuclear regions of Mrk~231 contain numerous blue star-forming knots with a range of ages ($\sim$10$^7 - 10^9$~yr).  These knots are especially dense in the southern region in which we detect a second continuum peak and emission-line spectrum, the `horseshoe' region of \citet{hk87}.

We observe redshifted emission in \nad\ in the spectrum 2.3~kpc north of the nucleus (see Figure \ref{mrk231_spec}).  This is not light contamination from the nuclear spectrum, since the emission is not broad and there is no emission in the southern spectra at the same distance from the nucleus.  The observed redshifted velocities match the lower range of blueshifted velocities; this emission may be resonant scattering from receding gas.

We also observe the blueshifted emission-line wings in \ntll\ claimed by \citet{hk87}.  There is some uncertainty in this result due to telluric absorption of the \ntl\ line and to broad H$\alpha$ from the nuclear spectrum that contaminates the near-nuclear bins.  Farther from the nucleus, the conclusion is more secure.  The velocities appear greater than or equal to those in absorption, as seen in other objects.

Finally, there is clear evidence of line-splitting in H$\alpha$ and \nt\ in the two northernmost bins, with blue- and red-shifted velocities of $100-200$~\kms.  The emission lines present a higher \nt/H$\alpha$ flux ratio ($1.1-1.2$) than elsewhere, suggesting the presence of shock ionization.  Accompanying the line-splitting, there is a faint blueshifted emission-line component in these spectra with velocity relative to systemic of $\sim$800~\kms\ (or larger, if it is not \nt\ $\lambda6548$).  There is a probable redshifted counterpart of similar velocity redward of \ntl.  A high-velocity blueshifted component is also seen south of the nucleus by \citet{hk87}.

\subsection{Interpretation}

What is the explanation for the gas dynamics in this source?  The blueshifted absorption lines point to an extended outflow.  Consistent with this scenario are the observed blue emission-line wings, line splitting, high-velocity emission-line components, and redshifted \nad\ emission.  The presence of comparable-velocity blue- and redshifted components in absorption and emission could arise on the approaching and receding sides of a spherical or biconical outflow.  The best interpretation is that of a large-scale, wide-angle wind.

The difference in north and south absorption-line velocities argues against a model for the wind where the absorption arises on the surface of the wind; in this model, the outflow velocity should be the same at a given radius north or south of the nucleus.  If we assume instead that the absorption arises in the walls of the outflow in the side of the outflow facing our line-of-sight, then the variations can be attributed to projection effects.  Consider a model of a constant-velocity, bipolar, wide-angle wind emerging perpendicular to the \ion{H}{1}/CO disk.  We can calculate the expected inclination of the disk and opening angle of the outflow from the ratio of the observed maximum velocities north and south of the nucleus.  A reasonable range for this ratio is $v_{south}/v_{north} \sim 0.2-0.4$, using the maximum observed values of $v_{north} = 1000-2000$~\kms\ and $v_{south} = 400$~\kms.  The resulting calculated disk inclinations ($i = 25-40\degr$) are consistent with those observed \citep{bs96,cwu98}, and the resulting wind opening angles ($65-90\degr$) are consistent with those expected based on observations of local starburst and Seyfert~2 outflows \citep{c_ea96a,vcb05}.

Because Mrk~231 has both a strong starburst and a strong AGN, the power source of this wind is ambiguous.  The maximum outflow velocity is larger than in any starburst or Seyfert~2 galaxy we observe.  If the wind is AGN-powered, the power could come from a wide-angle wind originating at scales of a few to tens of parsec, or could result from a jet dumping energy into the ISM of the host galaxy.  In 3C~305, neutral gas with comparable velocities and on similar spatial scales to the outflow in Mrk~231 arises as the radio jet interacts with the ISM \citep{m_ea05}.

In fact, there is a north-south radio jet in this galaxy on scales ranging from tens of pc to tens of kpc \citep{cwu98,uwc99}.  This jet is probably perpendicular to the kiloparsec gas disk seen in CO \citep{bs96,ds98}, \ion{H}{1} \citep{cwu98}, and radio continuum emission \citep{cwu98,t_ea99}.  The disk has an inclination of $\la$60$\degr$ \citep{bs96}, perhaps close to 45\degr\ \citep{cwu98}, with respect to the plane of the sky and a roughly east-west major axis position angle.  (However, \citealt{ds98} argue for $i \la 20\degr$.)  The northern half of the disk appears to be the near side \citep{cwu98}, meaning that the jet is approaching from the south and receding to the north.  This is inconsistent with the higher velocities observed north of the nucleus than south of the nucleus, suggesting that the gas dynamics we observe are not directly connected with the jet's motion.

\section{DISCUSSION} \label{discuss}

\subsection{Location of Absorbing Gas} \label{radius}

Our absorption-line data contain limited information about the galactocentric radius at which the absorption occurs.  A typical method in absorption-line studies is to use photoionization codes to determine the radius.  Since we have information on only one transition in a single atomic species, this technique is unavailable to us.  However, by using simple physical arguments and comparing to other data we can make some useful statements.

\subsubsection{Seyfert~1s} \label{radius_sy1}

There are several physical arguments that we can make that the broad
absorption lines observed in F07599$+$6508 and Mrk~231 arise at small
scales.  First, the high velocities observed ($\Delta v \sim
10^4$~\kms) are comparable to emission-line velocities observed in the
broad-line region, which extends $\sim$0.1~pc from the black hole.
Second, the bluest component in the nucleus of Mrk~231 is
time-variable in both velocity and equivalent width \citep{b_ea91,frm95,rvs02,g_ea05}.  Variability on short timescales can imply short distances to the ionizing source \citep{ckg03}.  Third, the mass and energy outflow rates from these nuclei are too large if we assume $r=1$~kpc.  For F07599$+$6508 and Mrk~231, this yields mass outflow rates of several thousand \smpy.  This will move over $10^{10}$~\msun\ of gas in only 10 million years, which is a substantial disruption.  If these winds are driven by black hole accretion, these mass outflow rates are at least 1000 times the Eddington accretion rate for a mass-to-energy conversion efficiency of 0.1 and a bolometric luminosity of 10$^{46}$~erg~s$^{-1}$ (as in Mrk~231).  Finally, the whole luminosity of the galaxy would be driving the outflow, since the energy outflow rate (10$^{46-47}$~erg~s$^{-1}$) would be comparable to the galaxy's luminosity!

These arguments do not apply to F11119$+$3257, whose outflow velocity is much smaller ($\la$1500~\kms).  Indirect evidence, including comparisons with other galaxies in our sample (\S\ref{sy1_ofprop}) and with UV and X-ray intrinsic absorbers in Seyfert~1s (\S\ref{intabs}), suggests that the absorbers are at small radius (i.e., tens of pc or less).

We conclude that the nuclear \nad\ absorbers in our Seyfert~1 subsample almost certainly arise on small scales.  In calculating the masses, momenta, and energies listed in Tables~\ref{objprop} and \ref{avgprop}, we therefore assume an absorbing radius of 10~pc (i.e., in the inner NLR).  However, we cannot {\it rule out} absorbers on larger scales in the nuclear spectra of these galaxies, since there are isolated cases of broad absorbers \citep{d_ea01} and narrow absorbers \citep{hb_ea01} arising on scales of hundreds of pc to tens of kpc.  In fact, larger scale outflows (powered either by a starburst or the AGN) probably occur in most of these galaxies, as in Mrk~231, but we cannot detect them due to limitations in spatial resolution and nuclear contamination.

\subsubsection{Seyfert~2s}

The absorbers in Seyfert~2s have different properties than those in Seyfert~1s.  They are similar to the absorbers in starbursts, which lie at kpc radii (Paper~II).

We directly observe extended, blueshifted absorption in three Seyfert~2s (F05189$-$2524, F13451$+$1232, and F15001$+$1433).  This absorption arises several kpc from the nucleus in F05189$-$2524 and up to radii of $10-15$~kpc in the other two galaxies.  For F05189$-$2524, two other lines of evidence suggest large-scale outflow.  A low-resolution spectrum, obtained with the Space Telescope Imaging Spectrograph (STIS) on {\it HST}, probes nuclear radii of $\la$200~pc \citep{f_ea05}.  This spectrum does not show the deep \nad\ absorption observed in our ground-based data; we measure equivalent widths of 0.7~\AA\ and 5.4~\AA\ for the {\it HST} and ground-based spectra, respectively.  This implies that the strong, blueshifted absorption occurs primarily at a projected radius of $\ga$200~pc.  Furthermore, the background source is not the core of the Seyfert nucleus, but rather a more extended continuum source (a result that is consistent with the strong stellar lines observed in the ground-based spectrum).  In our ground-based spectrum, we also observe redshifted \nad\ {\it in emission} at a distance of $\sim$2~kpc from the nucleus.  Similar behavior is seen in the outflow in NGC~1808; this emission is interpreted as resonance scattering by the receding part of the outflow \citep{p93}.  The velocity of this emission is consistent with this interpretation, as it appears to be redshifted roughly 100~\kms\ from systemic (matching the velocity of the blueshifted absorbing component nearest to systemic).  Another potential origin of this emission is scattered light from the broad line region of the nucleus, but we do not observe this signature in other emission lines in this object in the offset spectra.

Indirect evidence for kpc outflows in Seyfert~2s comes from emission/absorption line correlations.  If the \nad\ outflow were located in the inner NLR ($r \la 100$~pc), one might expect a correlation of high-velocity ionized gas (as parameterized by the full-width at 20\%\ maximum of the \otl\ line) with the maximum outflow velocity of the neutral gas (\dvmax).  We do not observe such a correlation (\S\ref{o3corr}).  Instead, we observe a correlation between the width of the core of the \otl\ line and the lower-velocity neutral gas, which is natural in the case of a wind arising in the `extended' NLR ($r \ga 0.1-1$~kpc).

\subsection{Dependence of Outflow Properties on Galaxy Mass and Luminosity} \label{v_host}

In Paper~II, we studied outflow properties as a function of galaxy properties in our sample of 78 infrared-luminous starbursts.  We also included \nad\ observations of four dwarf starbursts \citep{sm04}.  We found an increase in outflow velocity, mass, momentum, and energy as a galaxy's star formation rate, luminosity, and circular velocity (mass) increase.  The increases in outflow velocity, mass, etc. with SFR are natural in a model where the ram pressure of supernova-heated gas drives ambient material out of the starburst.  Radiation pressure probably also plays a role, given the high luminosity and dustiness of many of these galaxies.  However, its momentum input at a given infrared luminosity is an order of magnitude less than that from star formation.  Thus, supernovae-heated gas is {\it necessary} to power the outflows observed in many objects.

In Paper~II, we also showed that these trends of outflow properties with galaxy properties become shallow or disappear at high values of SFR, luminosity, and $v_{c}$ (i.e., in the ULIRG population).  We cannot conclusively demonstrate why this is so, though there are several possible explanations: (a) the entrainment of all available gas clouds; (b) a velocity threshold due to the finite velocity of the hot wind \citep{mqt05,m05}; or (c) a reduction in thermalization efficiency at high SFR, luminosity, and $v_c$.

Our sample of  Seyferts is too small to look  for correlations, but we
can  compare the phase-space  locations of  our Seyfert  galaxies with
those of the  starbursts.  In Figures~$\ref{dvmax_v}-\ref{dedt_v}$, we
plot maximum  velocity, mass outflow rate, momentum  outflow rate, and
energy outflow rate vs. infrared luminosity and circular velocity.

All  of  the Seyfert~2  points  fall in  the  regions  of these  plots
populated by  starburst ULIRGs.   This similarity reinforces  the idea
that  the  outflows in  Seyfert~2  and  starburst  ULIRGs are  largely
similar (see also \S\ref{sy2} and \S\ref{sb_v_agn}).  Furthermore, the
momentum  and energy  injected from  the starburst  are  sufficient to
explain the momentum  and energy in the outflows  in Seyfert~2 ULIRGs,
implying  that the  contribution  from  the AGN  is  comparable to  or
smaller than  that from  the starburst.  Finally,  Figure \ref{dedt_v}
again  illustrates   the  point  that,  on   average,  Seyfert~2s  and
starbursts have approximately the  same mechanical luminosity per unit
radiative luminosity (see \S\ref{sy2_ofprop}).

\subsection{Frequency of Occurrence and Global Covering Factor} \label{gcf}

As we discuss in Paper~II, the detection rate $D$ is a function of both the actual frequency of occurrence of winds $F$ and the global angular covering factor of the wind, $\Omega$.  The global angular covering factor, or solid angle subtended by the wind, is the product of the large-scale opening angle (given by \co, a fraction of $4\pi$) and local clumping (which we assume to be represented by \cf).  The opening angle is given by the detection rate if the frequency of occurrence of winds is 100\%, as in starbursts.  The global covering factor is thus $\Omega/4\pi = \co~\langle C_f \rangle = D~\langle C_f \rangle$ \citep[Paper~II; see also][]{ckg03}.  In the more general case, we can set lower limits to $F$ and $\co$ using our detection rate: $D < F \leq < 1$ and $D < \co \leq 1$.

Assuming $F = 1$ for Seyfert~2s (as in starbursts of comparable SFR; Paper~II), we measure $\co = D = 0.45 \pm 0.1$.  This corresponds to wind opening angles of $100-125\degr$, consistent with measurements of winds in local Seyfert~2s \citep[e.g.,][]{c_ea96a,vcb05}.  If instead we assume $F = D = 0.45$, then the projected wind opening angles of 360\degr\ are inconsistent with measurements.  We thus conclude that winds occur in most Seyfert~2s.  Using the value of $\langle C_f \rangle$ listed in Table \ref{avgprop}, we compute that the average global covering factor of neutral gas in these winds is $\Omega/4\pi = 0.19$, close to the value measured in IRGs (Paper~II).

For Seyfert~1s, the detection rate of \nad\ absorbers (both broad and narrow) is $\sim$50\%\ (based on this work and \citealt{bm92}), and the average $C_f$ is 0.7 (Table \ref{avgprop}).  Assuming $F = 1$ for Seyfert~1s, we compute $D = 0.5 \pm 0.1$ and $\Omega/4\pi = 0.4$.  This result is consistent with the value of $\Omega \sim 0.5$ measured from observations of intrinsic, narrow, high-ionization absorption lines in local Seyfert~1s \citep{ckg03}.  It is also consistent with the lack of observed broad \nad\ in polarized light in F07599$+$6508 \citep{hw95}, from which we can infer that the global covering factor of \nad\ absorbers in Seyfert~1 ULIRGs is not generically of order unity.

\subsection{Gas Escape Fraction}

There are several Seyfert~2 galaxies with very high velocities in our sample.  This means that some of this material, if it is at large enough radius and does not encounter substantial drag from interstellar gas, will escape the galaxy and enter the intergalactic medium.  It is difficult to estimate precise escape velocities for these galaxies.  However, we can make simplifying assumptions that allow us to calculate an `escape fraction' of gas.  We consider this illustrative rather than strictly quantitative due to the uncertainties.

As  in Paper~II, we  assume a  singular isothermal,  spherical density
distribution with a density that  truncates at a radius $r_{max}$.  If
the  gas  absorbs   at  radius  $r$,  then  the   escape  velocity  is
parameterized uniquely by  $v_c$ and $r_{max} / r$.   Our procedure to
calculate  $f_{esc}$ is  as follows:  (a)~to get  $v_c$,  use measured
values where possible; otherwise,  use an average from measurements of
other  ULIRGs (K.  Dasyra,  private communication);  (b)~use $v_c$  to
compute $v_{esc}$; (c) compute the mass outflow rate of gas that has a
velocity above  $v_{esc}$; (d)~sum $dM/dt$ and  $dM/dt_{esc}$ over all
galaxies with measured  $v_c$; and (e)~divide $dM/dt_{esc}^{total}$ by
$dM/dt^{total}$.  (Note  that $M$  and $dM/dt$ are  interchangeable in
this  algorithm;  the  values  of  $f_{esc}$ computed  using  $M$  are
comparable to those computed using $dM/dt$, but smaller by a factor of
2.  The  discrepancy is due to the  extra factor of $\Delta  v$ in the
definition of $dM/dt$; see Paper~II.)

Using this procedure, we measure $f_{esc} \la 60\%$ for the neutral outflowing gas in AGN-dominated ULIRGs, which is much higher then the value of $\la$20\%\ measured for starbursts.  Thus, the neutral, dusty outflowing gas in these galaxies may play a significant role in enriching the intergalactic medium at higher redshifts, where the number density of ULIRGs is high.  Furthermore, the hot free wind that drives the expansion of the large-scale winds in starbursts may also be present in these systems.  This gas is metal-enriched and has more specific energy than the neutral phase.  It is hence more likely to escape the galaxy and enter the IGM.

\subsection{Connection to Intrinsic Absorbers in Seyfert~1s} \label{intabs}

It is instructive to compare our results to those found in UV and X-ray absorption-line studies of local ($z < 0.1$), moderate-luminosity (log$[L_{\mathrm{bol}}/\lsun] \sim 10-12$) Seyfert~1s.  The observed intrinsic (or `warm' in the X-ray case) absorbers in these galaxies probably arise on scales of tens of pc or smaller and are thus powered by the AGN \citep[e.g.,][]{ckg03,ck05,myr05}.

Similarities in detection rate, column density, and velocity exist between the absorbers we detect in Seyfert ULIRGs and intrinsic absorbers.  The detection rate of intrinsic absorbers in Seyfert~1s is $50-70$\%\ \citep{ckg03}.  Measured UV and X-ray column densities of some absorbers \citep{ckg03,myr05} are also consistent with our measurements of $N$(H)~$\sim 10^{21-22}$~cm$^{-2}$.  Finally, the distribution of velocities in intrinsic absorbers (ranging up to 2300~\kms; \citealt{ck05,myr05}) is broadly consistent with what we measure in \nad\ in Seyfert~2s (Figure \ref{histdv}) and in one Seyfert~1, F11119$+$3257.  In the UV, the majority of observed components (75\%) have $\Delta v > -700$~\kms, and there are some redshifted components with velocities up to 200~\kms\ \citep{ck05}.  Out of 14 X-ray `warm absorbers' modeled in \citet{myr05}, 10 have velocities less than 1000~\kms, though there are no redshifted components.

However, the distribution of FWHM in UV absorbers is not consistent with our measurements.  Most of UV absorbers (80\%) have FHWM~$<$~100~\kms\ \citep{ck05}, while the average \nad\ FWHM in Seyfert~2 ULIRGs is 400~\kms\ (Table~\ref{avgprop} and Figure~\ref{histdv}).  In the X-ray the situation is less clear, as most components are unresolved (FWHM~$\la 200-300$~\kms; \citealt{myr05}).

This difference, as well as the conclusion that \nad\ absorbers in Seyfert~2s are located at galactocentric radii of $r \ga 1$~kpc (\S\ref{radius}), suggest that the neutral absorbers are not directly related to Seyfert~1 intrinsic absorbers.  However, the lines we observe in F11119$+$3257 could be an optical manifestation of intrinsic absorbers.  The \nad\ components in this galaxy are high in velocity and covering fraction and low in velocity width (\S\ref{sy1_ofprop} and Table~\ref{compprop}).

\subsection{Outflows and Merger Evolution}

Most ULIRGs are in the late phase of a merger between two massive, gas-rich galaxies \citep{vks02}.  We showed in Paper~II that the detection rate of outflows in ULIRGs does not obviously depend on merger phase.  There is some variation of outflow velocity with interaction class, but our sample size is too small to show that this is significant.  Adding the Seyfert~2s to the starburst sample does not change the conclusions of Paper~II in this regard.

Seyfert ULIRGs are on average later in the stages of merging than starburst-dominated ULIRGs \citep{vks02}.  We now discuss our observations in light of the mechanisms by which they may be removing the obscuring dust screen surrounding the AGN and evolving into optical quasars or radio galaxies.

\subsubsection{ULIRG Connection to Broad Absorption-Line Quasars} \label{balqsos}

Orientation-based models and evolutionary models compete to describe the properties of broad absorption lines (BALs) found in quasars.  The latter are meaningful in the context of ULIRGs, as BAL outflows could be a mechanism for removing the screen of dust surrounding ULIRG nuclei, converting them from obscured AGN to bright quasars.

Out of ten Seyfert~1 ULIRGs observed by us and \citet{bm92}, at least four are low-ionization BAL quasars (loBALQSOs; \citealt{w_ea91,bm92}).  Two or three of these ten possess BALs in \nad\ (F07599$+$6508, Mrk~231, and possibly F17002$+$5153), and two others possess narrow \nad\ absorption (F11119$+$3257 and F14026$+$4341).  LoBALQSOs are more common in IR-selected than optically-selected quasar samples (27\%\ vs. 1.4\%; \citealt{bm92}).  As we show, they are apparently even more common in ULIRG samples, with a detection rate of $\geq(40\pm10)\%$.

The loBALQSOs as a class have a number of unique properties.  These include weak \otl\ \citep{bm92}, strong \ion{Fe}{2} lines \citep{w_ea91}, red continua \citep{w_ea91}, and relatively weak X-ray flux and/or high absorbing columns \citep{gb_ea01}.  It has been hypothesized that loBALQSOs have high global covering factors, and that IR-luminous loBALQSOs may represent a stage in the life of quasar when the nucleus is enshrouded in dust and is in the process of destroying this screen via fast outflows.  Thus, loBALQSOs may be an evolutionary stage in the proposed scenario by which some quasars are formed from major mergers.  In this scenario dust-enshrouded ULIRGs are an intermediate stage between the merger and a quasar.  This scenario does not conflict with recent sub-mm observations showing that BALQSOs and QSOs have similar dust properties \citep{wrg03}, since most of the BALQSOs observed were not loBALQSOs.  The merger-to-loBALQSO-to-QSO model is also consistent with our measurement of the global covering factor of \nad -absorbing gas in Seyfert~1 ULIRGs ($C_\Omega \sim 0.4$).

\subsubsection{ULIRG Connection to Radio Galaxies} \label{radiogalaxies}

Galaxies may also undergo a radio-loud stage in the evolution from a merger remnant to an unobscured QSO.  Young, massive, and reddened stellar populations (ages $0.1-1$~Gyr) are observed in a number of radio galaxies on kpc scales \citep[e.g.,][]{t_ea05}.  The radio jets in these galaxies could be triggered by a massive merger that funnels gas to the nucleus and feeds the central black hole; given the observed stellar ages, the jet turns on late in the merger sequence \citep{t_ea05}.  In a number of radio galaxies, a jet interacts with the surrounding ISM at kpc or sub-kpc radii and powers massive [$N$(H)~$= 10^{21-22}$~cm$^{-2}$], high-velocity ($\Delta v \la 2000$~\kms) outflows of neutral and ionized gas (\citealt{m_ea03} [3C~293]; \citealt{m_ea04b,htm03} [4C~12.50]; \citealt{m_ea05} [3C~305]; \citealt{m_ea01} [PKS~1814-63 and 3C~459]).  This phenomenon is also observed in at least one Seyfert~2 galaxy with a radio jet, IC~5063 \citep{mot98,o_ea00,m_ea04a}.

These high-velocity outflows, observed in \ion{H}{1} 21~cm and optical emission lines, could have counterparts in our data.  In general, comparing our data to broad-band \ion{H}{1} absorption observations is a good way to confirm our column density measurements; radio-loud galaxies are the easiest sources in which to do this comparison, since they provide a bright background continuum source.

Seven objects in our sample have 1.4~GHz fluxes $>$100~mJy, based on FIRST \citep{bwh95} or NVSS \citep{cc_ea98} fluxes.  There has been no systematic search for radio jets in these seven galaxies, but three are known to possess radio jets: F12265$+$0219 (3C~273), Mrk~231, and F13451$+$1232:W (4C~12.50).  Two others may possess small-scale, low-power jets -- F13428$+$5608 (Mrk~273; \citealt{ct00,kb04}) and F23389$+$0300:N \citep{n_ea03}.  The two other galaxies with f(1.4~GHz) $>$100~mJy and without known jets are F11119$+$3257 and F17179$+$5444.  An eighth galaxy with a lower radio flux (F04210$+$0401) also possesses a radio jet.  Out of these eight galaxies with high radio fluxes and/or jets, we observe outflows in \nad\ in 4 objects, and in 3 of the 6 galaxies with known jets (see Table \ref{ofprop}).

4C~12.50 is interesting because it is the one Seyfert~2 galaxy in our sample with both a powerful, extended radio jet and a \nad\ outflow.  This galaxy has single-dish and very-long baseline interferometer (VLBI) \ion{H}{1} observations \citep{m_ea04b} and previous long-slit optical observations \citep{htm03} with which to compare our data.  The \ion{H}{1} component seen in the western nucleus with VLBI \citep{m_ea04b} and the average optical emission-line peak redshift agree to within a few tens of \kms.  This suggests that the high-velocity blue wings in the recombination and forbidden lines \citep{htm03} are in fact high-velocity outflowing gas.  We measure the peaks of these emission lines along a PA of 104\degr\ (the line connecting the two nuclei), and our results are in quantitative agreement with those of \citet{htm03}.

In our analysis (\S\ref{analysis}), we extracted the light in this galaxy using two apertures, one for each nucleus, yielding a $\dvmax = 360\kms$ outflow in the eastern nucleus.  An extraction with more apertures confirms this picture.  There is gas blueshifted by $\la$$400-500$~\kms\ from the emission-line peaks over a region extending at least 10~kpc eastward of the eastern nucleus.  These velocities do not reach those of the high-velocity \ion{H}{1} gas ($|\Delta v| = 700-1100$~\kms; \citealt{m_ea04b}) seen in single-dish data, so it is unclear whether they are related.  Our data, however, were not taken along the PA of the radio jet, which is 160\degr\ on scales of $\sim$100~pc \citep{s_ea97}.  There is no clear link between the jet and either the \ion{H}{1} outflow \citep{m_ea04a} or the \nad\ outflow; further observations are needed.

The Seyfert~1 Mrk~231 also possesses a radio jet on scales of tens of
pc to tens of kpc (see \S\ref{mrk231}).  This galaxy has neutral and
ionized gas outflows on kpc scales at velocities up to
$\sim$2100~\kms, along the same axis as the radio jet.  Thus, Mrk~231
may be a ULIRG analogue to the radio galaxies with young stellar
populations that are experiencing high-velocity outflow seen in
\ion{H}{1}, and could serve as an evolutionary link between the two.
It could represent a stage where the radio jet and corresponding
neutral gas outflow have recently removed the dust screen from the
galaxy's active nucleus.  Low-resolution {\it Very Large Array} observations of the nucleus at the velocities of the BALs have not revealed any \ion{H}{1} absorption in Mrk~231 \citep{cwu98}, but sensitive, broad-band, single-dish \ion{H}{1} data just blueward of systemic do not exist for this galaxy (i.e., to study the extended absorbing gas).

Clearly, a sensitive broad-band search for \ion{H}{1} absorption in radio-loud Seyfert ULIRGs is in order to better study the connection between \nad\ outflows in ULIRGs and \ion{H}{1} outflows in radio galaxies, as well as the implications for the evolutionary link between mergers, radio galaxies, and quasars.  Mrk~231 would be an excellent target to begin this search.

\subsection{Starburst- or AGN-Powered Winds?} \label{sb_v_agn}

Is there evidence from our data that AGN help to power large-scale, wide-angle outflows in Seyfert~1 and 2 ULIRGs?  The fact that the infrared luminosities of many of these galaxies may be dominated by (dust-reprocessed) radiation from an AGN rather than a starburst does not a priori imply that the outflows we observe in \nad\ are AGN-driven.  The star formation rates in these galaxies are still high ($\ga$100~\smpy), and thus a starburst should be able to power an outflow on its own.

\subsubsection{Seyfert~1s}

The broad absorption lines we observe in two Seyfert~1 galaxies (F07599$+$6508 and Mrk~231) are unique, as these types of profiles are not found in pure starbursts and belong to the class of BALs found only in quasars.  These lines are most likely produced in or near the broad-line region of the AGN (\S\ref{radius}).  The properties of the narrow, low-velocity ($|\Delta v| \la 1450$~\kms) lines in F11119$+$3257 are similar to UV and X-ray intrinsic absorbers in local Seyfert~1s and different from absorbers in starburst ULIRGs (\S\S\ref{sy1_ofprop} and \ref{intabs}).  They are thus probably powered by the AGN, as well.

The large-scale outflow components in Mrk~231 may be either starburst- or AGN-driven.  The high velocities observed ($\Delta v \la 2100$~\kms) and the presence of a strong jet suggest that the AGN may play some role, though there is certainly a starburst contribution, as well.

In short, the dominant contribution to \nad\ absorbers in nuclear spectra of Seyfert~1s is from the AGN; on larger scales, both the AGN and a nuclear starburst may play a role.  As we show for Seyfert~2s, disentangling the contribution of the two on these scales is difficult.

\subsubsection{Seyfert~2s}

We have shown above that  the properties of outflows in Seyfert~2s are
statistically   consistent   with    those   in   starburst   galaxies
(\S\ref{sy2_ofprop}).  There is a hint of AGN contribution to the wind
in the  higher velocities observed,  but this is not  conclusive.  The
two-dimensional phase-space distributions of starbursts and Seyfert~2s
relating    outflow   and    galaxy    properties   are    essentially
indistinguishable, as well (\S\ref{v_host}).   Thus, the effect of AGN
on outflows in Seyfert~2s is difficult to ascertain; more observations
are  required.  Certainly  we can  say  that the  momentum and  energy
contribution to the large-scale outflow  from the AGN is comparable to
or  smaller  than  the  energy  contribution  from  the  circumnuclear
starburst also present in Seyfert~2 ULIRGs.  (We assume that there are
already powerful starburst-driven winds in these galaxies, as expected
based  on their  star formation  rates and  the results  of Paper~II).
Furthermore,  the outflow  mechanical  energy per  unit luminosity  in
IR-selected starbursts and Seyfert~2 ULIRGs is the same on average.

Two arguments provide some evidence for energy injection into the wind
from an AGN.  (1) The outflow detection rate and median star formation
rate are similar in Seyfert~2 ULIRGs and the starburst IRGs, but their
outflow    velocity   distributions   are    significantly   different
(\S\ref{sy2_ofprop}).   (2) The  correlation between  the low-velocity
neutral gas and  the FWHM of \otl\ (\S\ref{emlprop})  is driven by the
Seyfert~2 galaxies in our  sample, indicating a connection between the
neutral outflow and the extended NLR.

\section{SUMMARY} \label{summary}

We have demonstrated that outflows are a common phenomenon in infrared-selected galaxies, both in starbursts (Papers I and II) and Seyferts (this work).  Here, we summarize our conclusions about outflows in infrared-luminous Seyferts.

{\bf(1)} {\bf Detection Rate.}  We find a $\sim$45\%\ detection rate in infrared-luminous galaxies (mostly ULIRGs) that are optically classified as Seyfert~2s.  This detection rate is lower than in starburst ULIRGs, but comparable to that in lower-luminosity starbursts.  We argue that the detection rate reflects the wind geometry, and that outflows are found in all Seyfert~2 ULIRGs.  Given that Seyfert~2 ULIRGs host strong starbursts, and that all infrared-luminous starbursts host large-scale outflows (Paper~II), this is expected.

The presence of strong starbursts in Seyfert~1s implies that extended outflows also exist in the majority of these galaxies, but the high luminosities of the nuclei prevent their detection (except in Mrk~231).  The \nad\ outflow detection rate in the nuclear spectra of Seyfert~1 ULIRGs is still high ($\sim$50\%), but these nuclear outflows are physically distinct from those in Seyfert~2s and starbursts.

{\bf(2)} {\bf Velocities.}  We measure typical outflow velocities and maximum velocities of $-220$ and $-450$~\kms, respectively, in Seyfert~2 ULIRGs.  The upper limit for $|\dvmax|$ is high, however, at 1550~\kms\ (in F05024$-$1941).  There are few statistically significant differences between the velocities of Seyfert~2 ULIRGs and starbursts of comparable luminosity.  However, some significant differences arise when comparing Seyfert~2s to starbursts of lower infrared luminosity.  When separating our entire sample by spectral type, Seyfert~2s and LINERs have no significant differences in velocity, while \ion{H}{2} galaxies differ significantly from both LINERs and Seyfert~2s.

We find very high (up to $\Delta v \sim 10^4$~\kms) velocities in two broad absorption-line (BAL) Seyfert~1s.  The velocities in the nuclear spectrum of one Seyfert~1, F11119$+$3257, are lower, however; these absorbers may be analogous to UV and X-ray intrinsic absorbers in local Seyfert~1s.  We also find extended, high-velocity ($\Delta v \la 2100$~\kms) absorption in Mrk~231, confirming a suggestion by \citet{hk87} that this galaxy hosts a large-scale outflow along with its nuclear BAL outflow.

We observe significant correlations between the velocity of the highest column density neutral gas and the FWHM of the \otl\ line in the combined starburst $+$ Seyfert~2 sample.  This implies that the ionized and neutral gas are physically connected and could indicate a connection of neutral outflows to the narrow-line region in Seyferts.  However, the highest-velocity ionized and neutral gas are not significantly correlated.

{\bf(3)} {\bf Radius.}  The nuclear absorbers in the two BAL Seyfert~1s in our sample and in F11119$+$3257 probably arise on scales of tens of parsec or less.  However, we also observe blueshifted absorption at scales up to 4~kpc from the nucleus of Mrk~231.  In Seyfert~2s, the absorbing gas likely arises on kpc scales.  We directly observe extended, blueshifted absorption in three Seyfert~2s, and other lines of evidence support this picture.

{\bf(4)}  {\bf   Starburst  or   AGN?}   The  lack   of  statistically
significant differences between outflows in Seyfert~2 ULIRGs and those
in starburst  ULIRGs implies that the  contribution of the  AGN to the
large-scale outflows in  these galaxies is comparable to  or less than
the  contribution of  the  starburst.  (This  assumes  that there  are
starburst-driven winds  in these galaxies,  a conclusion based  on the
high star formation rates of  Seyfert~2s and the results of Paper~II.)
Furthermore, we  find that the outflow mechanical  luminosity per unit
radiative  luminosity  is  the  same  for  Seyfert~2s  and  starbursts
($5\times 10^{-4}$ on average).

Evidence  for  some  AGN  contribution  to the  wind  comes  from  the
observation  that  there   are  significant  differences  between  the
velocities of  outflows in Seyfert~2 ULIRGs and  starburst IRGs, while
their outflow  detection rates and  star formation rates  are similar.
The  correlation between  low-velocity  neutral and  ionized gas  also
suggests a connection of the outflow to the extended NLR.

The outflowing  \nad\ absorbers we  observe in Seyfert~1s  are powered
dominantly  by the  central  black  hole engine  and  emerge on  small
scales.  This  is based  on physical arguments  and comparison  to the
properties of  intrinsic absorbers.  There are  most likely starburst-
or AGN-powered  outflows in the  host galaxies of most  Seyfert~1s, as
well, as  illustrated by Mrk~231.  The high  nuclear luminosity limits
detection of a large-scale outflow in other galaxies.

{\bf(5)} {\bf Context.}  ULIRGs are believed to be part of an evolutionary sequence in which two massive, gas-rich galaxies merge; a dusty, intense starburst is created; the obscuring dust is blown away to reveal a quasar; and the galaxy evolves into an elliptical \citep[e.g.,][]{s_ea88}.  The small- and large-scale starburst- and AGN-driven outflows that we observe in these galaxies may be a part of this process.  Several of the Seyfert galaxies in our sample reveal unique stages in this process.  Particularly, the Seyfert~1s that show low-ionization BALs may be instances where dusty material near the quasar is being blown away.  Two of the Seyfert galaxies in our sample (Mrk~231 and 4C~12.50) also host strong radio jets and outflows in \nad.  The latter contains a high-velocity outflow observed in \ion{H}{1} 21~cm \citep{m_ea04a}, which in other radio galaxies is observed to be powered by a radio jet.  Jets may turn on late in the merger sequence in some ULIRGs as they evolve into unobscured quasars.

The average fraction of neutral outflowing gas that escapes Seyfert~2s and enters the IGM is non-negligible, perhaps as high as 60\%.  Thus, AGN-dominated ULIRGs are more likely to pollute the intergalactic medium with gas and energy than the starburst-dominated ULIRGs.  These outflows will be metal-enriched because of the strong starbursts present in these Seyferts.

{\bf(6)} {\bf  Outlook.}  Since  outflows in Seyfert~2  ULIRGs receive
substantial  momentum  and   energy  injection  from  a  circumnuclear
starburst, it would clearly be  useful to observe a sample of Seyferts
in which the  starburst component is minimal in  order to better study
energy injection from the AGN.  Observing Seyferts selected by nuclear
X-ray, radio,  and/or optical luminosity  would allow us to  study the
absorption-line phase of outflows in a sample of Seyferts that is less
biased towards  starbursting galaxies.  It would  be especially useful
to compare  to a set  of galaxies which have  well-studied narrow-line
regions in emission and/or UV and X-ray absorbers.  Ideally, one would
also want  to select a sample  at similar redshifts  to those observed
here in order to minimize aperture effects.

Another follow-up study suggested by this work is to survey radio-loud Seyferts with \nad\ outflows using sensitive, broad-band \ion{H}{1} 21~cm observations.  This may be one way to directly connect neutral Na to outflowing \ion{H}{1} and confirm our outflow mass estimates, as well as to illuminate the physics of energy injection (since radio sources can be studied at high angular resolution).  Mrk~231 is a particularly appealing target for this type of study. 

\acknowledgments

We thank Kalliopi Dasyra, Dong-Chan Kim, and Barry McKernan for supplying useful data prior to publication.  DSR is supported by NSF/CAREER grant AST-9874973.  During this work, SV was partially supported by a Cottrell Scholarship awarded by the Research Corporation, NASA/LTSA grant NAG 56547, and NSF/CAREER grant AST-9874973.  This research has made use of the NASA/IPAC Extragalactic Database (NED), which is operated by JPL/Caltech under contract with NASA.  It also makes use of data products from the Two Micron All Sky Survey, which is a joint project of the University of Massachusetts and IPAC/Caltech, funded by NASA and NSF.  The authors wish to recognize and acknowledge the very significant cultural role and reverence that the summit of Mauna Kea has always had within the indigenous Hawaiian community.  We are most fortunate to have the opportunity to conduct observations from this mountain.

\appendix
\section{APPENDIX: COMMENTS ON INDIVIDUAL OBJECTS} \label{app}

Here we discuss unique properties of individual galaxies based on our spectra.

\begin{itemize}

\item{{\bf F05024$-$1941.}  The blueshifted \nad\ component observed at $-1550$~\kms\ is seen also in \ion{Ca}{2} H \& K \citep[see][]{rvs02}.}

\item{{\bf F08559$+$1053.}  There is an emission-line object $4\farcs9$ north of the galaxy, possibly related to the faint, compact object seen in the $R$-band image \citep{vks02}.  No continuum is visible, but several emission lines (elongated and tilted) are, yielding $z = 0.253$.  \ntl\ is not visible but H$\alpha$ is, suggesting flux(H$\alpha$) $\gg$ flux(\nt).  \otl\ is also brighter than H$\beta$ by an uncertain factor (not more than a few).}

\item{{\bf F13443$+$0802.}  The southwest nucleus of this triplet is listed in \citet{vks99b} as a Seyfert~2.  The Seyfert~2 nucleus is actually the eastern nucleus (and the correct line ratios are listed in \citealt{vks99b} under the label of the SW nucleus), while the other two nuclei are an unknown spectral type (SW) and an \ion{H}{2} galaxy (NE).}

\item{{\bf F13451$+$1232}.  \citet{vsk99c} claim to observe a broad Pa$\alpha$ line in this galaxy.  We observe similar asymmetric profiles in the optical in the western nucleus (which hosts most or all of the emission line flux), but they also appear to be present in the [\ion{O}{1}] $\lambda\lambda6300,~6364$ forbidden lines (as $2000-3000$~\kms\ blue wings), which suggests that they do not originate in the broad line region.  It is difficult to discern the situation in the east nucleus due to its small separation from the west nucleus, which makes separation of the emission lines from the two nuclei very hard.  However, the optical continuum appears to peak in the east nucleus.}

\end{itemize}

\clearpage

\begin{figure}
\plotone{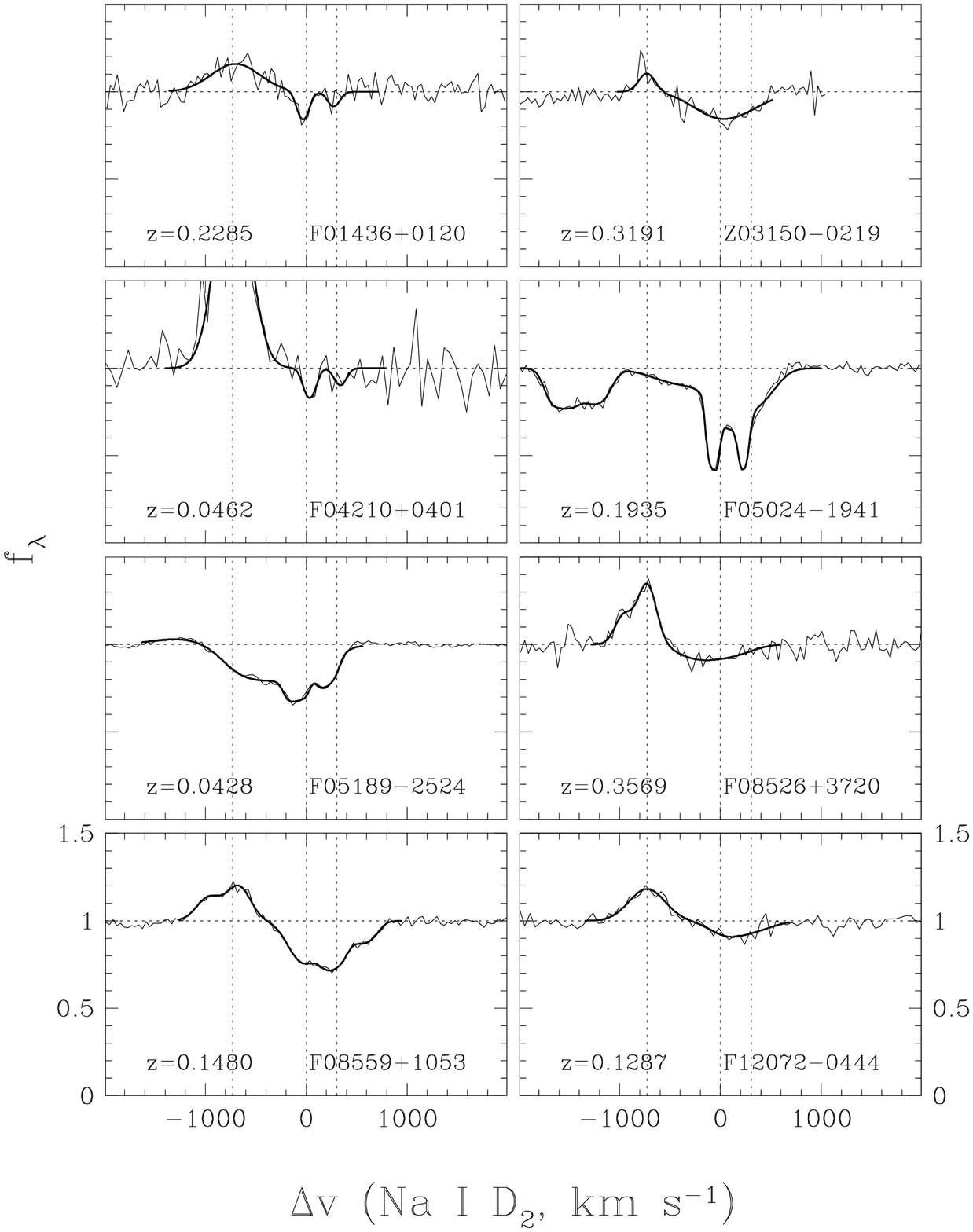}
\caption{Spectra of the \nad\ line in infrared-luminous Seyfert~2 galaxies.  The thin lines are the (smoothed) original spectra and the thick lines are fits to the data.  The vertical dotted lines locate the \nad~$\lambda\lambda5890,~5896$ doublet and \ion{He}{1} $\lambda5876$ emission line in the rest frame of the galaxy.  The diagonal hashed lines locate atmospheric absorption from O$_2$.}
\label{spec_s2}
\end{figure}
\setcounter{figure}{0} 
\begin{figure}
\plotone{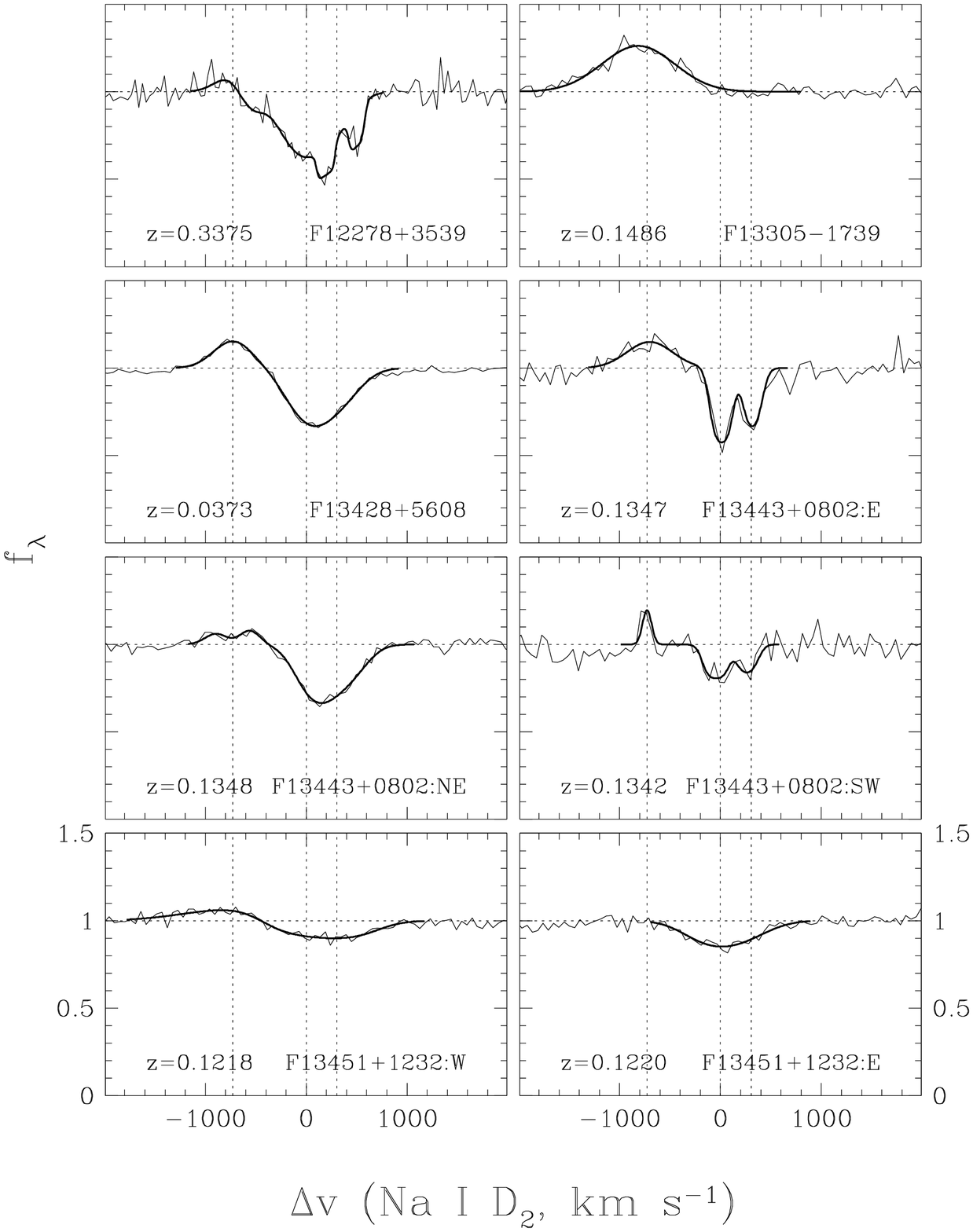}
\caption{\it{Continued.}}
\end{figure}
\setcounter{figure}{0}
\begin{figure}
\plotone{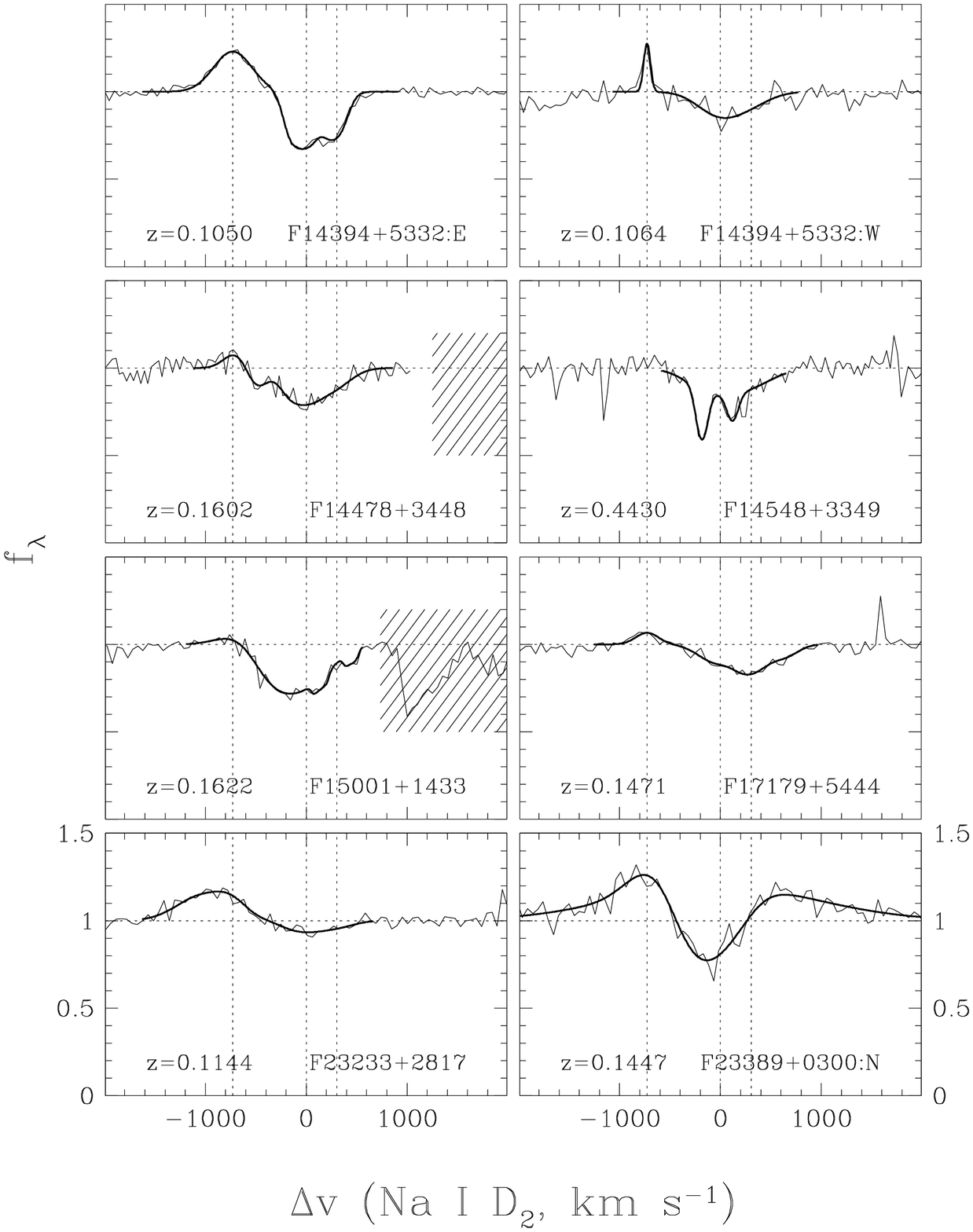}
\caption{\it{Continued.}}
\end{figure}
\setcounter{figure}{0}
\epsscale{0.60}
\begin{figure}
\plotone{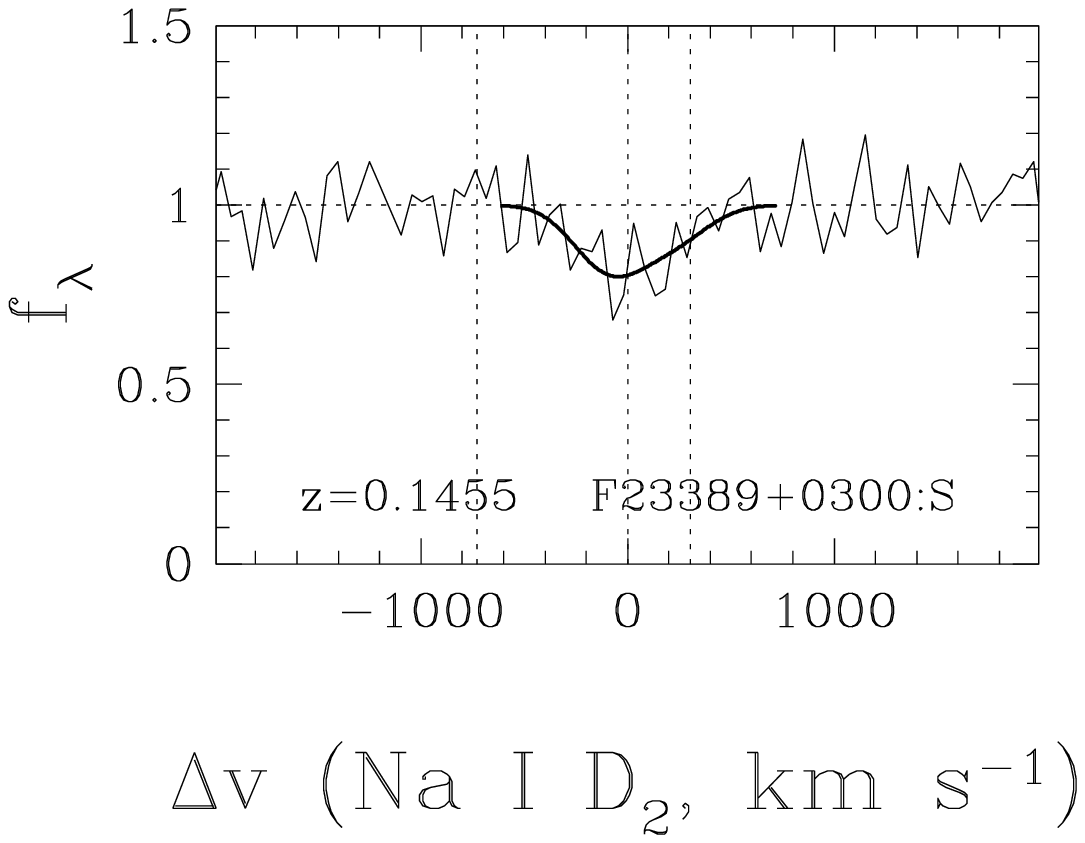}
\caption{\it{Continued.}}
\end{figure}
\epsscale{1.0}

\clearpage

\begin{figure}[t]
\plotone{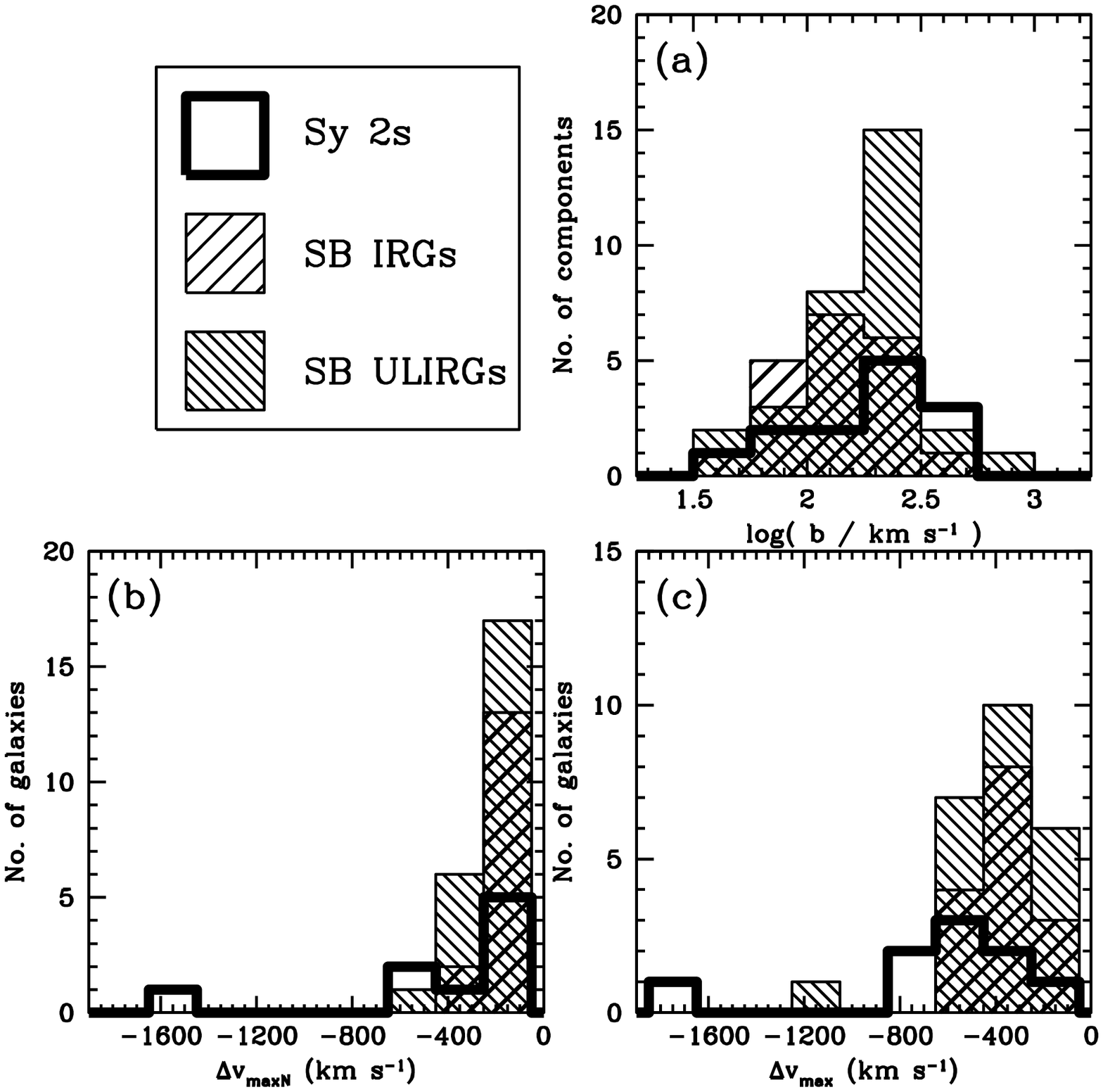}
\caption{Distributions of (a) Doppler parameter, (b) velocity of the highest column density gas, and (c) maximum velocity ($\dvmax = \Delta v - \mathrm{FWHM}/2$) in Seyfert~2 ULIRGs, infrared-luminous starbursts, and starburst ULIRGs.  The starburst data are from Paper~II.  We used K-S and Kuiper tests to look for significant differences among the different subsamples.  Statistically significant differences (P[null] $< 0.1$ in both tests) exist only in \dvtau, between the starburst ULIRGs and the starburst IRGs, and between Seyfert~2 ULIRGs and starburst IRGs.  See Table \ref{pnull} for the computed values of P(null), and \S\ref{sy2_ofprop} for further discussion.}
\label{histdv}
\end{figure}

\begin{figure}[t]
\plotone{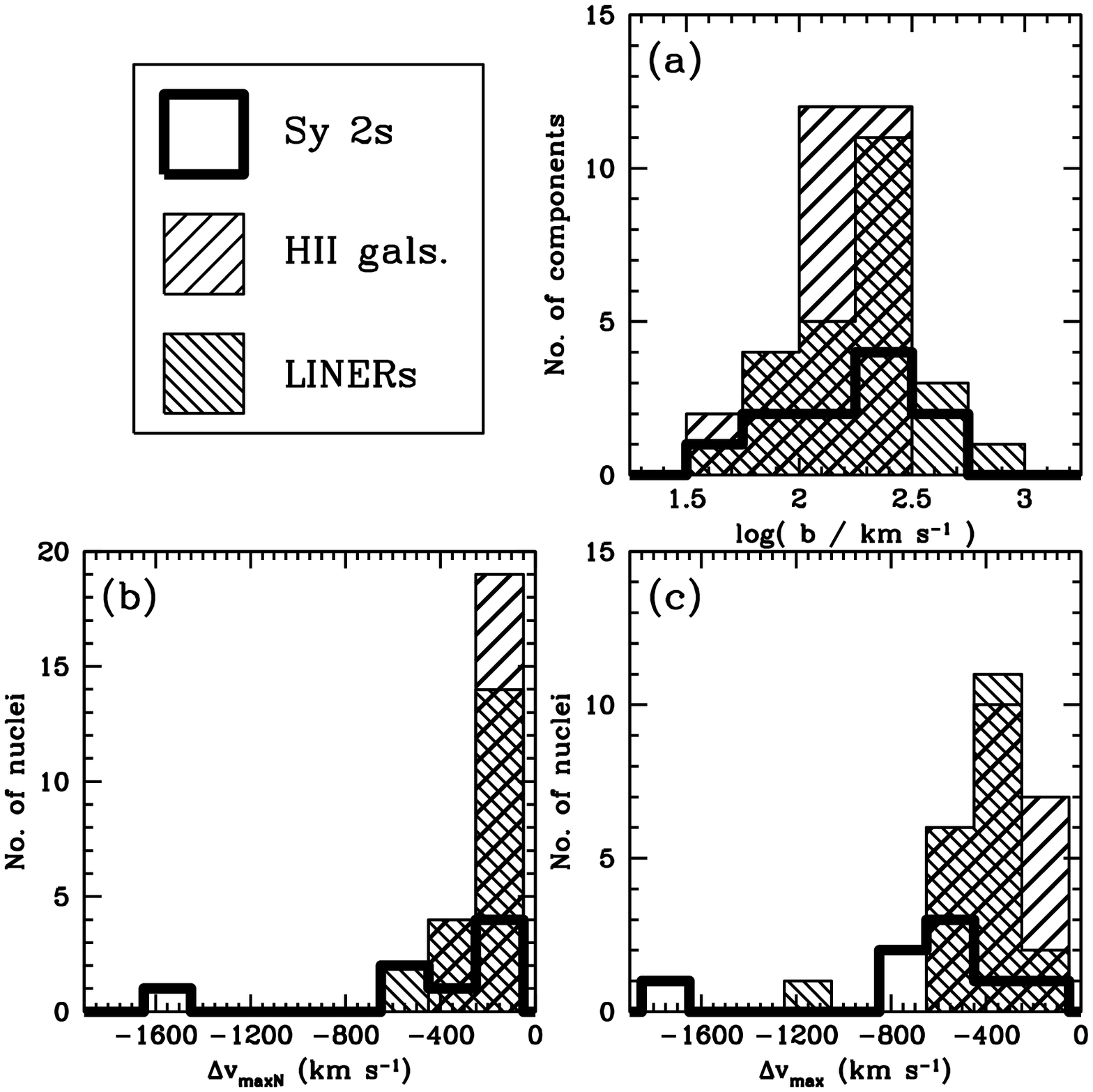}
\caption{Same as Figure \ref{histdv}, but the subsamples compared are the Seyfert~2s, \ion{H}{2} galaxies, and LINERs.}
\label{histdvhls}
\end{figure}

\epsscale{0.8}
\begin{figure}[t]
\plotone{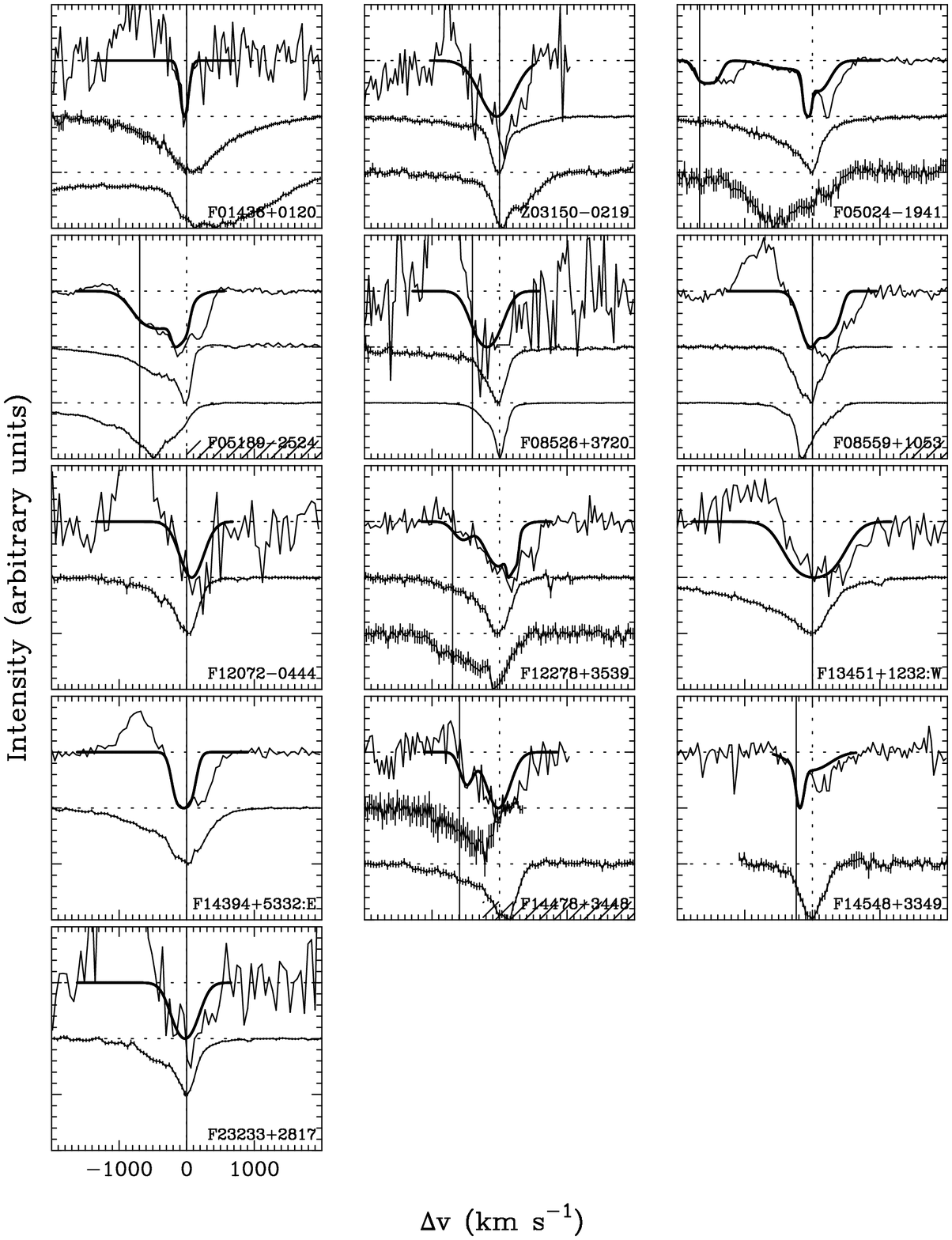}
\caption{\small{Emission-line spectra of 13 infrared-luminous Seyfert~2 galaxies from our sample which have noticeable emission-line asymmetries in the profile wings based on our moderate-resolution data.  The top spectrum (red) in each panel is \nad, with our fit to the \nad$_2$ $\lambda5890$ line superimposed (green).  The velocity scale is relative to systemic in this line.  The middle spectrum (blue) is \nt\ $\lambda6548$ for $\Delta v < 0$ and (properly scaled) \ntl\ for $\Delta v > 0$.  The bottom spectrum (orange) is \otl.  The vertical thick line locates the neutral gas maximum velocity, \dvmax.  The vertical error bars in the emission-line spectra are 2$\sigma$ errors.  Twelve of these galaxies have a blue emission-line wing (in \ot\ and/or \nt) with a higher maximum velocity and/or more flux than the red wing, plus another three based on low-resolution data, for a total of 15 of 20 Seyfert~2s.  See \S\ref{emlprop} for further discussion.}}
\label{emlspec}
\end{figure}
\epsscale{1.0}

\begin{figure}[t]
\plotone{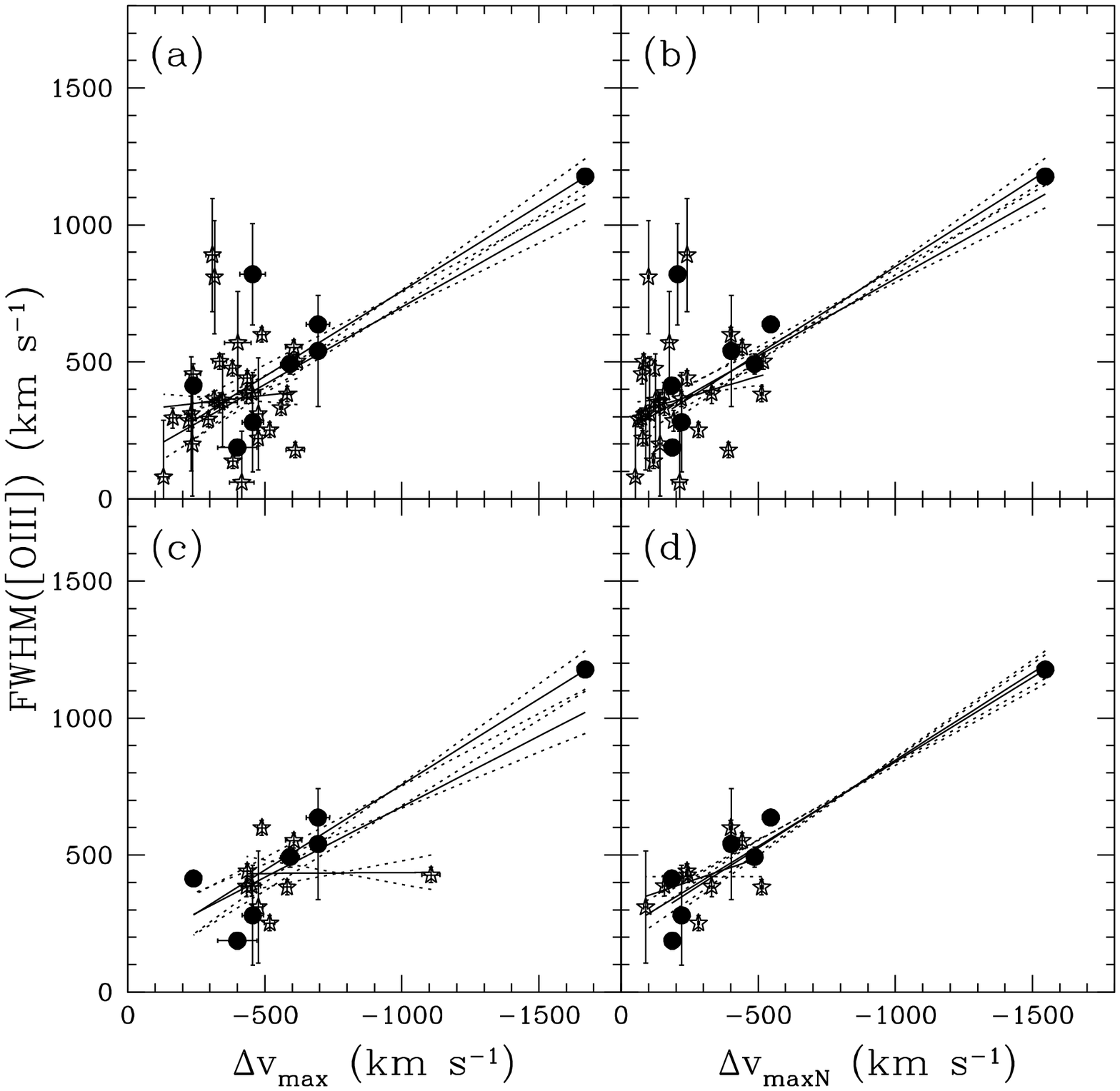}
\caption{Full-width at half-maximum of the \otl\ emission line as a function of (a) \dvmax\ and (b) \dvtau\ for all galaxies with \ot\ measurements and (c) \dvmax\ and (d) \dvtau\ for just those galaxies with obvious blue emission-line asymmetries (BELA) in \ot\ or \ntll.  Open stars (red) represent starbursts from Paper~II and filled circles (blue) represent Seyfert~2s.  The three solid lines are fits (with 1$\sigma$ errors shown as dashed lines) to the starbursts and Seyferts separately and to the whole sample.  We find no significant correlations in the starburst galaxies alone.  When the Seyfert~2s are added, correlations emerge between FWHM(\ot) and \dvtau, both in all the galaxies and in only those nuclei with BELA.  These correlations remain in the absence of the highest-velocity point (F05024$-$1941).  These plots indicate a connection of neutral outflows in Seyfert~2s to the `extended' narrow-line region (\S\ref{emlprop}).}
\label{fwhmcorr}
\end{figure}

\begin{figure}[t]
\plotone{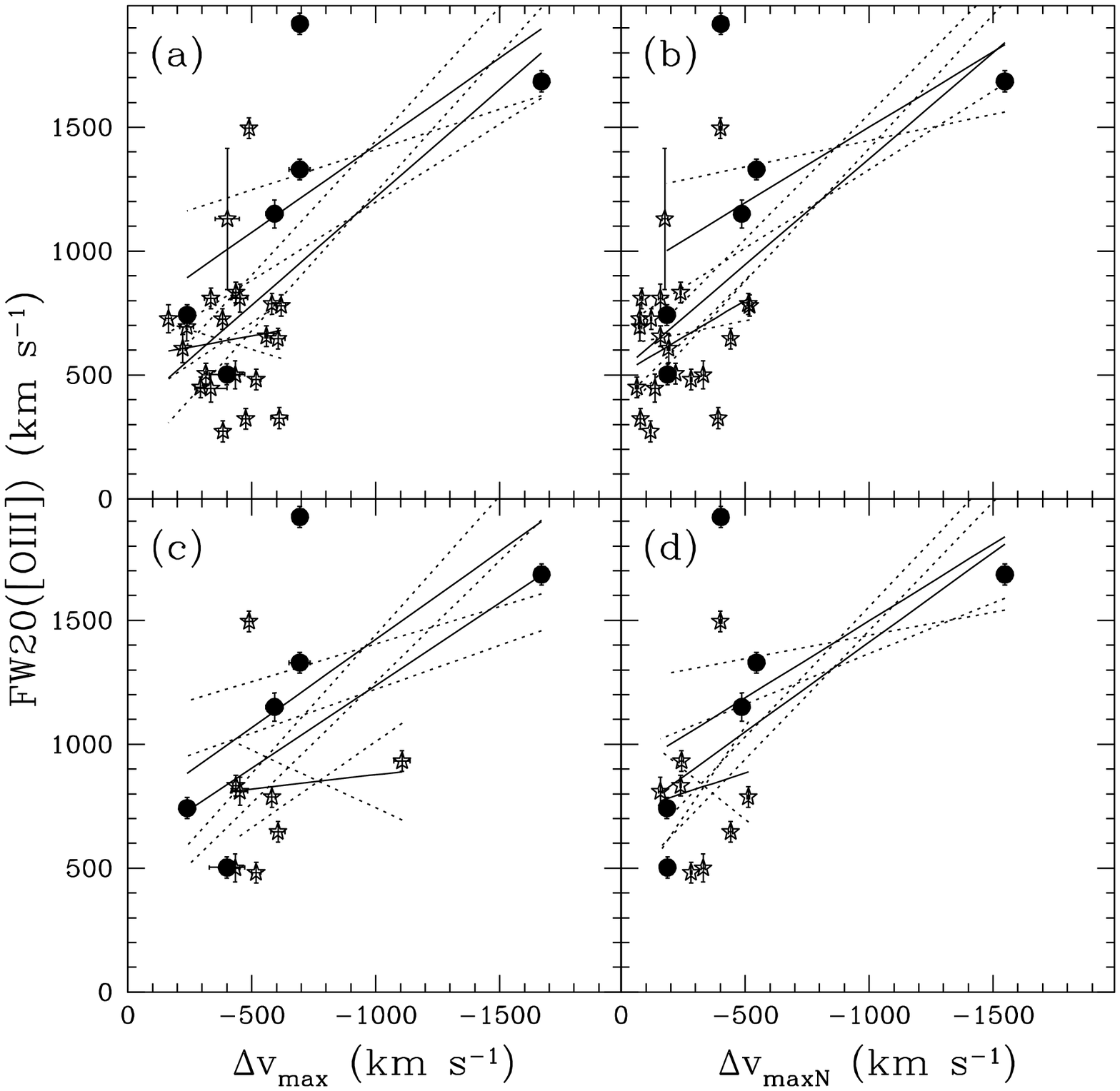}
\caption{Same as Figure~\ref{fwhmcorr}, but for the full-width at 20\%\ of maximum intensity of the \otl\ emission line.  There are no significant correlations.  This implies that the inner narrow-line region in Seyfert~2s, where the high-velocity ionized gas arises, is unconnected with the neutral outflows observed in \nad\ (\S\ref{emlprop}).}
\label{fwtmcorr}
\end{figure}

\clearpage

\begin{figure}
\plotone{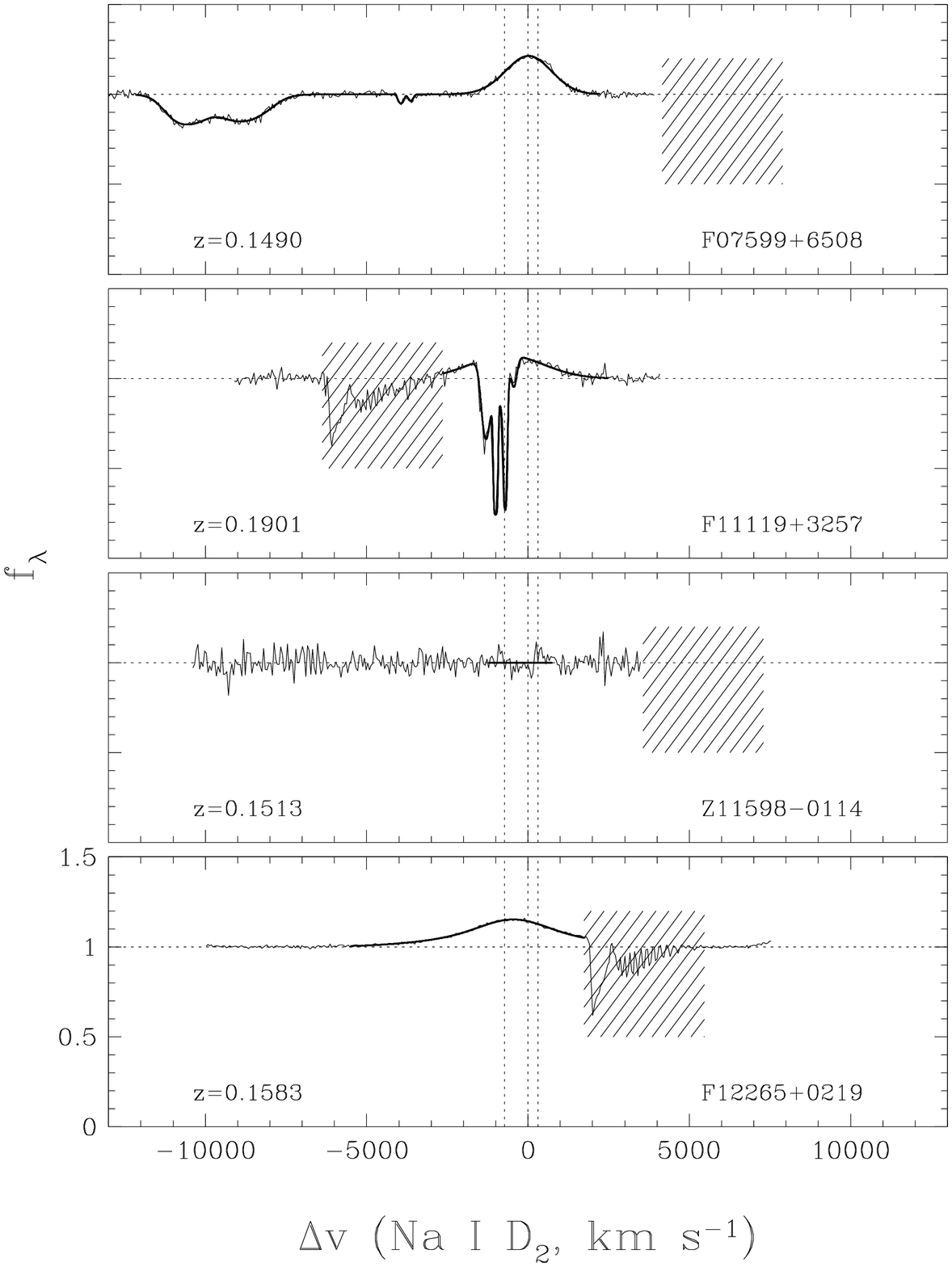}
\caption{Spectra of the \nad\ line in Seyfert~1 ULIRGs.  See Figure~\ref{spec_s2} for more details.}
\label{spec_s1}
\end{figure}
\setcounter{figure}{6} 
\begin{figure}
\plotone{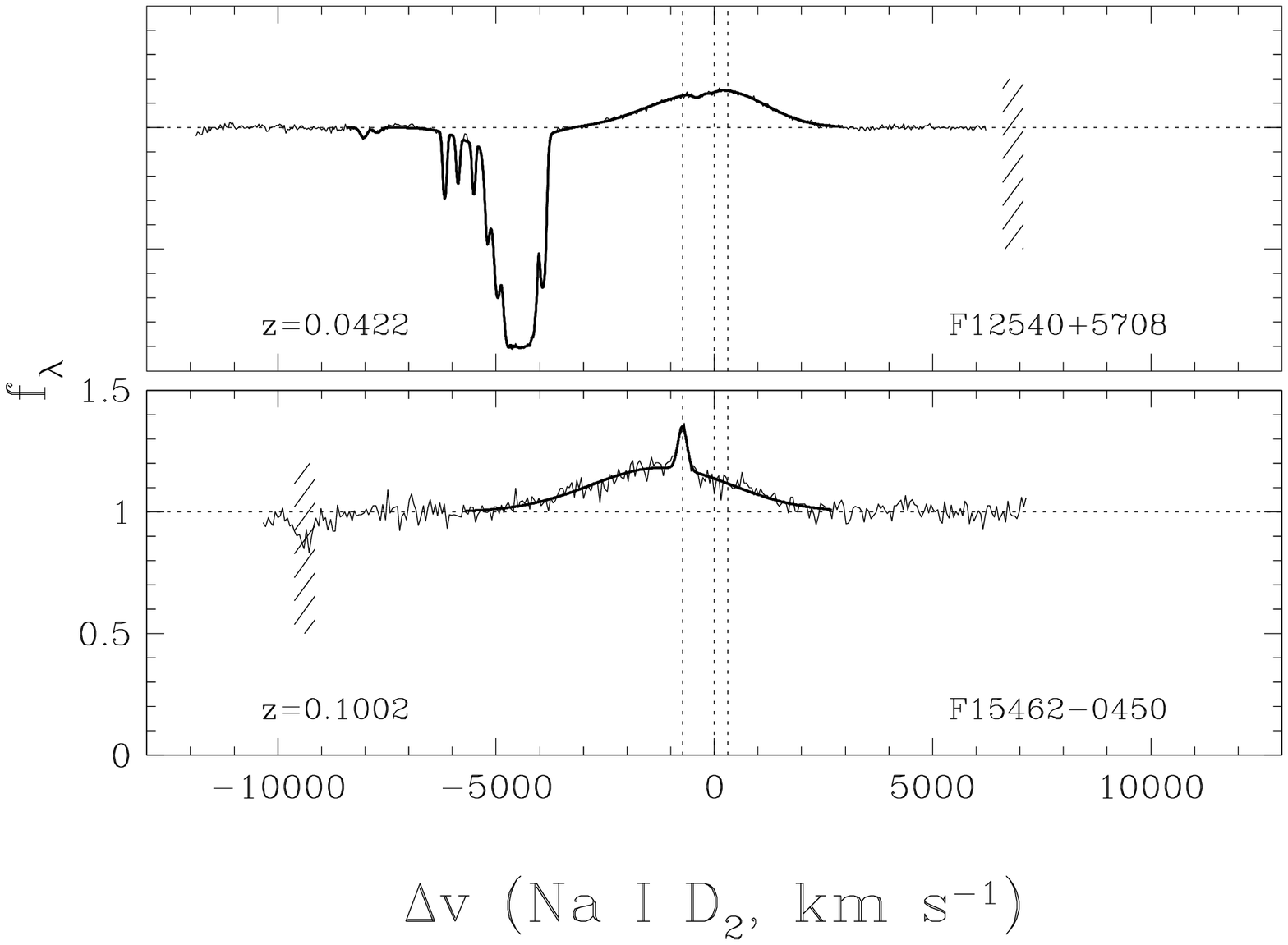}
\caption{\it{Continued.}}
\end{figure}

\clearpage

\begin{figure}[t]
\plotone{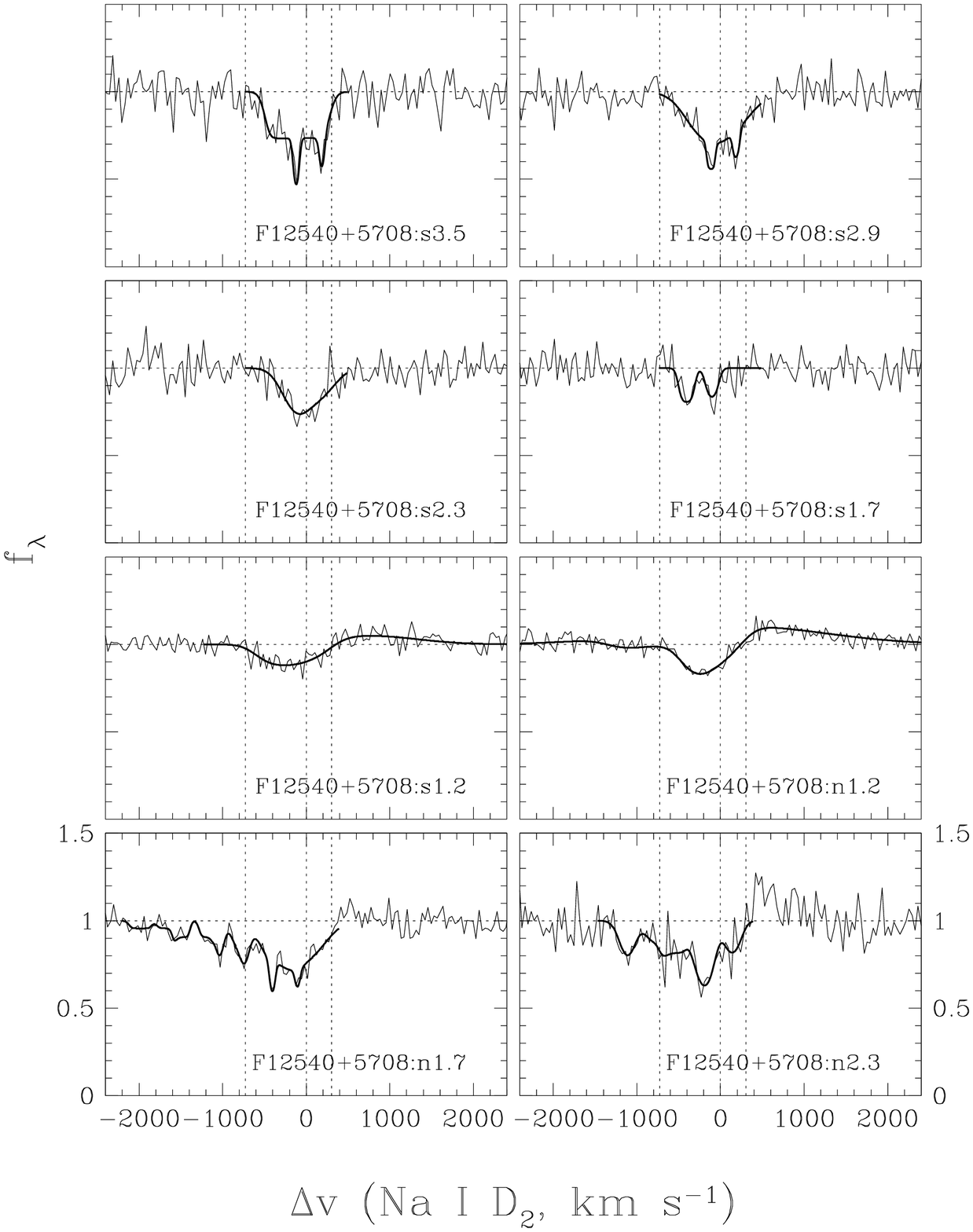}
\caption{Spectra of low-velocity, extended \nad\ emission in Mrk~231 along the north-south slit.  The aperture labels indicate position, in kpc, relative to the nucleus (s = south, n = north).  See Figure~\ref{spec_s2} for more details.  Further discussion of Mrk~231 is in \S\ref{mrk231}.}
\label{mrk231_spec}
\end{figure}

\begin{figure}[t]
\plotone{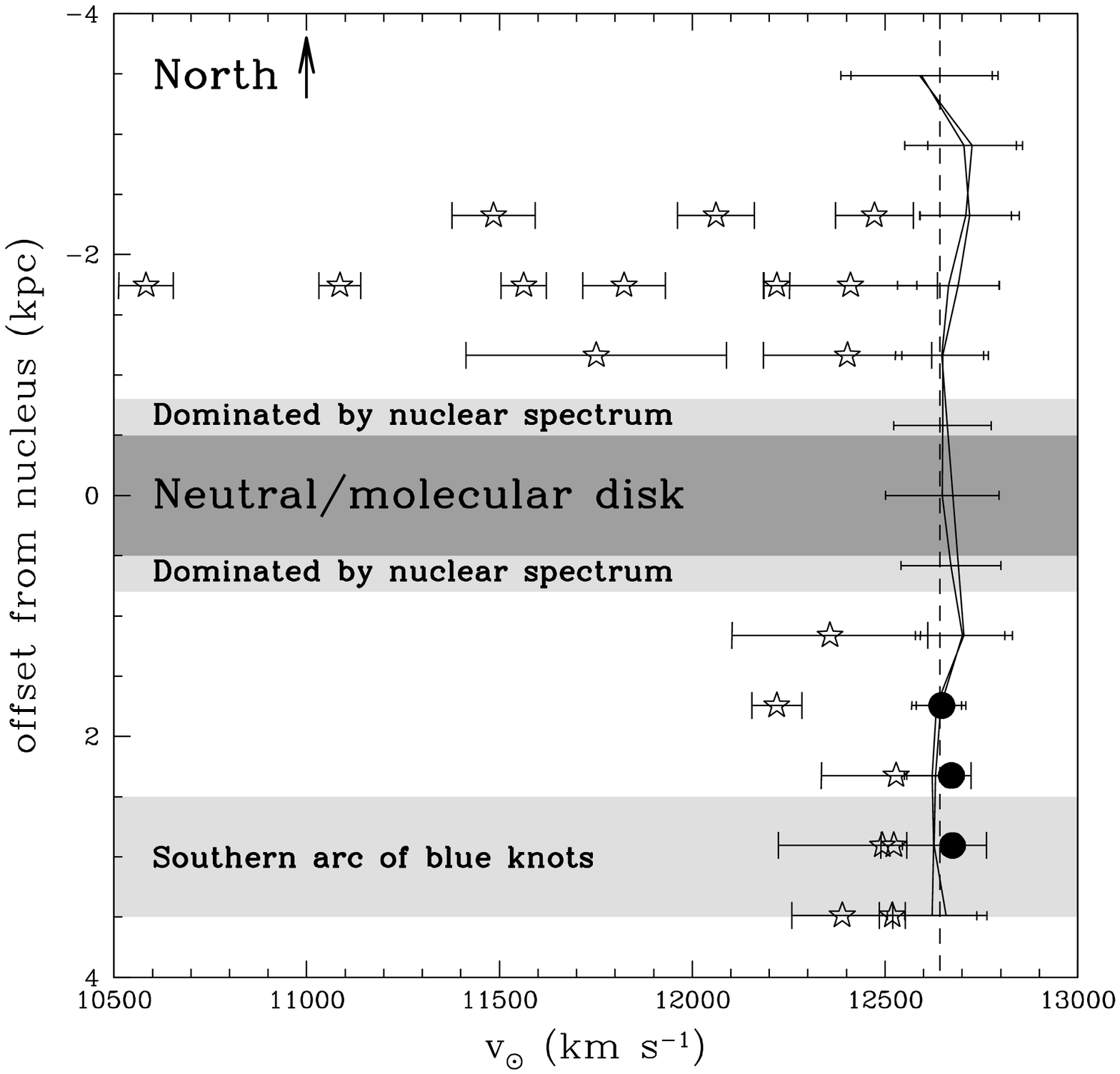}
\caption{Velocities of emission and absorption lines in Mrk~231 along the north-south slit (north is up).  The vertical dashed line is the systemic velocity of the nucleus from \ion{H}{1} and CO data \citep{cwu98,sss91}.  The jagged vertical lines show the velocities of the narrow component of H$\alpha$ (red) and [\ion{S}{2}] $\lambda\lambda6716,~6731$ (black).  The closed circles (green) show the velocity of the \ion{Ca}{2} triplet.  Open stars (blue) locate outflowing \nad\ components.  Error bars on the emission and absorption components show their FWHM.  The shaded regions locate the regions dominated by the nuclear spectrum in the three central bins, as well as the location and extent of the nuclear \ion{H}{1}/CO disk and the southern arc of blue star-forming knots.  The best model for these data as well as the emission lines in Mrk~231 is a wide-angle wind; see \S\ref{mrk231} for further discussion.}
\label{mrk231_vel}
\end{figure}

\clearpage

\begin{figure}[t]
\plotone{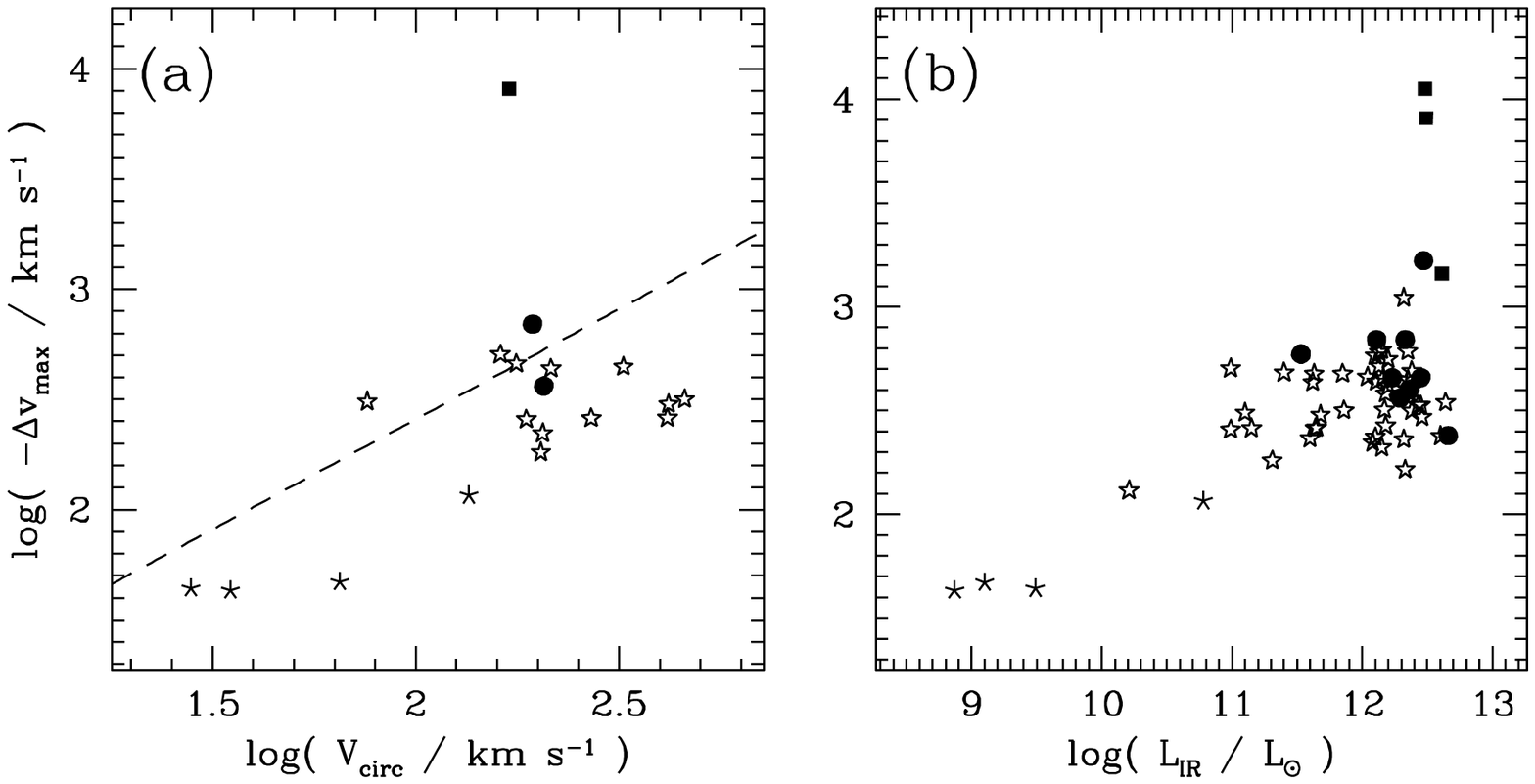}
\caption{Maximum outflow velocity vs. (a) circular velocity and (b) infrared luminosity.  Skeletal stars (red) are dwarf galaxies from \citet{sm04}; open stars (red) are infrared-luminous starbursts from Paper~II; closed circles (blue) are Seyfert~2s; and closed squares (black) are Seyfert~1s.  Note that the phase-space locations of starburst ULIRGs and Seyfert~2 ULIRGs are not significantly different.  The dashed line in (a) is the escape velocity of a singular isothermal sphere at a radius $r = 0.1 r_{max}$.  See \S\ref{v_host} for more discussion.}
\label{dvmax_v}
\end{figure}

\begin{figure}[t]
\plotone{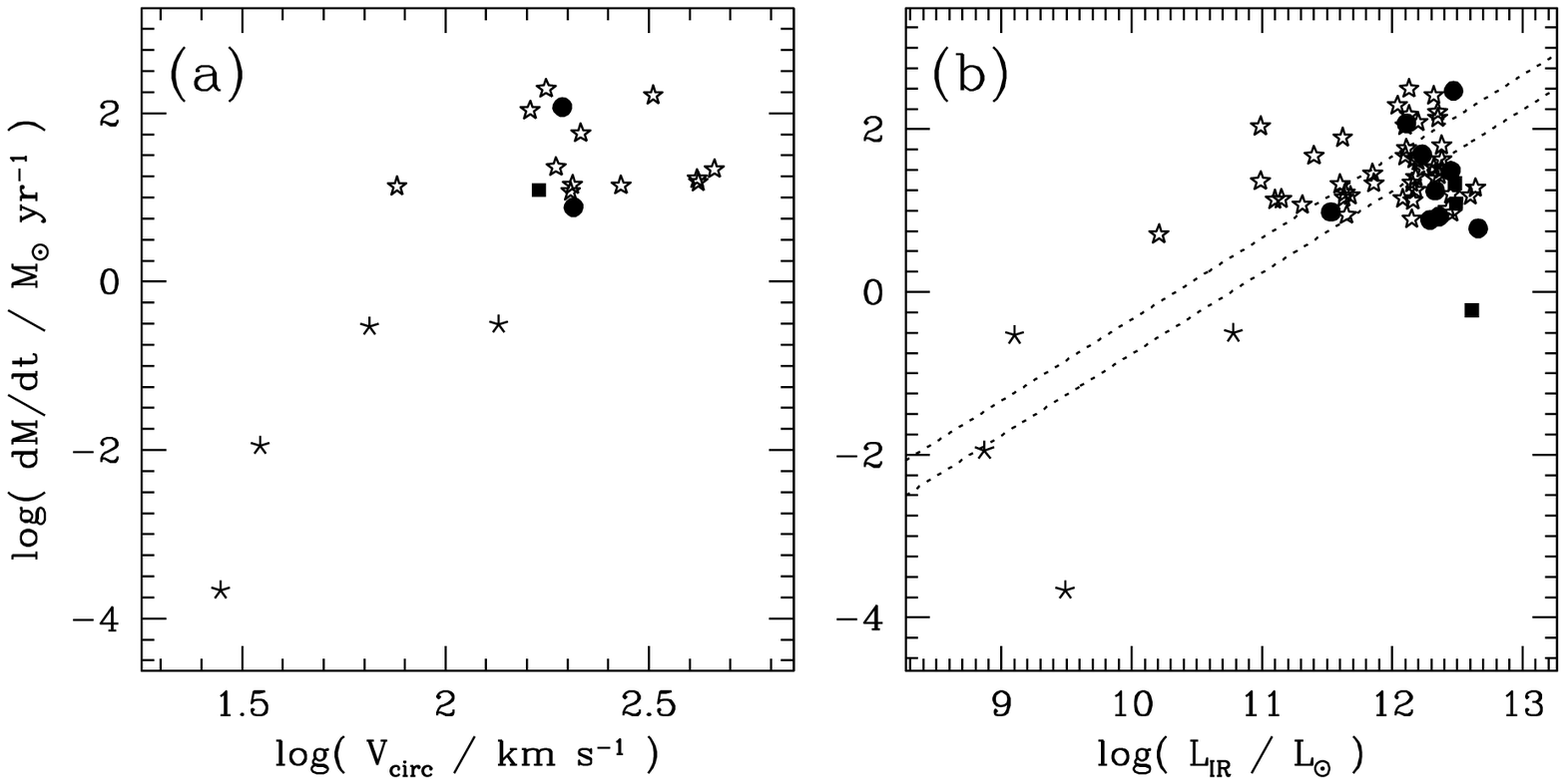}
\caption{Mass outflow rate vs.  (a) circular velocity and (b) infrared
  luminosity.   Symbols  are  as  in Figure~\ref{dvmax_v}.   The  mass
  outflow rates are  computed assuming an outflow radius  of 5 kpc for
  the Seyfert~2 ULIRGs and starbursts, 10 pc for the Seyfert~1s, and a
  varying  radius   for  the  dwarfs  \citep{sm04}.    Note  that  the
  phase-space locations  of starburst ULIRGs and  Seyfert~2 ULIRGs are
  not  significantly   different.   The  short-dashed   lines  in  (b)
  represent the injected hot gas  mass outflow rate from stellar winds
  and  supernovae  for  a  continuous starburst  of  age  $\ga$40~Myr,
  assuming twice solar metallicity  and a Salpeter IMF ($1-100$~\msun;
  \citealt{l_ea99}).  We  assume a conversion  from \lir\ to  SFR such
  that the starburst powers 80\%\ (for the starburst ULIRGs; top line)
  and  30\%\   (for  the  Seyfert   ULIRGs;  bottom  line)   of  \lir\
  (\S\ref{sfr}).  See \S\ref{v_host} for more discussion.}
\label{dmdt_v}
\end{figure}

\begin{figure}[t]
\plotone{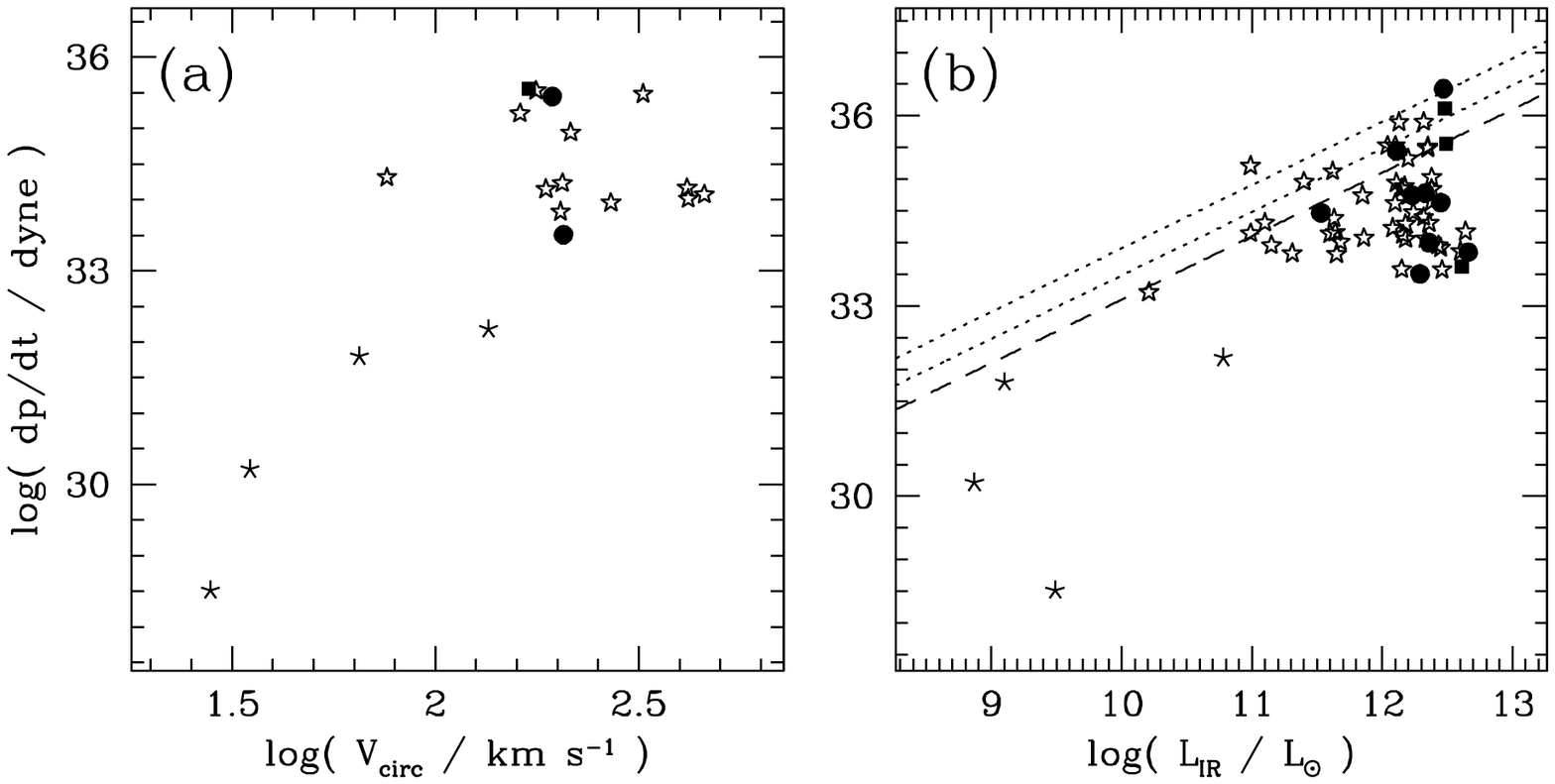}
\caption{Same as  Figure~\ref{dmdt_v}, but for  momentum outflow rate.
  The  long-dashed line  in (b)  is the  momentum injection  rate from
  radiation pressure, assuming  isotropic absorption of the bolometric
  luminosity of the  galaxy by optically thick clouds.   Note that the
  phase-space locations  of starburst ULIRGs and  Seyfert~2 ULIRGs are
  not significantly  different.  Note also that the  momentum from the
  starburst is in  each case (except perhaps one)  sufficient to power
  the outflow,  implying that the momentum injection  into the outflow
  from the  AGN is similar  to or less  than that from  the starburst.
  See \S\ref{v_host} for more discussion.}
\label{dpdt_v}
\end{figure}

\begin{figure}[t]
\plotone{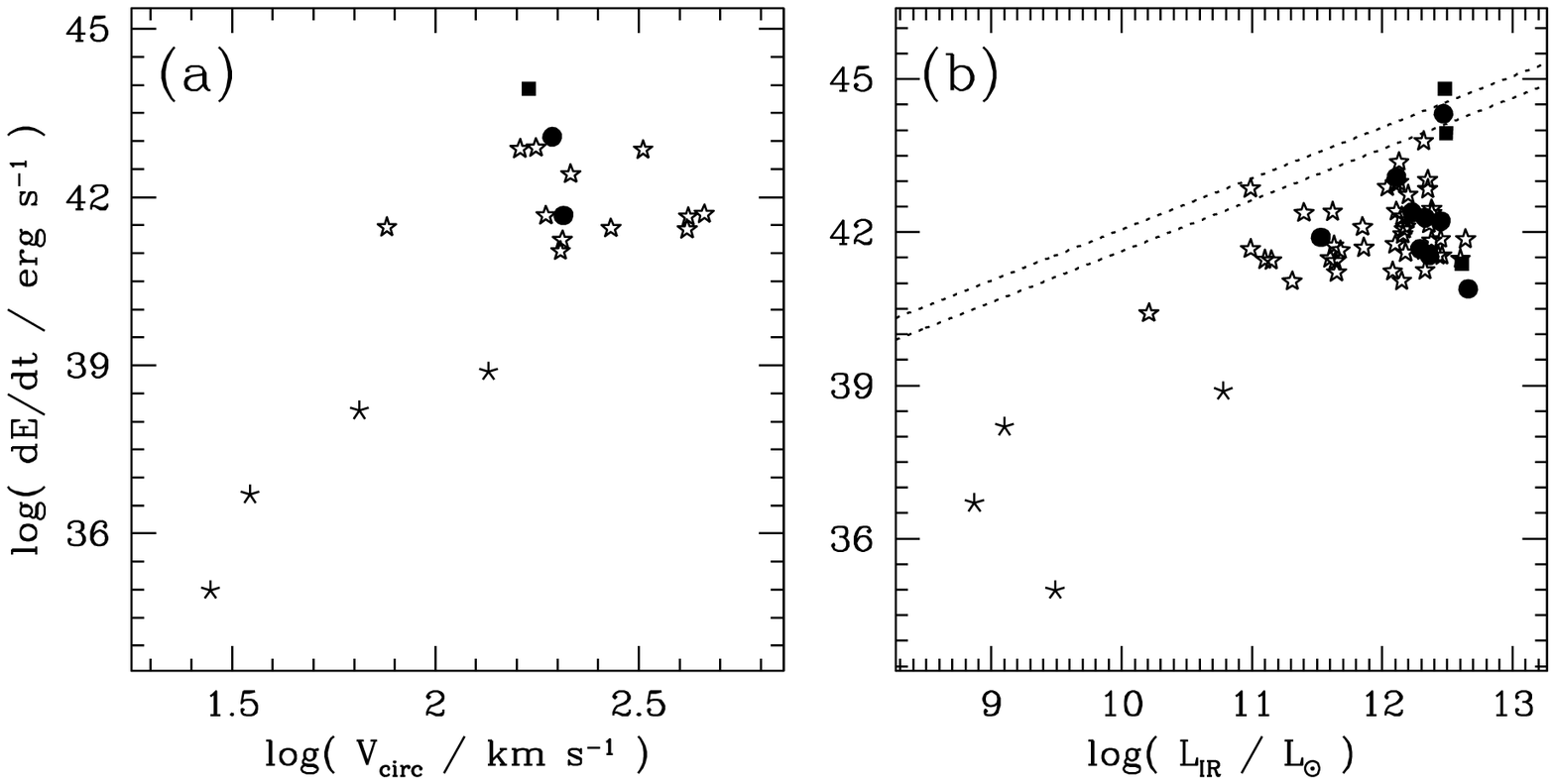}
\caption{Same  as Figure~\ref{dmdt_v},  but for  energy  outflow rate.
  Note  that  the  phase-space   locations  of  starburst  ULIRGs  and
  Seyfert~2  ULIRGs  are not  significantly  different,  and that  the
  energy  from  the  starburst  is  sufficient to  drive  the  outflow
  (implying that the energy in the outflow originating from the AGN is
  comparable to that originating from the starburst).  Furthermore, on
  average,  Seyfert~2s  and  starbursts  have approximately  the  same
  mechanical   luminosity   per   unit  radiative   luminosity.    See
  \S\ref{v_host} for more discussion.}
\label{dedt_v}
\end{figure}

\clearpage

\begin{deluxetable}{lcccrrrrcrcc}
\rotate
\tabletypesize{\footnotesize}
\tablecaption{Galaxy Properties \label{objprop}}
\tablewidth{0pt}
\tablehead{
\colhead{Name} & \colhead{Other} & \colhead{$z$} & \colhead{Type} & \colhead{\lir} & \colhead{SFR} & \colhead{$v_c$} & \colhead{$W_{eq}$} & \colhead{Run} & \colhead{$t_{exp}$} & \colhead{PA}  & \colhead{Refs} \\
\colhead{(1)} & \colhead{(2)} & \colhead{(3)} & \colhead{(4)} & \colhead{(5)} & \colhead{(6)} & \colhead{(7)} & \colhead{(8)} & \colhead{(9)} & \colhead{(10)} & \colhead{(11)} & \colhead{(12)}
}
\startdata
{\bf Seyfert~2s} & & & & & & & & & & & \\
\tableline
       F01436$+$0120 &        \nodata &  0.2285 &      S2 &   11.83 &      47 & \nodata &    0.58 &MMT/2002dec31 &    3600 &       0 &       2 \\
       Z03150$-$0219 &        \nodata &  0.3191 &      S2 &   12.30 &     138 & \nodata &    2.21 &Keck/2002jan17,MMT/2002dec31 &    9000 &       0 &       2 \\
       F04210$+$0401 &        \nodata &  0.0462 &      S2 &   11.14 &       7 & \nodata &    0.64 &KPNO/2003sep26 &    1800 &       0 &       4 \\
       F05024$-$1941 &        \nodata &  0.1935 &      S2 &   12.47 &     206 & \nodata &    8.87 &Keck/2001feb27 &    1800 &       0 &       1 \\
       F05189$-$2524 &        \nodata &  0.0428 &      S2 &   12.11 &      67 &     194 &    5.40 &Keck/2001feb28 &     900 &       0 &      13 \\
       F08526$+$3720 &        \nodata &  0.3569 &      S2 &   12.36 &     118 & \nodata &    1.24 &Keck/2002jan17 &    1800 &       0 &       2 \\
       F08559$+$1053 &        \nodata &  0.1480 &      S2 &   12.24 &     120 & \nodata &    3.80 &Keck/2001jan24 &    1200 &       0 &       1 \\
       F12072$-$0444 &        \nodata &  0.1287 &      S2 &   12.37 &     163 & \nodata &    1.04 &KPNO/2004apr14 &   10800 &       8 &       1 \\
       F12278$+$3539 &        \nodata &  0.3375 &      S2 &   12.33 &     147 & \nodata &    6.70 &Keck/2002mar15 &    3600 &       0 &       2 \\
       F13305$-$1739 &        \nodata &  0.1486 &      S2 &   12.25 &     122 & \nodata &    0.00 &KPNO/2004apr17 &    3600 &       0 &       1 \\
       F13428$+$5608 &        Mrk.273 &  0.0373 &      S2 &   12.09 &      84 &     399 &    4.43 &KPNO/2004apr15 &    6000 &      30 &       1 \\
     F13443$+$0802:E &        \nodata &  0.1347 &      S2 &   12.20 &     109 & \nodata &    3.02 &KPNO/2004apr13,KPNO/2004apr17 &   14400 &  166,62 &       1 \\
    F13443$+$0802:NE &        \nodata &  0.1348 &       H & \nodata & \nodata & \nodata &    4.18 & \nodata & \nodata & \nodata &       1 \\
    F13443$+$0802:SW &        \nodata &  0.1342 &       L & \nodata & \nodata & \nodata &    1.61 & \nodata & \nodata & \nodata &       1 \\
     F13451$+$1232:W &       4C.12.50 &  0.1218 &      S2 &   12.29 &     101 &     236 &    2.45 &KPNO/2004apr14 &   10800 &     104 &       1 \\
     F13451$+$1232:E &        \nodata &  0.1220 & \nodata & \nodata & \nodata &     206 &    2.35 & \nodata & \nodata & \nodata &       1 \\
     F14394$+$5332:E &        \nodata &  0.1050 &      S2 &   12.05 &      77 & \nodata &    3.57 &KPNO/2004apr15 &    9000 &      93 &       1 \\
     F14394$+$5332:W &        \nodata &  0.1064 & \nodata & \nodata & \nodata & \nodata &    1.90 & \nodata & \nodata & \nodata &       1 \\
       F14478$+$3448 &        \nodata &  0.1602 &      S2 &   11.53 &      23 & \nodata &    2.96 &MMT/2003jun04 &    7200 &     270 &       2 \\
       F14548$+$3349 &        \nodata &  0.4430 &      S2 &   12.66 &     315 & \nodata &    3.52 &Keck/2002feb16 &    3600 &       0 &       2 \\
       F15001$+$1433 &        \nodata &  0.1622 &      S2 &   12.45 &     195 & \nodata &    4.19 &KPNO/2004apr13,KPNO/2004apr16 &   10800 &      55 &       1 \\
       F17179$+$5444 &        \nodata &  0.1471 &      S2 &   12.26 &     124 & \nodata &    2.40 &KPNO/2003sep29 &    7200 &     100 &       1 \\
       F23233$+$2817 &        \nodata &  0.1144 &      S2 &   12.07 &      82 & \nodata &    0.82 &KPNO/2003sep29 &    7200 &      75 &       1 \\
     F23389$+$0300:N &      4C.03.60 &  0.1447 &      S2 &   12.23 &     117 & \nodata &    4.99 &KPNO/2003sep29 &    7200 &      12 &       1 \\
     F23389$+$0300:S &        \nodata &  0.1455 & \nodata & \nodata & \nodata & \nodata &    2.23 & \nodata & \nodata & \nodata &       1 \\
\tableline
{\bf Seyfert~1s} & & & & & & & & & & & \\
\tableline
       F07599$+$6508 &        \nodata &  0.1490 &      S1 &   12.48 &     157 & \nodata &    2.43 &KPNO/2004apr13 &    5400 &     168 &       1 \\
       F11119$+$3257 &        \nodata &  0.1901 &      S1 &   12.61 &     211 & \nodata &    6.85 &KPNO/2004apr13 &    9000 &      95 &      1a \\
       Z11598$-$0114 &        \nodata &  0.1513 &      S1 &   12.51 &     166 & \nodata &    0.00 &KPNO/2004apr16 &    3600 &      12 &       1 \\
       F12265$+$0219 &         3C.273 &  0.1583 &      S1 &   12.75 &     294 & \nodata &    0.00 &KPNO/2004apr16 &    2400 &      40 &       1 \\
       F12540$+$5708 &        Mrk.231 &  0.0422 &      S1 &   12.49 &     161 &     170 &   21.44 &Keck/2001feb28,KPNO/2004apr16 &    5100 &       0 &      13 \\
       F15462$-$0450 &        \nodata &  0.1002 &      S1 &   12.17 &      77 & \nodata &    0.00 &KPNO/2004apr14,KPNO/2004apr15 &    8400 &      30 &       1 \\
\enddata
\tablerefs{(1) \citealt{ks98,vks99b}; (2) \citealt{ssvd00}; (3) \citealt{sm_ea03}; (4) \citealt{k_ea95,v_ea95}; (a) \citealt{hlsa00}.
}
\tablecomments{Col.(1): {\it IRAS} Faint Source Catalog label, plus
  nuclear ID (e.g., N $=$ north).  Col.(2): Another name.  Col.(3):
  Heliocentric redshift.  Col.(4): Optical spectral type
  (\S\ref{spectype}).  Col.(5): Infrared luminosity, in logarithmic
  units of \lsun.  Col.(6): Star formation rate, computed from the
  infrared luminosity using a correction for AGN contribution to \lir\
  (\S\ref{sfr}).  Col.(7): Measured circular velocity, equal to
  $\sqrt{2\sigma^2 + v_{rot}^2}$ (K. Dasyra, private communication).
  Col.(8): Rest-frame equivalent width of \nad\, in \AA, as computed
  from our model fits.  Col.(9): Observing run (\S\ref{obs}).
  Instruments used were: ESI on Keck, the Red Channel Spectrograph on
  the MMT, and the R-C Spectrograph on the KPNO 4m.  Col.(10): Total exposure time in seconds.  Col.(11): Slit position angle.  Col.(12): Reference.  Numbered references are infrared survey references; lettered references are previous superwind surveys.}
\end{deluxetable}

\begin{deluxetable}{lrrlrlrlrlrr}
\rotate
\tablecaption{Outflow Component Properties \label{compprop}}
\tablewidth{0pt}
\tablehead{
\colhead{Name} & \colhead{$\lambda_{1,c}$} & \colhead{$\Delta v$} & & \colhead{$b$} & & \colhead{$\tau_{1,c}$} & & \colhead{\cf} & & \colhead{$N$(\nags)} & \colhead{$N$(H)}\\
\colhead{(1)} & \colhead{(2)} & \colhead{(3)} & & \colhead{(4)} & & \colhead{(5)} & & \colhead{(6)} & & \colhead{(7)} & \colhead{(8)}
}
\startdata
{\bf Seyfert~2s} & & & & & & & & & & & \\
\tableline
       F01436$+$0120 & 7244.37 &     -32 &($\pm$  5) &      67 &($\pm$ 16) &    0.09 &     ($^{+0.04}_{-0.01}$) &    1.00 &($^{+0.00}_{-0.00}$) &   12.32 &   19.61 \\
       Z03150$-$0219 & 7778.40 &     -41 &($\pm$  8) &     338 &($\pm$ 37) &    0.07 &     ($^{+0.03}_{-0.00}$) &    1.00 &($^{+0.00}_{-0.00}$) &   12.90 &   20.20 \\
       F04210$+$0401 & 6170.61 &      18 &($\pm$ 17) &      63 &($\pm$ 45) &    0.32 &   ($^{+\infty}_{-0.14}$) &    0.36 &($^{+0.44}_{-0.02}$) &$>$12.85 &$>$20.19 \\
       F05024$-$1941 & 7002.50 &   -1547 &($\pm$  2) &     146 &($\pm$  7) &    2.00 &     ($^{+0.18}_{-0.16}$) &    0.24 &($^{+0.02}_{-0.02}$) &   14.02 &   21.31 \\
             \nodata & 7030.36 &    -357 &($\pm$ 34) &     398 &($\pm$ 72) &    0.27 &     ($^{+0.20}_{-0.04}$) &    0.19 &($^{+0.15}_{-0.03}$) &   13.58 &   20.88 \\
             \nodata & 7036.97 &     -75 &($\pm$  1) &      38 &($\pm$  4) &    1.93 &     ($^{+0.23}_{-0.12}$) &    0.46 &($^{+0.05}_{-0.03}$) &   13.42 &   20.72 \\
             \nodata & 7041.06 &      99 &($\pm$  6) &     209 &($\pm$ 16) &    0.17 &     ($^{+0.02}_{-0.01}$) &    1.00 &($^{+0.00}_{-0.00}$) &   13.11 &   20.40 \\
       F05189$-$2524 & 6141.44 &    -402 &($\pm$  3) &     351 &($\pm$  7) &    1.21 &     ($^{+0.06}_{-0.05}$) &    0.22 &($^{+0.01}_{-0.01}$) &   14.18 &   21.54 \\
             \nodata & 6147.79 &     -92 &($\pm$  1) &      91 &($\pm$  3) &    2.48 &     ($^{+0.23}_{-0.20}$) &    0.16 &($^{+0.01}_{-0.01}$) &   13.90 &   21.26 \\
       F08526$+$3720 & 7997.42 &    -187 &($\pm$ 20) &     256 &($\pm$ 82) &    0.56 &   ($^{+\infty}_{-0.16}$) &    0.12 &($^{+0.21}_{-0.04}$) &$>$13.71 &$>$21.01 \\
       F08559$+$1053 & 6769.69 &     -31 &($\pm$  7) &     197 &($\pm$ 14) &    0.15 &     ($^{+0.04}_{-0.01}$) &    0.92 &($^{+0.06}_{-0.01}$) &   13.01 &   20.31 \\
             \nodata & 6776.83 &     285 &($\pm$  5) &     126 &($\pm$ 15) &    1.13 &     ($^{+0.24}_{-0.11}$) &    0.17 &($^{+0.04}_{-0.02}$) &   13.70 &   21.00 \\
       F12072$-$0444 & 6658.21 &      74 &($\pm$ 26) &     239 &($\pm$ 68) &    0.04 &     ($^{+0.12}_{-0.01}$) &    0.96 &($^{+0.02}_{-0.00}$) &   12.58 &   19.89 \\
       F12278$+$3539 & 7873.64 &    -546 &($\pm$ 31) &     178 &($\pm$ 35) &    0.07 &     ($^{+0.15}_{-0.01}$) &    1.00 &($^{+0.00}_{-0.00}$) &   12.65 &   19.95 \\
             \nodata & 7887.57 &     -16 &($\pm$ 10) &     246 &($\pm$ 30) &    0.31 &     ($^{+0.06}_{-0.03}$) &    0.72 &($^{+0.14}_{-0.07}$) &   13.44 &   20.74 \\
             \nodata & 7893.11 &     195 &($\pm$  2) &      42 &($\pm$  5) &    5.00 &   ($^{+\infty}_{-0.28}$) &    0.23 &($^{+0.01}_{-0.01}$) &$>$13.88 &$>$21.17 \\
       F13305$-$1739 & \nodata & \nodata &   \nodata & \nodata &   \nodata & \nodata &                  \nodata & \nodata &             \nodata & \nodata & \nodata \\
       F13428$+$5608 & 6118.06 &      29 &($\pm$  3) &     285 &($\pm$ 15) &    0.23 &     ($^{+0.03}_{-0.01}$) &    0.78 &($^{+0.09}_{-0.04}$) &   13.37 &   20.73 \\
     F13443$+$0802:E & 6692.21 &      11 &($\pm$  3) &      89 &($\pm$  9) &    1.28 &     ($^{+0.25}_{-0.18}$) &    0.46 &($^{+0.09}_{-0.06}$) &   13.61 &   20.91 \\
    F13443$+$0802:NE & 6695.18 &     105 &($\pm$  4) &     248 &($\pm$ 19) &    0.30 &     ($^{+0.05}_{-0.03}$) &    0.68 &($^{+0.11}_{-0.06}$) &   13.42 &   20.72 \\
    F13443$+$0802:SW & 6687.87 &     -43 &($\pm$ 14) &      99 &($\pm$ 33) &    1.56 &   ($^{+\infty}_{-0.60}$) &    0.20 &($^{+0.61}_{-0.08}$) &$>$13.74 &$>$21.04 \\
     F13451$+$1232:W & 6616.51 &      28 &($\pm$ 23) &     400 &($\pm$101) &    1.18 &     ($^{+1.05}_{-0.45}$) &    0.11 &($^{+0.10}_{-0.04}$) &   14.22 &   21.52 \\
     F13451$+$1232:E & 6615.59 &     -67 &($\pm$ 13) &     357 &($\pm$ 71) &    0.27 &     ($^{+0.21}_{-0.05}$) &    0.29 &($^{+0.22}_{-0.05}$) &   13.54 &   20.83 \\
     F14394$+$5332:E & 6515.87 &     -43 &($\pm$  3) &     134 &($\pm$  9) &    1.66 &     ($^{+0.21}_{-0.17}$) &    0.34 &($^{+0.04}_{-0.03}$) &   13.90 &   21.20 \\
     F14394$+$5332:W & 6524.51 &     -17 &($\pm$ 18) &     286 &($\pm$ 72) &    0.07 &     ($^{+0.08}_{-0.01}$) &    1.00 &($^{+0.00}_{-0.00}$) &   12.83 &   20.13 \\
       F14478$+$3448 & 6831.25 &    -487 &($\pm$ 14) &     127 &($\pm$ 37) &    0.06 &   ($^{+\infty}_{-0.01}$) &    0.99 &($^{+0.01}_{-0.00}$) &$>$12.41 &$>$19.75 \\
             \nodata & 6841.92 &     -19 &($\pm$ 13) &     228 &($\pm$ 37) &    0.12 &     ($^{+0.05}_{-0.02}$) &    0.85 &($^{+0.09}_{-0.03}$) &   13.00 &   20.34 \\
       F14548$+$3349 & 8504.94 &    -185 &($\pm$  2) &      66 &($\pm$  9) &    0.20 &     ($^{+0.05}_{-0.01}$) &    1.00 &($^{+0.00}_{-0.00}$) &   12.66 &   19.96 \\
             \nodata & 8509.28 &     -32 &($\pm$ 14) &     347 &($\pm$ 68) &    0.37 &     ($^{+0.28}_{-0.08}$) &    0.25 &($^{+0.19}_{-0.05}$) &   13.66 &   20.95 \\
       F15001$+$1433 & 6849.10 &    -220 &($\pm$ 14) &     283 &($\pm$ 43) &    0.46 &     ($^{+0.12}_{-0.07}$) &    0.42 &($^{+0.11}_{-0.06}$) &   13.67 &   20.97 \\
             \nodata & 6857.30 &     138 &($\pm$ 11) &      21 &($\pm$ 19) &    5.00 &   ($^{+\infty}_{-2.48}$) &    0.08 &($^{+0.02}_{-0.04}$) &$>$13.58 &$>$20.87 \\
       F17179$+$5444 & 6763.92 &     -52 &($\pm$ 43) &     232 &($\pm$100) &    0.06 &     ($^{+0.10}_{-0.02}$) &    0.80 &($^{+0.16}_{-0.03}$) &   12.73 &   20.03 \\
             \nodata & 6772.13 &     312 &($\pm$ 29) &     179 &($\pm$ 69) &    0.47 &   ($^{+\infty}_{-0.14}$) &    0.19 &($^{+0.32}_{-0.05}$) &$>$13.48 &$>$20.78 \\
       F23233$+$2817 & 6571.65 &     -24 &($\pm$ 26) &     231 &($\pm$ 77) &    0.47 &   ($^{+\infty}_{-0.11}$) &    0.10 &($^{+0.33}_{-0.02}$) &$>$13.59 &$>$20.91 \\
     F23389$+$0300:N & 6746.30 &    -206 &($\pm$ 11) &     300 &($\pm$ 54) &    0.24 &     ($^{+0.11}_{-0.02}$) &    0.80 &($^{+0.18}_{-0.04}$) &   13.41 &   20.81 \\
     F23389$+$0300:S & 6753.76 &     -84 &($\pm$ 31) &     232 &($\pm$ 84) &    0.10 &   ($^{+\infty}_{-0.02}$) &    0.97 &($^{+0.02}_{-0.00}$) &$>$12.93 &$>$20.33 \\
\tableline
{\bf Seyfert~1s} & & & & & & & & & & & \\
\tableline
       F07599$+$6508 & 6538.21 &  -10718 &($\pm$ 53) &     632 &($\pm$126) &    0.50 &     ($^{+0.65}_{-0.12}$) &    0.11 &($^{+0.15}_{-0.03}$) &   14.05 &   21.35 \\
             \nodata & 6562.50 &   -9608 &($\pm$116) &    1026 &($\pm$199) &    4.09 &     ($^{+4.03}_{-2.12}$) &    0.09 &($^{+0.09}_{-0.05}$) &   15.17 &   22.47 \\
             \nodata & 6575.95 &   -8995 &($\pm$ 95) &     581 &($\pm$196) &    0.39 &     ($^{+1.10}_{-0.13}$) &    0.10 &($^{+0.28}_{-0.03}$) &   13.91 &   21.20 \\
             \nodata & 6687.84 &   -3939 &($\pm$ 10) &      78 &($\pm$ 33) &    0.96 &     ($^{+6.40}_{-0.32}$) &    0.06 &($^{+0.39}_{-0.02}$) &   13.43 &   20.72 \\
       F11119$+$3257 & 6988.12 &   -1309 &($\pm$  3) &     167 &($\pm$  8) &    0.25 &     ($^{+0.05}_{-0.01}$) &    1.00 &($^{+0.00}_{-0.00}$) &   13.17 &   20.47 \\
             \nodata & 6995.27 &   -1002 &($\pm$  1) &      23 &($\pm$  4) &    2.84 &     ($^{+0.28}_{-0.17}$) &    0.71 &($^{+0.07}_{-0.04}$) &   13.37 &   20.67 \\
             \nodata & 7001.29 &    -744 &($\pm$  4) &     121 &($\pm$  9) &    0.17 &     ($^{+0.03}_{-0.02}$) &    1.00 &($^{+0.00}_{-0.00}$) &   12.86 &   20.15 \\
       Z11598$-$0114 & \nodata & \nodata &   \nodata & \nodata &   \nodata & \nodata &                  \nodata & \nodata &             \nodata & \nodata & \nodata \\
       F12265$+$0219 & \nodata & \nodata &   \nodata & \nodata &   \nodata & \nodata &                  \nodata & \nodata &             \nodata & \nodata & \nodata \\
       F12540$+$5708 & 5983.92 &   -8031 &($\pm$  3) &     103 &($\pm$  9) &    0.03 &     ($^{+0.00}_{-0.00}$) &    0.71 &($^{+0.09}_{-0.03}$) &   12.05 &   19.35 \\
             \nodata & 6021.10 &   -6175 &($\pm$  0) &      33 &($\pm$  1) &    0.92 &     ($^{+0.03}_{-0.03}$) &    0.33 &($^{+0.01}_{-0.01}$) &   13.03 &   20.33 \\
             \nodata & 6034.48 &   -5510 &($\pm$  0) &      21 &($\pm$  6) &    0.81 &     ($^{+0.09}_{-0.19}$) &    0.28 &($^{+0.03}_{-0.06}$) &   12.78 &   20.08 \\
             \nodata & 6041.45 &   -5164 &($\pm$  4) &     100 &($\pm$  8) &    0.42 &     ($^{+0.06}_{-0.03}$) &    0.61 &($^{+0.09}_{-0.05}$) &   13.18 &   20.47 \\
             \nodata & 6044.39 &   -5019 &($\pm$ 11) &     934 &($\pm$ 45) &    0.26 &     ($^{+0.03}_{-0.02}$) &    0.17 &($^{+0.02}_{-0.01}$) &   13.94 &   21.23 \\
             \nodata & 6045.34 &   -4972 &($\pm$  2) &      79 &($\pm$  8) &    0.68 &     ($^{+0.06}_{-0.08}$) &    0.84 &($^{+0.07}_{-0.10}$) &   13.29 &   20.58 \\
             \nodata & 6049.06 &   -4787 &($\pm$ 21) &      80 &($\pm$ 30) &    0.42 &     ($^{+0.50}_{-0.15}$) &    1.00 &($^{+0.00}_{-0.00}$) &   13.08 &   20.38 \\
             \nodata & 6050.89 &   -4696 &($\pm$ 23) &      94 &($\pm$ 22) &    0.70 &     ($^{+0.18}_{-0.46}$) &    1.00 &($^{+0.00}_{-0.00}$) &   13.37 &   20.67 \\
             \nodata & 6052.66 &   -4609 &($\pm$  9) &      38 &($\pm$ 34) &    1.32 &     ($^{+0.28}_{-0.48}$) &    0.64 &($^{+0.14}_{-0.23}$) &   13.25 &   20.55 \\
             \nodata & 6054.16 &   -4534 &($\pm$  8) &      27 &($\pm$ 20) &    1.30 &     ($^{+0.13}_{-0.64}$) &    0.84 &($^{+0.08}_{-0.42}$) &   13.09 &   20.39 \\
             \nodata & 6056.61 &   -4413 &($\pm$  1) &      45 &($\pm$  3) &    3.04 &     ($^{+0.18}_{-0.18}$) &    0.74 &($^{+0.04}_{-0.04}$) &   13.69 &   20.98 \\
             \nodata & 6060.23 &   -4234 &($\pm$  0) &      69 &($\pm$  2) &    2.72 &     ($^{+0.08}_{-0.10}$) &    0.69 &($^{+0.02}_{-0.02}$) &   13.83 &   21.12 \\
             \nodata & 6138.53 &    -386 &($\pm$ 10) &     149 &($\pm$ 28) &    0.14 &     ($^{+0.07}_{-0.01}$) &    0.08 &($^{+0.04}_{-0.01}$) &   12.86 &   20.16 \\
       F15462$-$0450 & \nodata & \nodata &   \nodata & \nodata &   \nodata & \nodata &                  \nodata & \nodata &             \nodata & \nodata & \nodata \\
\enddata
\tablecomments{Col.(2): Redshifted heliocentric wavelength, in vacuum, of the \nad$_1$ $\lambda5896$ line; in \AA.  Col.(3): Velocity relative to systemic, in \kms; $\Delta v \equiv v_{comp} - v_{sys}$.  Negative velocities are blueshifted.  Components with $\Delta v < -50$~\kms\ and $|\Delta v| > 2~\delta(\Delta v)$ are assumed to be outflowing.  1$\sigma$ errors are listed in parentheses; these only include measurement uncertainties in $v_{comp}$ (not $v_{sys}$).  Col.(4): Doppler parameter, in \kms.  1$\sigma$ measurement uncertainties are listed in parentheses.  Col.(5): Central optical depth of the \nad$_1$ $\lambda5896$ line; the optical depth of the D$_2$ line is twice this value.  1$\sigma$ measurement uncertainties are listed in parentheses.  Col.(6): Covering fraction of the gas.  1$\sigma$ measurement uncertainties are listed in parentheses.  Col.(7-8): Logarithm of column density of \nags\ and H, respectively, in cm$^{-2}$.}
\end{deluxetable}

\begin{deluxetable}{crrrrrrrr}
\tablecaption{Outflow Properties \label{ofprop}}
\tablewidth{0pt}
\tablehead{
\colhead{Name} & \colhead{\dvtau} & \colhead{\dvmax} & \colhead{$M$} & \colhead{$dM/dt$} & \colhead{$p$} & \colhead{$dp/dt$} & \colhead{$E$} & \colhead{$dE/dt$} \\
\colhead{(1)} & \colhead{(2)} & \colhead{(3)} & \colhead{(4)} & \colhead{(5)} & \colhead{(6)} & \colhead{(7)} & \colhead{(8)} & \colhead{(9)}
}
\startdata
{\bf Seyfert~2s} & & & & & & & & \\
\tableline
       F05024$-$1941 &   -1547 &   -1669 &    9.19 &    2.47 &   50.46
       &   36.43 &   58.33 &   44.32 \\
       F05189$-$2524 &    -402 &    -694 &    9.26 &    2.07 &   50.06
       &   35.45 &   57.67 &   43.08 \\
       F08526$+$3720 &    -187 &    -400 & $>$8.35 & $>$0.93 &$>$48.92
       &$>$34.00 &$>$56.47 &$>$41.55 \\
       F12278$+$3539 &    -546 &    -694 &    8.20 &    1.24 &   49.23
       &   34.78 &   56.73 &   42.28 \\
       F13451$+$1232 &     -67 &    -364 &    8.75 &    0.88 &   48.87
       &   33.51 &   57.04 &   41.67 \\
       F14478$+$3448 &    -487 &    -592 & $>$7.99 & $>$0.98 &$>$48.97
       &$>$34.47 &$>$56.40 &$>$41.90 \\
       F14548$+$3349 &    -185 &    -240 &    8.21 &    0.78 &   48.77
       &   33.85 &   55.81 &   40.89 \\
       F15001$+$1433 &    -220 &    -456 &    8.83 &    1.49 &   49.48
       &   34.63 &   57.06 &   42.21 \\
       F23389$+$0300 &    -206 &    -456 & $>$9.18 & $>$1.69 &$>$49.68
       &$>$34.75 &$>$57.33 &$>$42.39 \\
\tableline
{\bf Seyfert~1s} & & & & & & & & \\
\tableline
       F07599$+$6508 &   -9608 &  -11244 &    4.34 &    1.33 &   46.62
       &   36.12 &   55.31 &   44.81 \\
       F11119$+$3257 &   -1002 &   -1448 &    3.73 &   -0.23 &   45.06
       &   33.62 &   52.81 &   41.39 \\
       F12540$+$5708 &   -5019 &   -8117 &    4.41 &    1.09 &   46.38
       &   35.56 &   54.75 &   43.93 \\

\enddata
\tablecomments{Col.(2): Velocity of the highest column density gas in the outflow, \dvtau, in \kms.  Col.(3): Maximum velocity in the outflow, $\dvmax \equiv \Delta v - \mathrm{FWHM}/2$, in \kms.  Col.(4): Log of mass, in \msun.  Col.(5): Log of mass outflow rate, in \smpy.  Col.(6): Log of momentum, in dyne~s.  Col.(7): Log of momentum outflow rate, in dyne.  Col.(8): Log of total kinetic energy, in erg.  Col.(9): Log of energy outflow rate, in erg~s$^{-1}$.}
\end{deluxetable}

\begin{deluxetable}{ccccc}
\tablecaption{Subsample Average Properties \label{avgprop}}
\tablewidth{0pt}
\tablehead{
\colhead{Quantity} & \colhead{IRGs}  & \colhead{low-$z$ ULIRGs} & \colhead{Seyfert~2s} & \colhead{Seyfert~1s} \\
\colhead{(1)} & \colhead{(2)} & \colhead{(3)} & \colhead{(4)} & \colhead{(5)}
}
\startdata
Number of Galaxies & 35 & 30 & 20 & 6 \\
Detection Rate (\%) & 42$\pm$8 & 80$\pm$7 & 45$\pm$11 & 50$\pm$20 \\
\tableline
Galaxy Properties & & & \\
\tableline
$ z $ & $ 0.031^{+0.04}_{-0.02} $ & $ 0.129^{+0.07}_{-0.04} $ & $ 0.148^{+0.14}_{-0.07} $ & $ 0.150^{+0.11}_{-0.06} $ \\
$ \mathrm{log} [\lir/\lsun] $ & $11.36\pm0.4$ & $12.21\pm0.2$ & $12.24\pm0.3$ & $12.50\pm0.2$ \\
$ \mathrm{SFR}  \: (\smpy) $ & $40^{+55}_{-23}$ & $225^{+95}_{-67}$ & $118^{+151}_{-66}$ & $164^{+91}_{-59}$ \\
$ \Delta v_{max}  \: (\kms) $ & $301^{+145}_{-98}$ & $408^{+224}_{-144}$ & $456^{+330}_{-191}$ & $8110^{+16291}_{-5414}$ \\
$ \Delta v_{maxN}  \: (\kms) $ & $104^{+80}_{-45}$ & $167^{+122}_{-70}$ & $220^{+316}_{-130}$ & $5023^{+11068}_{-3455}$ \\
$ \mathrm{log} [N$(\ion{Na}{1})/cm$^{-2}$]$  \: $ & $13.8\pm0.2$ & $13.8\pm0.4$ & $13.5\pm0.7$ & $14.5\pm0.8$ \\
$ \mathrm{log} [N$(H)/cm$^{-2}$]$  \: $ & $21.2\pm0.3$ & $21.2\pm0.4$ & $20.9\pm0.7$ & $21.8\pm0.8$ \\
$ \mathrm{log} [M/\msun]  \: $ & $8.8\pm0.2$ & $9.1\pm0.3$ & $8.8\pm0.5$ & $4.3\pm0.4$ \\
$ dM/dt  \: (\smpy) $ & $17^{+20}_{-9}$ & $42^{+82}_{-28}$ & $18^{+50}_{-13}$ & $12^{+73}_{-10}$ \\
$ \mathrm{log} [p/\mathrm{dyne\;s}]  \:$ & $49.2\pm0.3$ & $49.6\pm0.5$ & $49.2\pm0.6$ & $46.4\pm0.8$ \\
$ \mathrm{log} [dp/dt/\mathrm{dyne}]  \:$ & $34.1\pm0.5$ & $34.7\pm0.7$ & $34.6\pm0.9$ & $35.6\pm1.3$ \\
$ \mathrm{log} [E/\mathrm{erg}]  \:$ & $56.7\pm0.4$ & $57.2\pm0.5$ & $57.0\pm0.7$ & $54.8\pm1.3$ \\
$ \mathrm{log} [dE/dt/\mathrm{erg\;s^{-1}}]  \:$ & $41.6\pm0.6$ & $42.2\pm0.7$ & $42.2\pm1.0$ & $43.9\pm1.8$ \\
\tableline
Component Properties & & & \\
\tableline
$ \tau $ & $1.06^{+1.4}_{-0.6}$ & $0.85^{+2.4}_{-0.6}$ & $0.27^{+0.7}_{-0.2}$ & $0.69^{+1.6}_{-0.5}$ \\
$ b  \: (\kms) $ & $152^{+109}_{-64}$ & $196^{+170}_{-91}$ & $232^{+244}_{-119}$ & $87^{+199}_{-61}$ \\
$ C_f $ & $0.37^{+0.2}_{-0.1}$ & $0.40^{+0.5}_{-0.2}$ & $0.42^{+0.5}_{-0.2}$ & $0.67^{+1.2}_{-0.4}$ \\
\enddata
\tablecomments{For most quantities we list the median and 1$\sigma$ dispersions, under the assumption of a Gaussian distribution in the log of the quantity.  For the detection rate errors, we assume a binomial distribution.  Statistics for all quantities except $z$, \lir, and SFR are computed only for galaxies or velocity components with outflows.}
\end{deluxetable}

\begin{deluxetable}{clll}
\tablecaption{Average Velocities by Spectral Type \label{avgprop_hls}}
\tablewidth{0pt}
\tablehead{\colhead{Quantity} & \colhead{\ion{H}{2}} & \colhead{LINER} & \colhead{Seyfert~2}}
\startdata
\dvmax (\kms) & $-267\pm133$ & $-393\pm194$ & $-524\pm439$ \\
\dvtau (\kms) & $-119\pm95$  & $-229\pm135$ & $-311\pm457$ \\
\enddata
\tablecomments{For each quantity, we list the median and 1$\sigma$ dispersions.}
\end{deluxetable}

\begin{deluxetable}{crr}
\tablecaption{Comparing Velocity Distributions of Seyfert~2s and Starbursts \label{pnull}}
\tablewidth{0pt}
\tablehead{\colhead{Samples} & \colhead{P(null,K-S)} & \colhead{P(null,Kuiper)}}
\startdata
\dvmax & & \\
\tableline
IRGs vs. SB ULIRGs              &   0.11 &   0.28 \\
SB ULIRGs vs. Sy2 ULIRGs        &   0.25 &   0.75 \\
Sy2 ULIRGs vs. IRGs             &   0.04 &   0.21 \\
\tableline
{\bf \ion{H}{2} vs. LINER}      &{\bf$<$0.01}&{\bf 0.04}\\
LINER vs. Sy2                   &   0.14 &   0.45 \\
Sy2 vs. \ion{H}{2}              &   0.05 &   0.25 \\
\tableline
\dvtau & & \\
\tableline
{\bf IRGs vs. SB ULIRGs}                &{\bf 0.01}&{\bf 0.09}\\
SB ULIRGs vs. Sy2 ULIRGs        &   0.13 &   0.16 \\
{\bf Sy2 ULIRGs vs. IRGs}               &{\bf 0.01}&{\bf 0.09}\\
\tableline
\ion{H}{2} vs. LINER            &   0.07 &   0.27 \\
LINER vs. Sy2                   &   0.14 &   0.55 \\
{\bf Sy2 vs. \ion{H}{2}}                &{\bf$<$0.01}&{\bf  0.01}\\
\tableline
{\it Doppler parameter} & & \\
\tableline
IRGs vs. SB ULIRGs              &   0.19 &   0.55 \\
SB ULIRGs vs. Sy2 ULIRGs        &   0.52 &   0.59 \\
Sy2 ULIRGs vs. IRGs             &   0.14 &   0.33 \\
\tableline
{\bf \ion{H}{2} vs. LINER}      &{\bf0.02}&{\bf$<$0.01}\\
LINER vs. Sy2                   &   0.99 &   0.99 \\
Sy2 vs. \ion{H}{2}              &   0.13 &   0.06 \\
\enddata
\tablecomments{P(null) is the probability that the two observed distributions arise from the same intrinsic distribution.  Comparisons in which both values of P(null) are less than 0.10 are printed in bold face.}
\end{deluxetable}

\begin{deluxetable}{ccrrcc}
\tablecaption{Emission- and Absorption-Line correlations \label{emlcorr}}
\tablewidth{0pt}
\tablehead{
\colhead{Axes} & \colhead{Sample} & \colhead{$N$} & \colhead{$a\pm\delta a$} & \colhead{$P(r_p=0)$} & \colhead{$P(r_s=0)$} \\
\colhead{(1)} & \colhead{(2)} & \colhead{(3)} & \colhead{(4)} & \colhead{(5)} & \colhead{(6)}
}
\startdata
FWHM vs. \dvtau & all galaxies                  & 38 & $-0.56\pm0.07$ & 0.00 & 0.01 \\
FWHM vs. \dvtau & all galaxies with BELA        & 16 & $-0.62\pm0.07$ & 0.00 & 0.01 \\
FWHM vs. \dvtau & Seyfert~2s with BELA          &  7 & $-0.64\pm0.07$ & 0.00 & 0.01 \\
\enddata
\tablecomments{Col.(1): Variables to be fit.  The line widths are of the \otl\ line.  Col.(2): Subsample of galaxies in fit; full sample consists of all galaxies with neutral gas outflows and low-to-moderate resolution \ot\ linewidths.  BELA $=$ blue emission-line asymmetry (\S\ref{emlprop}).  Col. (3): Number of nuclei in fit.  Col.(4): Slope and 1$\sigma$ error; $X = (a\pm\delta a)Y + Y_0$.  Col.(5): Probability that Pearson's (parametric) correlation coefficient is zero.  Col.(6): Probability that Spearman's (non-parametric) correlation coefficient is zero.}
\end{deluxetable}


\clearpage

\begin{thebibliography}{}
\bibitem[Baum et~al.(1993)]{b_ea93} Baum, S.~A., O'Dea, C.~P., Dallacassa, D., de Bruyn, A.~G., \& Pedlar, A.  1993, \apj, 419, 553 
\bibitem[Becker et~al(1995)Becker, White, \& Helfand]{bwh95} Becker, R.~H., White, R.~L., \& Helfand, D.~J.  1995, \apj, 450, 559
\bibitem[Bicknell et~al.(2000)]{b_ea00} Bicknell, G.~V., Sutherland, R.~S., van Breugel, W.~J.~M., Dopita, M.~A., Dey, A., \& Miley, G.~K.  2000, \apj, 540, 678 
\bibitem[Boksenberg et~al.(1977)]{b_ea77} Boksenberg, A., Carswell, R.~F., Allen, D.~A., Fosbury, R.~A.~E., Penston, M.~V., \& Sargent, W.~L.~W.  1977, \mnras, 178, 451
\bibitem[Boroson et~al.(1991)]{b_ea91} Boroson, T.~A., Meyers, K.~A., Morris, S.~L., \& Persson, S.~E.  1991, \apj, 370, L19
\bibitem[Boroson \& Meyers(1992)]{bm92} Boroson, T.~A., \& Meyers, K.~A.  1992, \apj, 397, 442
\bibitem[Bryant \& Scoville(1996)]{bs96} Bryant, P.~M., \& Scoville, N.~Z.  1996, \apj, 457, 678 
\bibitem[Carilli \& Taylor(2000)]{ct00} Carilli, C.~L., \& Taylor, G.~B.  2000, \apj, 532, 95 
\bibitem[Carilli et~al.(1998)Carilli, Wrobel, \& Ulvestad]{cwu98} Carilli, C. L., Wrobel, J. M., \& Ulvestad, J. S. 1998, \aj, 115, 928 
\bibitem[Cecil et~al.(1990)Cecil, Bland, \& Tully]{cbt90} Cecil, G., Bland, J., \& Tully, R.~B.  1990, \apj, 355, 70 
\bibitem[Colina et~al.(1999)Colina, Arribas, \& Borne]{cab99} Colina, L., Arribas, S., \& Borne, K.~D.  1999, \apj, 527, 13L 
\bibitem[Crenshaw \& Kraemer(2005)]{ck05} Crenshaw, D.~M., \& Kraemer, S.~B.  2005, \apj, 625, 680
\bibitem[Crenshaw et~al.(2003)Crenshaw, Kraemer, \& George]{ckg03} Crenshaw, D.~M., Kraemer, S.~B.,  \& George, I.~M.  2003, \araa, 41, 117 
\bibitem[Colbert et~al.(1996a)]{c_ea96a} Colbert, E.~J.~M., Baum, S.~A., Gallimore, J.~F., O'Dea, C.~P., \& Christensen, J.~A.  1996a, \apj, 467, 551 
\bibitem[Colbert et~al.(1996b)]{c_ea96b} Colbert, E.~J.~M., Baum, S.~A., Gallimore, J.~F., O'Dea, C.~P., Lehnert, M.~D., Tsvetanov, Z.~I., Mulchaey, J.~S., \& Caganoff, S.  1996b, \apjs, 105, 75 
\bibitem[Colbert et~al.(1998)]{cb_ea98} Colbert, E.~J.~M., Baum, S.~A., O'Dea, C.~P., \& Veilleux, S.  1998, \apj, 496, 786 
\bibitem[Condon et~al.(1998)]{cc_ea98} Condon, J.~J., Cotton, W.~D., Greisen, E.~W., Yin, Q.~F., Perley, R.~A., Taylor, G.~B., \& Broderick, J.~J.  1998, \aj, 115, 1693
\bibitem[de Kool et~al.(2001)]{d_ea01} de Kool, M., Arav, N., Becker, R.~H., Gregg, M.~D., White, R.~L., Laurent-Muehleisen, S.~A., Price, T., \& Korista, K.~T.  2001, \apj, 548, 609 
\bibitem[De Robertis \& Osterbrock(1984)]{do84} De Robertis, M.~M., \& Osterbrock, D.~E.  1984, \apj, 286, 171 
\bibitem[De Robertis \& Osterbrock(1986)]{do86} De Robertis, M.~M., \& Osterbrock, D.~E.  1986, \apj, 301, 727 
\bibitem[Downes \& Solomon(1998)]{ds98} Downes, D., \& Solomon, P.~M.  1998, \apj, 507, 615
\bibitem[Fan et~al.(2004)]{f_ea04} Fan, X., et~al.  2004, \aj, 128, 515 
\bibitem[Farrah et~al.(2005)]{f_ea05} Farrah, D., Surace, J.~A., Veilleux, S., Sanders, D.~B., \& Vacca, W.~D.  2005, \apj, 626, 70
\bibitem[Forster et~al.(1995)Forster, Rich, \& McCarthy]{frm95} Forster, K., Rich, R.~M., \& McCarthy, J.~K.  1995, \apj, 450, 74
\bibitem[Fragile et~al.(2004)]{fm_ea04} Fragile, P.~C., Murray, S.~D., Anninos, P., \& van Breugel, W.  2004, \apj, 604, 74
\bibitem[Gallagher et~al.(2001)]{gb_ea01} Gallagher, S.~C., Brandt, W.~N., Laor, A., Elvis, M., Mathur, S., Wills, B.~J., \& Iyomoto, N.  2001, \apj, 546, 795 
\bibitem[Gallagher et~al.(2005)]{g_ea05} Gallagher, S.~C., Schmidt, G.~D., Smith, P.~S., Brandt, W.~N., Chartas, G., Hylton, S., Hines, D.~C., \& Brotherton, M.~S.  2005, \apj, in press (astro-ph/0506616) 
\bibitem[Garc\'{i}a-Lorenzo et~al.(2001)Garc\'{i}a-Lorenzo, Arribas, \& Mediavilla]{gam01} Garc\'{i}a-Lorenzo, B., Arribas, S., \& Mediavilla, E. 2001, \aap, 378, 787 
\bibitem[Genzel et~al.(1998)]{g_ea98} Genzel, R., et~al.  1998, \apj, 498, 579
\bibitem[Gonz\'{a}lez Delgado et~al.(1998a)]{gd_ea98a} Gonz\'{a}lez Delgado, R.~M., Leitherer, C., Heckman, T., Lowenthal, J.~D., Ferguson, H.~C., \& Carmelle, R.  1998a, \apj, 495, 698 
\bibitem[Gonz\'{a}lez Delgado et~al.(1998b)]{gd_ea98b} Gonz\'{a}lez Delgado, R.~M., Heckman, T., Leitherer, C., Meurer, G., Krolik, J., Wilson, A.~S., Kinney, A., \& Koratkar, A.  1998b, \apj, 505, 174 
\bibitem[Hamilton \& Keel(1987)]{hk87} Hamilton, D., \& Keel, W.~C.  1987, \apj, 321, 211
\bibitem[Hamann et~al.(2001)]{hb_ea01} Hamann, F.~W., Barlow, T.~A., Chaffee, F.~C., Foltz, C.~B., \& Weymann, R.~J.  2001, \apj, 550, 142 
\bibitem[Heckman et~al.(2000)]{hlsa00} Heckman, T.~M., Lehnert, M.~D., Strickland, D.~K., \& Armus, L.  2000, \apjs, 129, 493
\bibitem[Hines \& Wills(1995)]{hw95} Hines, D.~C., \& Wills, B.~J.  1995, \apj, 448, L69 
\bibitem[Holt et~al.(2003)Holt, Tadhunter, \& Morganti]{htm03} Holt, J., Tadhunter, C.~N., \& Morganti, R.  2003, \mnras, 342, 227 
\bibitem[Iwasawa et~al.(2003)]{i_ea03} Iwasawa, K., Wilson, A.~S., Fabian, A.~C., \& Young, A.~J.  2003, \mnras, 345, 369 
\bibitem[Kauffmann et~al.(2003)]{k_ea03} Kauffmann, G., et~al.  2003, \mnras, 341, 33
\bibitem[Kennicutt(1998)]{k98} Kennicutt, R. C. 1998, \apj, 498, 541
\bibitem[Kim \& Sanders(1998)]{ks98} Kim, D.-C., \& Sanders, D.~B.  1998, \apjs, 119, 41
\bibitem[Kim et~al.(1995)]{k_ea95} Kim, D.-C., Sanders, D.~B., Veilleux, S., Mazzarella, J.~M., \& Soifer, B.~T.  1995, \apjs, 98, 129
\bibitem[Kim et~al.(1998)Kim, Veilleux, \& Sanders]{kvs98} Kim, D.-C., Veilleux, S., \& Sanders, D.~B.  1998, \apj, 484, 92
\bibitem[Kl\"{o}ckner \& Baan(2004)]{kb04} Kl\"{o}ckner, H.-R., \& Baan, W.~A.  2004, \aap, 419, 887 
\bibitem[Lanzetta et~al.(1993)Lanzetta, Turnshek, \& Sandoval]{lts93} Lanzetta, K.~M., Turnshek, D.~A, \& Sandoval, J.  1993, \apjs, 84, 109 
\bibitem[Laureijs et~al.(2000)]{l_ea00} Laureijs, R.~J., et~al. 2000, \aap, 359, 900
\bibitem[Leitherer et~al.(1999)]{l_ea99} Leitherer, C., et~al.  1999, \apjs, 123, 3 
\bibitem[Levenson et~al.(2001a)Levenson, Weaver, \& Heckman]{lwh01a} Levenson, N.~A., Weaver, K.~A., \& Heckman, T.~M.  2001a, \apjs, 133, 269
\bibitem[Levenson et~al.(2001b)Levenson, Weaver, \& Heckman]{lwh01b} Levenson, N.~A., Weaver, K.~A., \& Heckman, T.~M.  2001b, \apj, 550, 230
\bibitem[Lindblad(1999)]{l99} Lindblad, P.~O.  1999, \aapr, 9, 221
\bibitem[L\'{i}pari(1994)]{l94} L\'{i}pari, S.  1994, \apj, 436, 102 
\bibitem[Lutz et~al.(1999)Lutz, Veilleux, \& Genzel]{lvg99} Lutz, D., Veilleux, S., \& Genzel, R.  1999, \apj, 517, L13
\bibitem[Martin(2005)]{m05} Martin, C.~L. 2005, \apj, 621, 227
\bibitem[McKernan et~al.(2005)McKernan, Yaqoob, \& Reynolds]{myr05} McKernan, B., Yaqoob, T., \& Reynolds, C.~S.  2005, \apj, in press
\bibitem[Morganti et~al.(2001)]{m_ea01} Morganti, R., Oosterloo, T.~A., Tadhunter, C.~N., van Moorsel, G., Killeen, N., \& Wills, K.~A.  2001, \mnras, 323, 331 
\bibitem[Morganti et~al.(2004a)]{m_ea04a} Morganti, R., Oosterloo, T.~A., Emonts, B.~H.~C., Tadhunter, C.~N., Holt, J.  2004a, in IAU 217, Recycling Intergalactic and Interstellar Matter, eds. P.~A. Duc, J. Braine, \& E. Brinks (IAU) 
\bibitem[Morganti et~al.(2003)]{m_ea03} Morganti, R., Oosterloo, T.~A., Emonts, B.~H.~C., van der Hulst, J.~M., \& Tadhunter, C.~N.  2003, \apj, 593, 69 
\bibitem[Morganti et~al.(2005)]{m_ea05} Morganti, R., Oosterloo, T.~A., Tadhunter, C.~N., van~Moorsel, G., \& Emonts, B.  2005, \aap, in press (astro-ph/0505365)
\bibitem[Morganti et~al.(2004b)]{m_ea04b} Morganti, R., Oosterloo, T.~A., Tadhunter, C.~N., Vermeulen, R., Pihlstr\"{o}m, Y.~M., van Moorsel, G., \& Wills, K.~A.  2004b, \aap, 424, 119 
\bibitem[Morganti et~al.(1998)]{mot98} Morganti, R., Oosterloo, T.~A., \& Tsvetanov, Z.  1998, \aj, 115, 915 
\bibitem[Morganti et~al.(1999)]{m_ea99} Morganti, R., Tsvetanov, Z.~I., Gallimore, J., \& Allen, M.~G.  1999, \aaps, 137, 457 
\bibitem[Mould et~al.(2000)]{m_ea00}  Mould, J.~R., et~al.  2000, \apj, 536, 266 
\bibitem[Murray et~al.(2005)Murray, Quartaert, \& Thompson]{mqt05}  Murray, N., Quartaert, E., \& Thompson, T.~A.  2005, \apj, 618, 569
\bibitem[Nagar et~al.(2005)Nagar, Falcke, \& Wilson]{nfw05} Nagar, N.~M., Falcke, H., \& Wilson, A.~S.  2005, \aap, in press
\bibitem[Nagar et~al.(2003)]{n_ea03} Nagar, N.~M., Wilson, A.~S., Falcke, H., Veilleux, S., \& Maiolino, R.  2003, \aap, 409, 115
\bibitem[Nelson \& Whittle(1996)]{nw96} Nelson, C.~H., \& Whittle, M.  1996, \apj, 465, 96
\bibitem[Osterloo et~al.(2000)]{o_ea00} Osterloo, T.~A., Morganti, R., Tzioumis, A., Reynolds, J., King, E., McCulloch, P., \& Tsvetanov, Z.  2000, \aj, 119, 2085 
\bibitem[Phillips(1993)]{p93} Phillips, A.~C. 1993, \aj, 105, 486
\bibitem[Rejkuba et~al.(2002)]{r_ea02} Rejkuba, M., Minniti, D., Courbin, F., \& Silva, D.~R.  2002, \apj, 564, 688 
\bibitem[Reynolds et~al.(2002)Reynolds, Heinz, \& Begelman]{rhb02} Reynolds, C.~S., Heinz, S., \& Begelman, M.~C.  2002, \mnras, 332, 271 
\bibitem[Roychowdhury et~al.(2004)]{r_ea04} Roychowdhury, S., Ruszkowski, M., Nath, B.~B., \& Begelman, M.~C.  2004, \apj, 615, 681 
\bibitem[Rupke et~al.(2002)Rupke, Veilleux, \& Sanders]{rvs02} Rupke, D.~S., Veilleux, S., \& Sanders, D.~B.  2002, \apj, 570, 588
\bibitem[Rupke et~al.(2005a)Rupke, Veilleux, \& Sanders]{rvs05a}
  Rupke, D.~S., Veilleux, S., \& Sanders, D.~B.  2005a, \apjs, in press
\bibitem[Rupke et~al.(2005b)Rupke, Veilleux, \& Sanders]{rvs05b}
  Rupke, D.~S., Veilleux, S., \& Sanders, D.~B.  2005b, \apjs, in press
\bibitem[Salzer et~al.(2005)]{s_ea05}  Salzer, J.~J., Lee, J.~C., Melbourne, J., Hinz, J.~L., Alonso-Herrero, A., \& Jangren, A. 2005, \apj, 624, 661 
\bibitem[Sanders et~al.(2003)]{sm_ea03} Sanders, D.~B., Mazzarella, J.~M., Kim, D.-C., Surace, J.~A., \& Soifer, B.~T.  2003, \aj, 126, 1607
\bibitem[Sanders et~al.(1991)Sanders, Scoville, \& Soifer]{sss91} Sanders, D. B., Scoville, N. Z., \& Soifer, B. T. 1991, \apj, 370, 158 
\bibitem[Sanders et~al.(1988)]{s_ea88} Sanders, D.~B., Soifer, B.~T., Elias, J.~H., Madore, B.~F., Matthews, K., Neugebauer, G., \& Scoville, N.~Z.  1988, \apj, 325, 74
\bibitem[Schiano(1985)]{s85} Schiano, A.~V.~R.  1985, \apj, 299, 24 
\bibitem[Schwartz \& Martin(2004)]{sm04} Schwartz, C.~M., \& Martin, C.~L.  2004, \apj, 610, 201
\bibitem[Sheinis et~al.(2002)]{s_ea02} Sheinis, A.~I., Bolte, M., Epps, H.~W., Kibrick, R.~I., Miller, J.~S., Radovan, M.~V., Bigelow, B.~C., \& Sutin, B.~M.  2002, \pasp, 114, 851 
\bibitem[Springel et~al.(2005)Springel, Di Matteo, \& Hernquist]{sdh05} Springel, V., Di Matteo, T., \& Hernquist, L.  2005, \apj, 520, L79 
\bibitem[Stanford et~al.(2000)]{ssvd00} Stanford, S.~A., Stern, D., van Breugel, W., \& De Breuck, C.  2000, \apjs, 131, 185
\bibitem[Stanghellini et~al.(1997)]{s_ea97} Stanghellini, C., O'Dea, C.~P., Baum, S.~A., Dallacasa, D., Fanti, R., \& Fanti, C.  1997, \aap, 325, 943 
\bibitem[Steffen et~al.(1996a)Steffen, Holloway, \& Pedlar]{shp96a}  Steffen, W., Holloway, A.~J., \& Pedlar, A.  1996a, \mnras, 282, 130 
\bibitem[Steffen et~al.(1996b)Steffen, Holloway, \& Pedlar]{shp96b}  Steffen, W., Holloway, A.~J., \& Pedlar, A.  1996b, \mnras, 282, 1203 
\bibitem[Surace et~al.(1998)]{s_ea98} Surace, J.~A., Sanders, D.~B., Vacca, W.~D., Veilleux, S., \& Mazzarella, J.~M.  1998, \apj, 492, 116 
\bibitem[Tadhunter et~al.(2005)]{t_ea05} Tadhunter, C., Robinson, T.~G., Gonz\'{a}lez Delgado, R.~M., Wills, K., \& Morganti, R.  2005, \mnras, 356, 480 
\bibitem[Taylor et~al.(1999)]{t_ea99} Taylor, G.~B., Silver, C.~S., Ulvestad, J.~S., \& Carilli, C.~L.  1999, \apj, 519, 185 
\bibitem[Ulvestad et~al.(1999)Ulvestad, Wrobel, \& Carilli]{uwc99} Ulvestad, J.~S., Wrobel, J.~M., \& Carilli, C.~L.  1999, \apj, 516, 127 
\bibitem[Veilleux(1991a)]{v91a} Veilleux, S.  1991a, \apjs, 75, 357
\bibitem[Veilleux(1991b)]{v91b} Veilleux, S.  1991b, \apjs, 75, 383
\bibitem[Veilleux(1991c)]{v91c} Veilleux, S.  1991c, \apj, 369, 331
\bibitem[Veilleux et~al.(1999a)]{v_ea99a} Veilleux, S., Bland-Hawthorn, J., Cecil, G., Tully, R.~B., \& Miller, S.~T.  1999a, \apj, 520, 111 
\bibitem[Veilleux et~al.(2005)Veilleux, Cecil, \& Bland-Hawthorn]{vcb05} Veilleux, S., Cecil, G., \& Bland-Hawthorn, J.  2005, \araa, in press (astro-ph/0504435)
\bibitem[Veilleux et~al.(1999b)Veilleux, Kim, \& Sanders]{vks99b} Veilleux, S., Kim, D.-C., \& Sanders, D.~B.  1999b, \apj, 522, 113
\bibitem[Veilleux et~al.(2002)Veilleux, Kim, \& Sanders]{vks02} Veilleux, S., Kim, D.-C., \& Sanders, D.~B.  2002, \apj, 143, 315
\bibitem[Veilleux et~al.(1995)]{v_ea95} Veilleux, S., Kim, D.-C., Sanders, D.~B., Mazzarella, J.~M., \& Soifer, B.~T.  1995, \apjs, 98, 171
\bibitem[Veilleux et~al.(1999c)Veilleux, Sanders, \& Kim]{vsk99c} Veilleux, S., Sanders, D.~B., \& Kim, D.-C.  1999c, \apj, 522, 139
\bibitem[Veilleux et~al.(2001)Veilleux, Shopbell, \& Miller]{vsm01} Veilleux, S., Shopbell, P.~L., \& Miller, S.~T.  2001, \aj, 121, 198 
\bibitem[Veilleux et~al.(2003)]{v_ea03} Veilleux, S., Shopbell, P.~L., Rupke, D.~S., Bland-Hawthorn, J., \& Cecil, G.  2003, \aj, 26, 2185
\bibitem[Weiner et~al.(2005)]{w_ea05} Weiner, B.~J., et~al.  2005, \apj, 620, 595 
\bibitem[Weymann et~al.(1991)]{w_ea91} Weymann, R.~J., Morris, S.~L., Foltz, C.~B., \& Hewett, P.~C.  1991, \apj, 373, 23 
\bibitem[Whittle(1992a)]{w92a} Whittle, M.  1992a, \apjs, 79, 49
\bibitem[Whittle(1992b)]{w92b} Whittle, M.  1992b, \apj, 387, 109
\bibitem[Whittle(1992c)]{w92c} Whittle, M.  1992c, \apj, 387, 121
\bibitem[Willott et~al.(2003)Willott, Rawlings, \& Grimes]{wrg03} Willott, C.~J., Rawlings, S., \& Grimes, J.~A.  2003, \apj, 598, 909
\bibitem[Worrall \& Birkinshaw(2004)]{wb04} Worrall, D.~M., \& Birkinshaw, M.  2004, in Lecture Notes in Physics, Physics of Active Galactic Nuclei at all Scales, eds. D. Alloin, R. Johnson, \& P. Lira (Springer Verlag), astro-ph/0410297
\bibitem[Zensus(1997)]{z97} Zensus, J.~A.  1997, \araa, 35, 607
\end{thebibliography}
\end{document}